\pdfoutput=1
\documentclass[a4paper,11pt]{article}

\usepackage{epsfig}
\usepackage{graphicx}
\usepackage{url}
\usepackage{float}
\usepackage{amsmath}
\usepackage[bf, sf, FIGTOPCAP, nooneline, tight]{subfigure}
\usepackage{graphicx}
\usepackage{dcolumn}
\usepackage{bm}
\usepackage{latexsym}
\usepackage{amssymb}
\usepackage{jcappub} 

\usepackage[T1]{fontenc} 

\newcommand{\be}{\begin{equation}}
\newcommand{\ee}{\end{equation}}
\newcommand{\bq}{\begin{eqnarray}}
\newcommand{\eq}{\end{eqnarray}}

\def\({\left(}
\def\){\right)}

\title{Comparing dark energy models with current observational data}

\author[a]{Sixiang Wen,}
\author[a,1]{Shuang Wang,\note{Corresponding author.}}
\author[a]{Xiaolin Luo}

\affiliation[a]{School of Physics and Astronomy, Sun Yat-Sen University, Guangzhou 510297, P. R. China}

\emailAdd{wensx@mail2.sysu.edu.cn}
\emailAdd{wangshuang@mail.sysu.edu.cn}
\emailAdd{luoxl23@mail2.sysu.edu.cn}

\abstract{
We make a comparison for {\bf thirteen} dark energy (DE) models by using current cosmological observations, including type Ia supernova, baryon acoustic oscillations, and cosmic microwave background. To perform a systematic and comprehensive analysis, we consider three statistics methods of SNIa, including magnitude statistic (MS), flux statistic (FS), and improved flux statistic (IFS), as well as two kinds of BAO data. In addition, Akaike information criteria (AIC) and Bayesian information criteria (BIC) are used to assess the worth of each model. We find that: (1) The thirteen models can be divided into four grades by performing cosmology-fits. The cosmological constant model, which is most favored by current observations, belongs to grade one;  $\alpha$DE, constant $w$ and generalized Chaplygin gas models belong to grade two; Chevalliear-Polarski-Linder (CPL) parametrization, Wang parametrization, doubly coupled massive gravity, new generalized Chaplygin gas and holographic DE models belong to grade three; {\bf agegraphic DE, Dvali-Gabadadze-Porrati, Vacuum metamorphosis and Ricci DE models, which are excluded by current observations, belong to grade four.} (2) For parameter estimation, adopting IFS yields the biggest $\Omega_m$ and the smallest $h$ for all the models. In contrast, using different BAO data does not cause significant effects. (3) IFS has the strongest constraint ability on various DE models. For examples, adopting IFS yields the smallest value of $\Delta$AIC for all the models; in addition, making use of this technique yields the biggest figure of merit for CPL and Wang parametrizations.
}

\keywords{Cosmology: dark energy, observations, cosmological parameters}

\begin{document}
\maketitle
\flushbottom

\section{Introduction}\label{sec:intro}

Since the discovery of cosmic acceleration \cite{Riess:1998cb,Perlmutter1999},
dark energy (DE) has become one of the most important issues in the modern cosmology~\cite{Sahni2000,Padmanabhan03,Sahni2006,Frieman2008,Li2011,Bamba2012,Li2013}.
In order to explore the nature of DE,
vast amounts of theoretical models are proposed.
In this paper, we only focus on five classes of DE models:
\begin{itemize}
\item Cosmological constant model.

The cosmological constant model \cite{Einstein1917} (also called $\Lambda$CDM), is the simplest DE model.
For this model, the equation-of-state (EOS) always satisfies $w=-1$.
EOS is defined as $w \equiv p/\rho$, where $p$ and $\rho$ are the pressure and the energy density of DE.

\item DE models with parameterized EOS.

The simplest way of studying the dynamical DE is to consider the parameterized EOS.
There are many parameterization models, including $w$CDM ($w$ = const.),
Chevalliear-Polarski-Linder (CPL)  parameterization
\cite{CPL,Linder:2002et}, Wang  parameterization~\citep{YunWang2008}, e.g.
\footnote{For other popular parametrizations, see, \cite{Huterer:2000mj,Wetterich:2004pv,Jassal:2004ej,Linder:2006sv,Barboza2008,Li:2012via}.}

\item Chaplygin gas models.

Chaplygin gas models describe a background fluid with
$p\propto\rho^{-\xi}$ that is commonly viewed as arising from the
d-brane theory.
Chaplygin gas has several versions, including old Chaplygin gas model \cite{cg}, generalized Chaplygin gas (GCG) model \cite{gcg},
and new  generalized  Chaplygin gas models (NGCG) \cite{ngcg}.

\item Holographic dark energy models.

Holographic dark energy paradigm arises from a theoretical attempt of applying the holographic principle \cite{t Hooft1993,Susskind1995} to the DE problem.
Based on the holographic principle and the dimensional analysis, the energy density of holographic DE can be written as \cite{wwl2017} $\rho_{\rm{de}}=3C^2M_{pl}^2L^{-2}$,
where $C$ is a constant parameter, $M_{pl}$ is the reduced
Planck mass, $L$ is the infrared cutoff length scale. Different choices of the $L$ lead to different holographic DE models,
such as original holographic DE (HDE) \cite{hde}, new agegraphic DE (ADE) model \cite{ade} and  Ricci DE
(RDE) model \cite{rde}.

\item  Modified gravity models.

The key idea of modified gravity (MG) theory is to modify the Einstein's tensor $G_{\mu \nu}$ in the left hand side of the Einstein's field equation \cite{Rham2014,Nojiri2017}.
Some popular MG models include Dvali-Gabadadze-Porrati (DGP) model
\cite{DGP}, DGP's phenomenological extension (namely,
the $\alpha$DE model)\cite{alphaDE} and double coupled massive gravity (CMG) \cite{Heisenberg2016a}. {\bf In addition, the vacuum metamorphosis (VM) \cite{Parker2000,Parker2004,Caldwell2006}, which has a phase transition in the nature of the vacuum, is also considered in this work.} For convenience, in this paper we treat these MG models as DE models.

\end{itemize}

Since there are so many DE candidates, it is
crucial to find out which one is more favored by the observational data.
A lot of research works have been done to test the DE models against the observational data \cite{Davis:2007na,WangZhang2008,WangZhangXia2008,Lim2010,wll2010,Zhang2012,WLZL2016,zhangxin2016}.
By using the $\chi^2$ statistics
and adopting the information criterion, such as
Akaike information criterion (AIC) \cite{AIC} and Bayesian information criterion (BIC) \cite{BIC},
one can find out that which model is more favored by cosmological observations.

Recently, Xu and Zhang \cite{zhangxin2016} made a comparison for ten DE models, by using the  JLA sample of type Ia supernovae (SNIa),
the Planck 2015 distance priors of cosmic microwave background observation (CMB), the baryon acoustic oscillations measurements (BAO),
and the direct measurement of the Hubble constant.
In this paper, we will perform a more systematic and more comprehensive analysis
by considering the following two factors that are ignored in Ref. \cite{zhangxin2016}.

First, the effect of different statistics methods of SNIa data.
As is well known, people always calculate the $\chi^2$ function of SNIa data
by comparing the observed values and theoretical values of the distance modulus $\mu$.
In this paper we call this statistics method as ``magnitude statistic'' (MS).
Unfortunately, MS always suffers from lots of systematic uncertainties of SNIa \citep{Mohlabeng2014,WangWang2013,Hu2016}.
In 2000, Wang\cite{Wang2000} proposed a ``flux-averaged'' (FA) technique,
whose key idea is to average the observed flux of SNIa at a series of uniformly divided redshift bins.
Here we call the statistic method based on the FA technique as ``flux statistic'' (FS).
FS can reduce the systematic uncertainties of SNIa \citep{WangTegmark05,YunWang2009,Wang12CM}, but is at the cost of giving worse DE constraints.
In 2013, Wang and Wang \cite{WangWang2013} developed an improved FA technique, which introduces a new quantity: the redshift cut-off $z_{cut}$.
The key idea of the improved FA technique is as follows: for the SN samples at $z < z_{cut}$,  the MS technique is used to compute $\chi^2$ function;
for the SN samples at $z \geq z_{cut}$, the FS technique is used to compute $\chi^2$ function.
Therefore, this new method can reduce systematic uncertainties and give tighter DE constraints at the same time \cite{Wang2016}.
In this paper we call the statistic method based on the improved FA technique as  ``improved flux statistics'' (IFS).
In most research works (e.g. \cite{zhangxin2016}), people only make use of the MS technique to analyse the SNIa data. In order to present a more comprehensive analysis, in this paper we take into account all the three SNIa analysis technique.

Second, the effect of different BAO measurements. It should be mentioned that Xu andi Zhang \cite{zhangxin2016} only used two old BAO data extracted from the  SDSS DR7 and the BOSS DR11. But in Ref.\cite{Alam2016}, the BAO data of BOSS DR12 was released. More importantly, in Ref. \cite{Alam2016},a different formula was used to estimate the theoretical value of the sound horizon $r_s(z_d)$. So it is interesting to compare the effect of adopting different BAO measurements.

In this work, we make a comparison of {\bf thirteen} DE  models.
It needs to point out that, compared with the Ref. \cite{zhangxin2016} that only considered ten DE models,
in this paper we take into account two new models, Wang parametrization and CMG model.
We perform a combined constraint, by using current cosmological observations, including the JLA SNIa data, the BAO measurements and the CMB distance priors derived from the 2015 Planck data \cite{Planck201514}.
For JLA data, we take into account all the three analysis technique of SNIa (i.e. MS, FS, IFS).
For the BAO measurement, we consider previous BAO measurements and current BAO measurements.
In Section~\ref{sec:method}, we discuss the methodology.  In Section \ref{sec:model}, we describe {\bf thirteen}
DE models, and give the correspondent fitting results. Tn Section~\ref{sec:concl},
we summarize the results of model comparison, and discuss the effect of adopting different statistic methods of SNIa.
In Section~\ref{sec:conc2}, we  discuss the conclusions of this work, and briefly describe the related future works.

\section{Methodology}\label{sec:method}

In the following, we will introduce how to assess different DE models by using various selection criteria.
It is important to stress that, when assessing different DE models one can not only use the $\chi^2$ statistics,
because different DE models have different parameter numbers.
Therefore, in this paper, we assess these models by employing the information criteria (IC)
that can include the effect of the model parameter numbers.
The most widely adopted selection criteria are AIC \cite{AIC} and BIC \cite{BIC}.

AIC \cite{AIC} is defined as
\begin{equation}
{\rm AIC}=-2\ln{\cal L}_{max}+2k,
\end{equation}
where ${\cal L}_{max}$ is the maximum likelihood, $k$ is the number
of parameters. For Gaussian errors, $\chi_{min}^2=-2\ln{\cal L}_{max}$.
Note that only the relative value of AIC between different models is important, and the difference in AIC can be written as $\Delta{\rm AIC}=\Delta\chi_{min}^2+2\Delta k$.
Generally speaking, the models with $\Delta {\rm AIC}=5$ and $\Delta {\rm AIC}=10$ are considered as
strong and very strong evidence against the weaker model, respectively.
As mentioned in  Ref.\cite{Liddle2007}, there is a version of
$AIC$ corrected for small number of data points $N$,
\begin{equation}
AIC_c = AIC + \frac{2k(k-1)}{(N-k-1)},
\end{equation}
which is important for $N-k<40$. In this paper, for the case of FS, we use this formula.

BIC, also known as the Schwarz information criterion \cite{BIC},
is given by
\begin{equation}
{\rm BIC}=-2\ln{\cal L}_{max}+k\ln N,
\end{equation}
where $N$ is the number of data points used in the fit.
Similarly, only the relative value of BIC between different models is important,
and the difference in BIC can be simplified to $\Delta{\rm BIC}=\Delta\chi_{min}^2+\Delta k\ln N$.
The models with $\Delta {\rm BIC}\geq2$ and $\Delta {\rm BIC}\geq6$ are considered as positive evidence and strong evidence against the weaker model, respectively.

In addition, figure of merit (FoM) is designed to assess the ability of constraining DE of an experiment project.
It is firstly defined to be the inverse of the area enclose by the $95\%$ confidence level (C.L.) contour of $(w_0,w_a)$ of CPL parametrization \citep{Albrecht}.
In this work, we adopt a relative generalized FoM \citep{YunWang2008} given by
\begin{equation}
FoM=\frac{1}{\sqrt{det~Cov(f_1,f_2,f_3,\cdots)}},
\end{equation}
where $Cov(f_1,f_2,f_3,\cdots)$ is the covariance matrix of the chosen set of DE parameters.
It is clear that larger FoM indicates better accuracy.

In the following, we describe
how to calculate the $\chi^2$ functions of SNIa, BAO and CMB.

\subsection{Type Ia supernovae}

In this section, firstly, we introduce how to calculate the $\chi^2$ function of JLA data by using the usual ``magnitude statistics''.
Then, we introduce how to use the ``flux statistics''  to deal with the JLA data.

\subsubsection{Magnitude statistics}
Theoretically, the distance modulus $\mbox{\bf $\mu$}_{th}$ in a flat universe can be written as
\be
  \mbox{\bf $\mu$}_{th} = 5 \log_{10}\bigg[\frac{d_L(z_{hel},z_{cmb})}{Mpc}\bigg] + 25,
\ee
where $z_{cmb}$ and $z_{hel}$ are the CMB restframe and heliocentric redshifts of SN.
The luminosity distance ${d}_L$ is given by
\be
  {d}_L(z_{hel},z_{cmb}) = (1+z_{hel}) r(z_{cmb}),
\ee

Note that $r(z)$ is given by
\be \label{eq:rz}
r(z)= cH_0^{-1}\int_0^z\frac{dz'}{E(z')},
\ee
where $c$ is the speed of light, $H_0$ is the present-day value of the Hubble parameter $H(z)$, and $E(z)\equiv H(z)/ H_0$.

The observation of distance modulus $\mbox{\bf $\mu$}_{obs}$ is given by an
empirical linear relation:
\be
  \mbox{\bf $\mu$}_{obs}= m_{B}^{\star} - M_B + \alpha_0 \times X_1
  -\beta_0 \times {\cal C},
\ee
where $m_B^{\star}$ is the observed peak magnitude in the rest-frame
\text{of the} $B$ band,
$X_1$ describes the time stretching of light-curve, and ${\cal C}$ describes the
supernova color at maximum brightness.
Note that $\alpha_0$ and $\beta_0$ are SN stretch-luminosity parameter and SN color-luminosity parameter, respectively
\footnote{Recent research works show that there is a strong evidence for the evolution of $\beta_0$ \citep{WLZ2014,WWGZ2014,WWZ2014,WGHZ2015,li2016}. For simplicity, in this paper we just treat $\beta_0$ as a constant parameter}.
In addition, $M_B$ is the absolute B-band magnitude,
which relates to the host stellar mass $M_{stellar}$ via a simple step function \citep{Betoule2014}
\begin{equation}
  \label{eq:mabs}
    M_B = \left\lbrace
   \begin{array}{ll}
    M^1_B &\quad \text{if}\quad  M_{stellar} < 10^{10} M_{\odot}\,,\\
    M^2_B &\quad \text{otherwise.}
    \end{array}
    \right.
\end{equation}
where $M_{\odot}$ is the mass of sun.

The $\chi^2$ of JLA data can be calculated as
\be
\label{eq:chi2_SN}
\chi^2_{SNIa} = \Delta \mbox{\bf $\mu$}^T \cdot \mbox{\bf Cov}^{-1} \cdot \Delta\mbox{\bf $\mu$},
\ee
where $\Delta \mbox{\bf $\mu$}\equiv \mbox{\bf $\mu$}_{obs}-\mbox{\bf $\mu$}_{th}$
is the data vector and $\mbox{\bf Cov}$ is the total covariance matrix, which can be calculated as
\be
\mbox{\bf Cov}=\mbox{\bf D}_{\rm stat}+\mbox{\bf C}_{\rm stat}
+\mbox{\bf C}_{\rm sys}.
\ee
Here $\mbox{\bf D}_{\rm stat}$ is the diagonal part of the statistical
uncertainty, which is given by \citep{Betoule2014},
\begin{eqnarray}
\mbox{\bf D}_{\rm stat,ii}&=&\left[\frac{5}{z_i \ln 10}\right]^2 \sigma^2_{z,i}+
  \sigma^2_{\rm int} +\sigma^2_{\rm lensing} + \sigma^2_{m_B,i} +\alpha_0^2 \sigma^2_{X_1,i}+\beta_0^2 \sigma^2_{{\cal C},i}\nonumber\\
&&+ 2 \alpha_0 C_{m_B X_1,i} - 2 \beta_0 C_{m_B {\cal C},i}  -2\alpha_0\beta_0 C_{X_1 {\cal C},i},
\end{eqnarray}
where the first three terms account for the uncertainty in redshift due to peculiar velocities,
the intrinsic variation in SN magnitude and the variation of magnitudes caused by gravitational lensing.
$\sigma^2_{m_B,i}$, $\sigma^2_{X_1,i}$, and $\sigma^2_{{\cal C},i}$
denote the uncertainties of $m_B$, $X_1$ and ${\cal C}$ for the $i$-th SN.
In addition, $C_{m_B X_1,i}$, $C_{m_B {\cal C},i}$ and $C_{X_1 {\cal C},i}$
are the covariances between $m_B$, $X_1$ and ${\cal C}$ for the $i$-th SN.
Moreover, $\mbox{\bf C}_{\rm stat}$ and $\mbox{\bf C}_{\rm sys}$
are the statistical and the systematic covariance matrices, given by
\begin{equation}
\mbox{\bf C}_{\rm stat}+\mbox{\bf C}_{\rm sys}=V_0+\alpha_0^2 V_a + \beta_0^2 V_b +
2 \alpha_0 V_{0a} -2 \beta_0 V_{0b} - 2 \alpha_0\beta_0 V_{ab},
\end{equation}
where $V_0$, $V_{a}$, $V_{b}$, $V_{0a}$, $V_{0b}$ and $V_{ab}$ are six matrices given  in Ref. \cite{Betoule2014}.
The reader can refer to the original JLA paper \cite{Betoule2014}, as well as their publicly released code, for more details of
calculating JLA data's $\chi^2$ function.
This process of analysing SN data is the so-called ``magnitude statistics''.

\subsubsection{Flux statistics}

Now, let us turn to the ``flux statistics'' of JLA data.
Flux statistics based on the FA technique, which is very useful to reduce the systematic uncertainties of SNIa \citep{Wang2004,WangTegmark05,YunWang2009}.
The original FA method divides the whole redshift region of SNIa into a lot of bins,
where the redshift interval of each bin is $\Delta z$.
Then, the specific steps of FA are as follows \citep{Wang12CM}:

(1) Convert the distance modulus of SNIa into ``fluxes'',
\be
\label{eq:flux}
F(z_l) \equiv 10^{-(\mu_0^{\rm obs}(z_l)-25)/2.5} =
\left( \frac{d_L^{\rm obs}(z_l)} {\mbox{Mpc}} \right)^{-2}.
\ee
Here $z_l$ represent the CMB restframe redshift of SN.

(2) For a given set of cosmological parameters $\{ {\bf s} \}$,
obtain ``absolute luminosities'', \{${\cal L}(z_l)$\},
\be
\label{eq:lum}
{\cal L}(z_l) \equiv d_L^2(z_l |{\bf s})\,F(z_l).
\ee

(3) Flux-average the ``absolute luminosities'' \{${\cal L}^i_l$\}
in each redshift bin $i$ to obtain $\left\{\overline{\cal L}^i\right\}$:
\be
 \overline{\cal L}^i = \frac{1}{N_i}
 \sum_{l=1}^{N_i} {\cal L}^i_l(z^{(i)}_l),
 \hskip 1cm
 \overline{z_i} = \frac{1}{N_i}
 \sum_{l=1}^{N_i} z^{(i)}_l.
\ee

(4) Place $\overline{\cal L}^i$ at the mean redshift $\overline{z}_i$ of
the $i$-th redshift bin, now the binned flux is
\be
\overline{F}(\overline{z}_i) = \overline{\cal L}^i /
d_L^2(\overline{z}_i|\mbox{\bf s}).
\ee
with the corresponding flux-averaged distance modulus:
\be
\overline\mu^{obs}(\overline{z}_i) =-2.5\log_{10}\overline{F}(\overline{z}_i)+25.
\ee

(5) Compute the covariance matrix of $\overline{\mu}(\overline{z}_i)$
and $\overline{\mu}(\overline{z}_j)$:
\begin{equation}
\begin{aligned}
\mbox{Cov}&\left[\overline{\mu}(\overline{z}_i),\overline{\mu}(\overline{z}_j)\right]
=\frac{1}{N_i N_j \overline{\cal L}^i \overline{\cal L}^j}\\
 &\sum_{l=1}^{N_i} \sum_{m=1}^{N_j} {\cal L}(z_l^{(i)})
{\cal L}(z_m^{(j)}) \langle \Delta \mu_0^{\rm obs}(z_l^{(i)})\Delta
\mu_0^{\rm obs}(z_m^{(j)})
\rangle
\end{aligned}
\end{equation}
where $\langle \Delta \mu_0^{\rm obs}(z_l^{(i)})\Delta \mu_0^{\rm obs}(z_m^{(j)})\rangle $
is the covariance of the measured distance moduli of the $l$-th SNIa
in the $i$-th redshift bin, and the $m$-th SNIa in the $j$-th
redshift bin. ${\cal L}(z)$ is defined by Eqs.(\ref{eq:flux}) and (\ref{eq:lum}).

(6) For the flux-averaged data, $\left\{\overline{\mu}(\overline{z}_i)\right\}$,
compute
\be
\label{eq:chi2_SN_fluxavg}
\chi^2_{SNIa} = \sum_{ij} \Delta\overline{\mu}(\overline{z}_i) \,
\mbox{Cov}^{-1}\left[\overline{\mu}(\overline{z}_i),\overline{\mu}(\overline{z}_j)
\right] \,\Delta\overline{\mu}(\overline{z}_j)
\ee
where
\be
\Delta\overline{\mu}(\overline{z}_i) \equiv
\overline{\mu}^{obs}(\overline{z}_i) - \mu^p(\overline{z}_i|\mbox{\bf s}),
\ee
and
\be
\overline\mu^p(\overline{z}_i) =-2.5\log_{10} F^p(\overline{z}_i)+25.
\ee
with $F^p(\overline{z}_i|\mbox{\bf s})=
\left( d_L(\overline{z}_i|\mbox{\bf s}) /\mbox{Mpc} \right)^{-2}$.
This process of analysing SN data is the so-called ``flux statistics''.
\subsubsection{Improved flux statistics}

As mentioned above, the improved FA method  \citep{WangWang2013} introduces a new quantity: the redshift cut-off $z_{cut}$.
For the SN samples at $z < z_{cut}$, the $\chi^2$ is computed by using the usual ``magnitude statistics'' (i.e., Eq. \ref{eq:chi2_SN});
for the SN samples at $z \geq z_{cut}$, the $\chi^2$ is computed by using the ``flux statistics'' (i.e., Eq. \ref{eq:chi2_SN_fluxavg}).
This new method includes the advantages of MS and FS, and thus can reduce systematic uncertainties and give tighter DE constraints at the same time.

In Ref. \citep{Wang2015}, Wang and Dai applied this improved FA method to explore the JLA data,
and found that it can give tighter constraints on DE.
But in Ref. \citep{Wang2015}, only one kind of FA recipe, $(z_{cut} = 0.5, \Delta z=0.04)$, was considered.
In a recent paper \citep{Wang2016}, we scanned the whole $(z_{cut}, \Delta z)$ plane,
and found that adopting the FA recipe, $(z_{cut} = 0.6, \Delta z=0.06)$, yielded the tightest DE constraints.
So in this paper, we will use the IFS technique with the best FA recipe $(z_{cut} = 0.6, \Delta z=0.06)$.

The details of these three statistic methods of SNIa  are listed in Table \ref{SN}.

\begin{table*}
\caption{\label{SN} Summary of statistics method of SNIa data}\centering
\begin{tabular*}{\textwidth}{@{}l*{15}{@{\extracolsep{0pt plus12pt}}l}}
\hline\hline
Statistics method    & Abbreviation   & FA recipe &   Number of SNIa samples \\
\hline
Magnitude statistics  &   MS    &N/A &  740 \\
Flux statistics     &  FS    & $z_{cut} = 0.0, \Delta z=0.06$ &  21 \\
Improved flux statistics    &   IFS    & $z_{cut} = 0.6, \Delta z=0.06$ &   606 \\
\hline
\end{tabular*}
\end{table*}

\subsection{Baryon acoustic oscillations}

The BAO matter clustering provides a "standard ruler" for length scale in cosmology.
The BAO signals can be used to measure the Hubble parameter $H(z)$ and angular diameter distance
$D_A(z)=r(z)/(1+z)$ in the radial and tangential directions, respectively.

\subsubsection{Current BAO measurement}

In this paper, the so-called ``current BAO measurement'' refers to the BAO data of BOSS DR12 \cite{Alam2016},
which includes the combinations
$H(z)r_s(z_d)/r_{s,fid}$ and $D_M(z)r_{s,fid}/r_s(z_d)$.
Here $r_{s,fid} = 147.78$Mpc is the sound horizon of the fiducial model, and $D_M(z)=(1+z)D_A(z)$ is the comoving angular diameter distance.
$r_s(z_d)$ is the sound horizon at the drag epoch $z_d$, defined by
\be
\label{eq:rd}
r_s(z_d)=\int_{z_d}^{\infty}\frac{c_s(z)}{H(z)}dz,
\ee
where $c_s(z)= 3^{-1/2}c[1+\frac{3}{4}\rho_{b}(z)/\rho_{r}(z)]^{-1/2}$ is the sound speed in the photon-baryon fluid.
In Ref.\cite{Alam2016}, $r_s(z_d)$ is approximated by \cite{Aubourg2015},
\begin{equation}
r_s(z_d) =
 \frac{55.154exp[-72.3(\omega_v+0.0006)^2]}{\omega_{b}^{0.12807}\omega_{cb}^{0.25351}}Mpc,
\label{eq:zd}
\end{equation}
where $\omega_v = 0.0107(\sum m_v/1.0$eV) is the density parameter of neutrinos;
$\omega_{b} = \Omega_bh^2$ is the density parameter of baryons, and $\omega_{cb}=\Omega_mh^2 - \omega_v $ is the density parameters of baryons and (cold) dark matter.
Following the process of Ref. \cite{Aubourg2015}, we set $\sum m_v=0.06$ for all the models we considered.

There are 6 BAO data points given in Table $7$ of Ref. \cite{Alam2016}:
\bq
& p_{1}=D_M(0.38)r_{s,fid}/r_s(z_d), &p_{1}^{data}=  1512,\nonumber\\
& p_{2}= H(0.38)r_s(z_d)/r_{s,fid}, &p_{2}^{data}= 81.2,\nonumber\\
& p_{3}=D_M(0.51)r_{s,fid}/r_s(z_d), &p_{3}^{data}= 1975, \nonumber\\
& p_{4}=H(0.51)r_s(z_d)/r_{s,fid}, &p_{4}^{data}= 90.9,\nonumber\\
& p_{5}=D_M(0.61)r_{s,fid}/r_s(z_d), &p_{5}^{data}=2307,\nonumber\\
& p_{6}=H(0.61)r_s(z_d)/r_{s,fid}, &p_{6}^{data}= 99.0.
\eq
Therefore, the $\chi^2$ function for current BAO data can be expressed as
\be
\label{eq:chi2bao}
\chi^2_{BAO}=\Delta p_{i} \left[ {\rm Cov}^{-1}_{BAO}(p_{i},p_{j})\right]
\Delta p_{j},
\hskip .5cm
\Delta p_{icur}= p_{i} - p_{i}^{data}.
\ee
The covariance matrix ${\rm Cov}_{BAO}$ can be taken from the on-line files of Ref. \cite{Alam2016}.

\subsubsection{Previous BAO measurement}
In addition to current BAO measurements, we also use previous BAO measurements, which include the individual measurements
of $H(z)r_s(z_d)/c$ and $D_A(z)/r_s(z_d)$ from the two-dimensional two-point correlation
function measured at z=0.35 \cite{CW12} and z=0.57 \cite{Anderson14}.
For previous BAO data, we stress that $r_s(z)$ is calculated as follows
\be~\label{eq:rs}
r_s(z) = cH_0^{-1}\int_{0}^{a}\frac{da^{\prime}}{\sqrt{3(1+\overline{R_b}a^\prime){a^\prime}^4E^2(z^\prime)}},
\ee
where $\overline{R_b}=31500\Omega_{b}h^2(T_{cmb}/2.7K)^{-4}$, and $\Omega_{b}$ is the present fractional density of baryon.
Here the $z_d$ is approximated by \cite{EisenHu98}
\begin{equation}
z_d  =
 \frac{1291(\Omega_mh^2)^{0.251}}{1+0.659(\Omega_mh^2)^{0.828}}
\left[1+b_1(\Omega_bh^2)^{b2}\right],
\label{eq:zd}
\end{equation}
where
\begin{eqnarray}
  b_1 &= &0.313(\Omega_mh^2)^{-0.419}\left[1+0.607(\Omega_mh^2)^{0.674}\right],\\
  b_2 &= &0.238(\Omega_mh^2)^{0.223}.
\end{eqnarray}

There are 2 data points extracted from SDSS-DR7 \cite{CW12}
\bq
&\hat p_1=D_A(0.35)/r_s(z_d),& \hat p_1^{data}= 6.60, \nonumber\\
&\hat p_2= H(0.35)r_s(z_d)/c,& \hat p_2^{data}= 0.0433.
\eq

The $\chi^2_{0.35}$ function of these BAO data can be expressed as
\be
\label{eq:chi2bao}
\chi^2_{0.35}=\Delta \hat p_i \left[ {\rm Cov}^{-1}_{0.35}(\hat p_i,\hat p_j)\right]
\Delta \hat p_j,
\hskip .5cm
\Delta \hat p_i= \hat p_i - \hat p_i^{data},
\ee
where the covariance matrix ${\rm Cov}^{-1}_{0.35}$ is given in Ref. \cite{CW12}

There are 2 data points extracted from BOSS-DR11 ~\cite{Wang2015}
\bq
&\tilde p_1=D_A(0.57)/r_s(z_d), &\tilde p_1^{data}= 9.27, \nonumber\\
&\tilde p_2=H(0.57)r_s(z_d)/c, &\tilde p_2^{data}= 0.04947.
\eq
The $\chi^2_{0.57}$ function of these BAO data can be expressed as
\be
\label{eq:chi2bao}
\chi^2_{0.57}=\Delta \tilde p_i \left[ {\rm Cov}^{-1}_{0.57}(\tilde p_i,\tilde p_j)\right]
\Delta \tilde p_j,
\hskip .5cm
\Delta \tilde p_i= \tilde p_i - \tilde p_i^{data},
\ee
where the covariance matrix ${\rm Cov}^{-1}_{0.57}$ is given in Ref. ~\cite{Wang2015}

Thus, the total $\chi^2_{BAO}$ function of the two BAO data can be expressed as
\begin{equation}
\label{eq:chi2bao}
\chi^2_{BAO}=\chi^2_{0.35}+\chi^2_{0.57}.
\end{equation}

{\bf In fact, BAO is affected by the assumption of the $\Lambda CDM$ model. However, the BOSS Collaboration had shown that BAO distance priors are relative robust against the assumptions of fiducial cosmologies or other kinds of systematics(see Ref.\citep{SAlam2017}). Therefore, the effect of directly using BAO data is very small.}

\subsection{Cosmic microwave background }

CMB gives us the comoving distance to the photon-decoupling surface $r(z_*)$ and the comoving sound horizon at photon-decoupling epoch $r_s(z_*)$.
In this
paper, we use the distance priors data extracted from Planck 2015~\citep{Planck201514}.
This includes the ``shift parameter'' $R$, the
``acoustic scale'' $l_A$, and the
redshift of the decoupling epoch of photons $z_*$.

The shift parameter $R$ is given by ~\citep{Wang2007}:
\be
R \equiv \sqrt{\Omega_{m} H_0^2} \,r(z_*)/c,
\ee
where $r(z_*)$ is the comoving distance given in \ref{eq:rz}. $z_*$ is the redshift of the photon
decoupling epoch estimated by \cite{Hu:1995en}:
\begin{equation}
\label{zstareq} z_*=1048[1+0.00124(\Omega_b
h^2)^{-0.738}][1+g_1(\Omega_m h^2)^{g_2}],
\end{equation}
here
\begin{equation}
g_1=\frac{0.0783(\Omega_b h^2)^{-0.238}}{1+39.5(\Omega_b
h^2)^{0.763}},\quad g_2=\frac{0.560}{1+21.1(\Omega_b h^2)^{1.81}}.
\end{equation}

The acoustic scale $l_A$ is defined as
\begin{equation}
\label{ladefeq} l_A\equiv \pi r(z_*)/r_s(z_*),
\end{equation}
where $r_s(z_*)$ is the comoving sound
horizon at $z_*$. The  $r_s(z)$ is given by \ref{eq:rs}.
These two distance priors, together with $\omega_b\equiv\Omega_bh^2$, provide an efficient summary of CMB data.

The $\chi^2$ function for the CMB distance prior data can be expressed as
\be
\label{eq:chi2CMB}
\chi^2_{CMB}=\Delta q_i \left[ \mbox{Cov}^{-1}_{CMB}(q_i,q_j)\right]
\Delta q_j,
\hskip .2cm
\Delta q_i= q_i - q_i^{data},
\ee
where $q_1=R(z_*)$, $q_2=l_a(z_*)$, and $q_3= \omega_b$.
The covariance matrix for $(q_1, q_2, q_3)$ is given by
\be
\mbox{Cov}_{CMB}(q_i,q_j)=\sigma(q_i)\, \sigma(q_j) \,\mbox{NormCov}_{CMB}(q_i,q_j),
\label{eq:CMB_cov}
\ee
where $\sigma(q_i)$ is the 1$\sigma$ error of observed quantity $q_i$,
$\mbox{NormCov}_{CMB}(q_i, q_j)$ is the corresponding normalized covariance matrix, which are listed in Table $4$ of Ref.~\cite{Planck201514}.

The Planck 2015 data are
\bq
&& q_{1}^{data} = 1.7382\pm0.0088, \nonumber\\
&& q_{2}^{data} = 301.63\pm0.15, \nonumber\\
&& q_{3}^{data} = 0.02262\pm0.00029.
\label{eq:CMB_mean_planck}
\eq

{\bf Using distance priors data of CMB has the issue of assuming $\Lambda$CDM model. So far, the most rigorous method of utilizing the CMB data is to adopt Markov Chain Monte Carlo global fit technique. However, some previous studies (see Ref.\citep{HLi2012,YHLi2013})  has indicated that the difference between using distance priors data and adopting global fit technique is very small. Therefore, CMB distance priors data are widely used in the literature (see Ref.\citep{YazhouHu2014,xinzhang2016}). It must be emphasized that, in this work we mainly focus on the analysis technique of SNIa data. So for simplicity, in this manuscript we just use the CMB distance prior data.}

\subsection{The total $\chi^2$ function }
We use the combined $\chi^2$ functions: $\chi^2=\chi^2_{SNIa}+\chi^2_{CMB}+\chi^2_{BAO}$.
We perform a MCMC likelihood analysis \citep{Lewis2002} to obtain $O(10^6)$ samples for each set of results presented in this paper.

\section{Dark Energy Models and Their Cosmological Constrains}\label{sec:model}

In a spatially flat universe \footnote{The assumption of flatness is motivated by the inflation scenario.
For a detailed discussion of the effects of spatial curvature, see \citep{Clarkson07}}, the Friedmann equation can be written as
\be
\label{Fried}
H=H_0\sqrt{\Omega_{\rm{r}}(1+z)^4+\Omega_{\rm{m}}(1+z)^3+\Omega_{\rm{de}}X(z)}.
\ee
Here $\Omega_{\rm{m}}$, $\Omega_{\rm{r}}$ and $\Omega_{\rm{de}}$ are the present fractional
densities of dust matter, radiation and dark energy, respectively.
Note that $X(z)\equiv \rho_{de}(z)/\rho_{de}(0)$ is given by the specific
dark energy models. This equation is usually rewritten as

\begin{equation}
\begin{aligned}
E(z)^2\equiv\ \left(\frac{H(z)}{H_0}\right)^2&= \Omega_{\rm{m}}(1+z)^{3}+\Omega_{\rm{r}}(1+z)^{4} \\
&+(1-\Omega_{\rm{m}}-\Omega_{\rm{r}})X(z).
\end{aligned}
\end{equation}

Here the radiation density parameter $\Omega_r$ is given by \citep{WangyunWangshuang2013},
\be
\Omega_{r}=\Omega_{m}/(1+z_{\rm eq}),
\ee
where $z_{\rm eq}=2.5\times10^4\Omega_{m}h^2(T_{\rm cmb}/2.7\,{\rm K})^{-4}$, $T_{\rm cmb}=2.7255\,{\rm K}$.

In this paper, we  analyze {\bf thirteen} popular DE models.  We divide these models into five classes:
\begin{itemize}

\item
Cosmological constant model.
\item
DE models with parameterized EOS
\item
Chaplygin gas models.
\item
Holographic dark energy models.
\item
Modified gravity models.
\end{itemize}
As mentioned above, here we just treat all the MG models as DE models.
It needs to point out that, compared with the Ref. \cite{zhangxin2016} that only considered ten DE models,
in this paper two new models,  Wang parametrization and CMG model, are taken into account.
For the convenience of readers, all the models  and the corresponding parameters are summarized in Table \ref{tab:model}. Note
that there are three additional parameters $(\alpha_0, \beta_0, \Omega_{b} h^2)$ for the cosmic-fits,
which are taken into account when calculating AIC and BIC.

We constrain these models with
the observational data mentioned above, and then make a comparison for them by using the information criteria.


\begin{table*}
\caption{\label{tab:model} Summary of models. Note that the additional parameters
$\alpha_0$, $\beta_0$ and $\Omega_{b} h^2$ appearing in the data fits are not considered as a model
parameter.}\centering
\begin{tabular*}{\textwidth}{@{}l*{15}{@{\extracolsep{0pt plus12pt}}l}}
\hline\hline
Model   &   Abbreviation   &  Model parameters   &   Number of model parameters\\
\hline
Cosmological constant   &   $\Lambda CDM$   &   $\Omega_m$, $h$   &   2\\
Constant $w$   &   $w CDM$   &   $\Omega_m$, $w$, $h$      &   3\\
Chevallier-Polarski-Linder   &  CPL  &  $\Omega_m$, $w_0$, $w_a$, $h$   &  4\\
Wang   &  Wang  &  $\Omega_m$, $w_0$, $w_a$, $h$   &  4\\
Generalized Chaplygin gas    &  GCG  &  $A_s$, $\xi$, $h$   &  3\\
New Generalized Chaplygin gas   &  NGCG  &  $\Omega_m$, $\zeta$, $\eta$, $h$   &  4\\
Holographic dark energy & HDE & $\Omega_m$, $c$, $h$  & 3\\
Agegraphic dark energy & ADE & $n$, $h$  & 2 \\
Ricci dark energy  & RDE & $\Omega_m$, $\gamma$, $h$  & 3\\
Dvali-Gabadadze-Porrati  & DGP & $\Omega_m$, $h$  & 2\\
Phenomenological extension of DGP  & $\alpha$DE & $\Omega_m$, $\alpha$, $h$  & 3\\
Doubly Coupled Massive Gravity    &  CMG  &  $\Omega_m$, $c_2$, $c_3$, $h$   &  4\\
Vacuum metamorphosis&VM&$M$,$h$&2\\
\hline
\end{tabular*}
\end{table*}

\subsection {Cosmological constant model}
Cosmological constant model is also called
$\Lambda$CDM.
Its EOS satisfies $w=-1$ all the times.
In a flat universe, we have,
\be
\label{LCDM}
E(z)=\sqrt{\Omega_{m}(1+z)^3+\Omega_{r}(1+z)^4+ 1-\Omega_{m}-\Omega_{r}}.
\ee

\begin{figure*}\centering
\includegraphics[width=7cm]{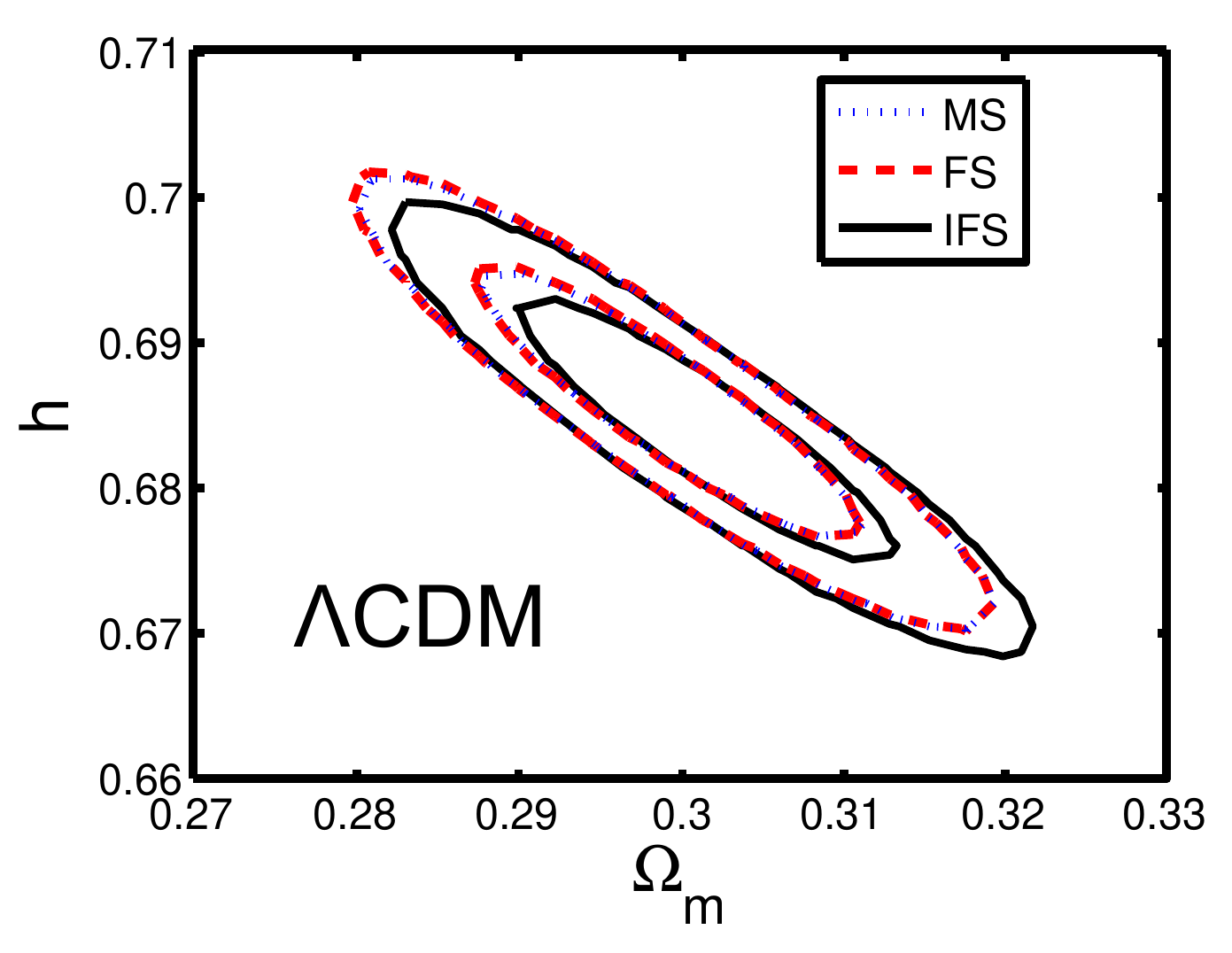}
\includegraphics[width=7cm]{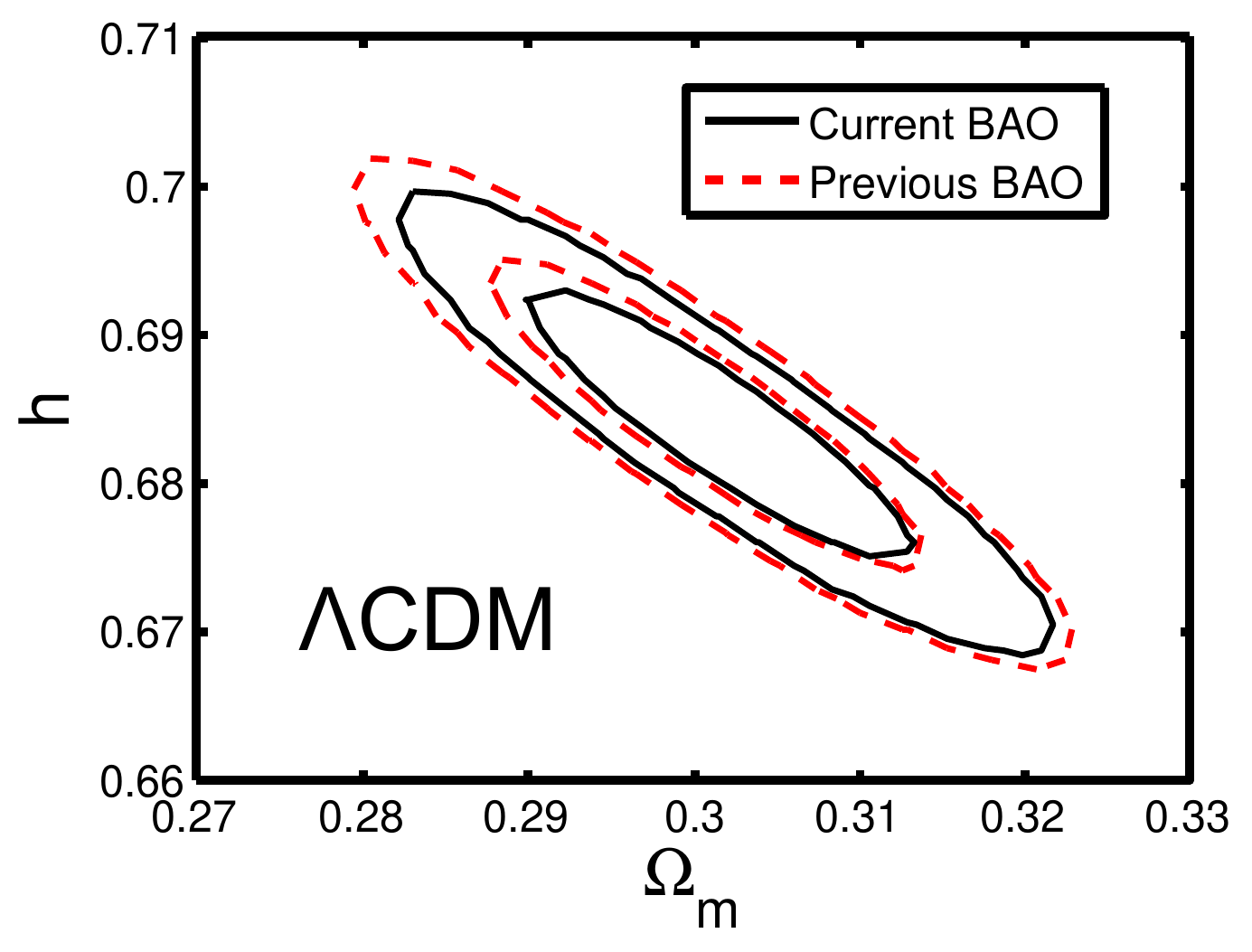}
\caption{\label{f11}$\Lambda $CDM: 1$\sigma$ and 2$\sigma$ confidence regions in the $\Omega_{\rm{m}}$--$h$ plane.
The left panel shows the effect of different statistic methods of SNIa, where current BAO data is used.
The blue dotted lines denote the results of MS, the red dashed lines represent the results of FS, and the black solid lines are the results of IFS.
Right panel shows the effect of different BAO data, where the IFS is used.
The black solid lines denote the results of current BAO data, and the red dashed lines represent the results of previous BAO data.}
\end{figure*}

In Fig.~\ref{f11}, for $\Lambda CDM$, we plot the 1$\sigma$ and 2$\sigma$ confidence regions in the $\Omega_{\rm{m}}$--$h$ plane.
The left panel shows the effect of different statistic methods of SNIa, where current BAO data is used in the analysis.
We find that, for the best-fit results, IFS yields a bigger $\Omega_{\rm{m}}$ and a smaller $h$. This is consistent with the result of Ref. \cite{Wang2016}.
The right panel shows the effect of different BAO data, where the IFS is used in the analysis.
We find that, compared with the results of previous BAO data,
adopting current BAO data can give a tighter constraint for $\Lambda CDM$, but will not have significant effects on the best-fit values of $\Omega_{\rm{m}}$ and $h$.

\subsection{Dark energy models with equation of state parameterized}
Here, we consider $w$CDM parametrization, CPL parametrization and Wang parametrization.

\subsubsection {Constant $w$ parametrization}

For $w$CDM parametrization, its EOS $w$ is
a constant all the time, so we have,
\begin{equation}
\begin{aligned}
E(z)^{2}&=\Omega_{\rm{m}}(1+z)^{3}+\Omega_{\rm{r}}(1+z)^{4}\\
&+(1-\Omega_{\rm{m}}-\Omega_{\rm{r}})(1+z)^{3(1+w)}.
\end{aligned}
\end{equation}

\begin{figure*}\centering
\includegraphics[width=7cm]{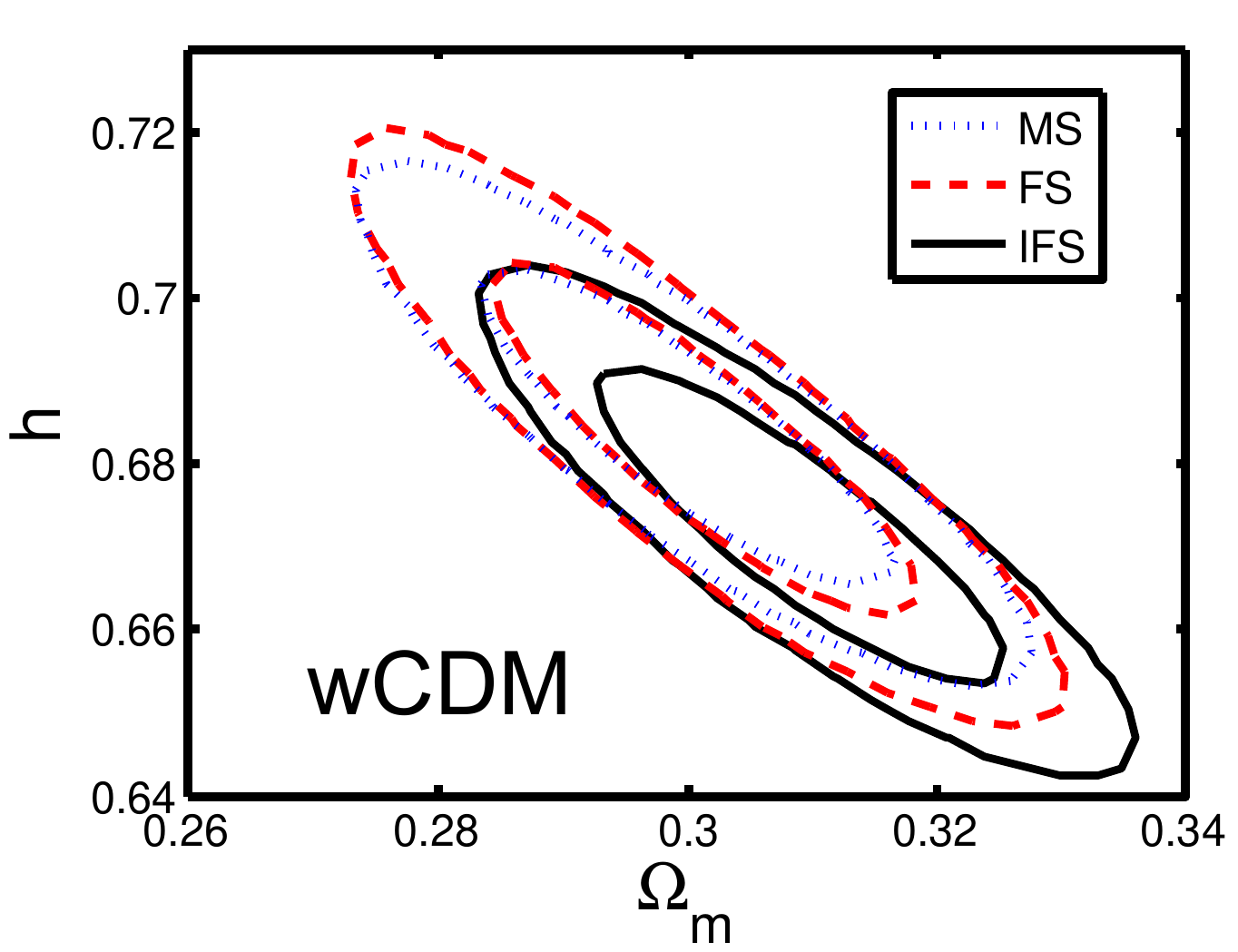}
\includegraphics[width=7cm]{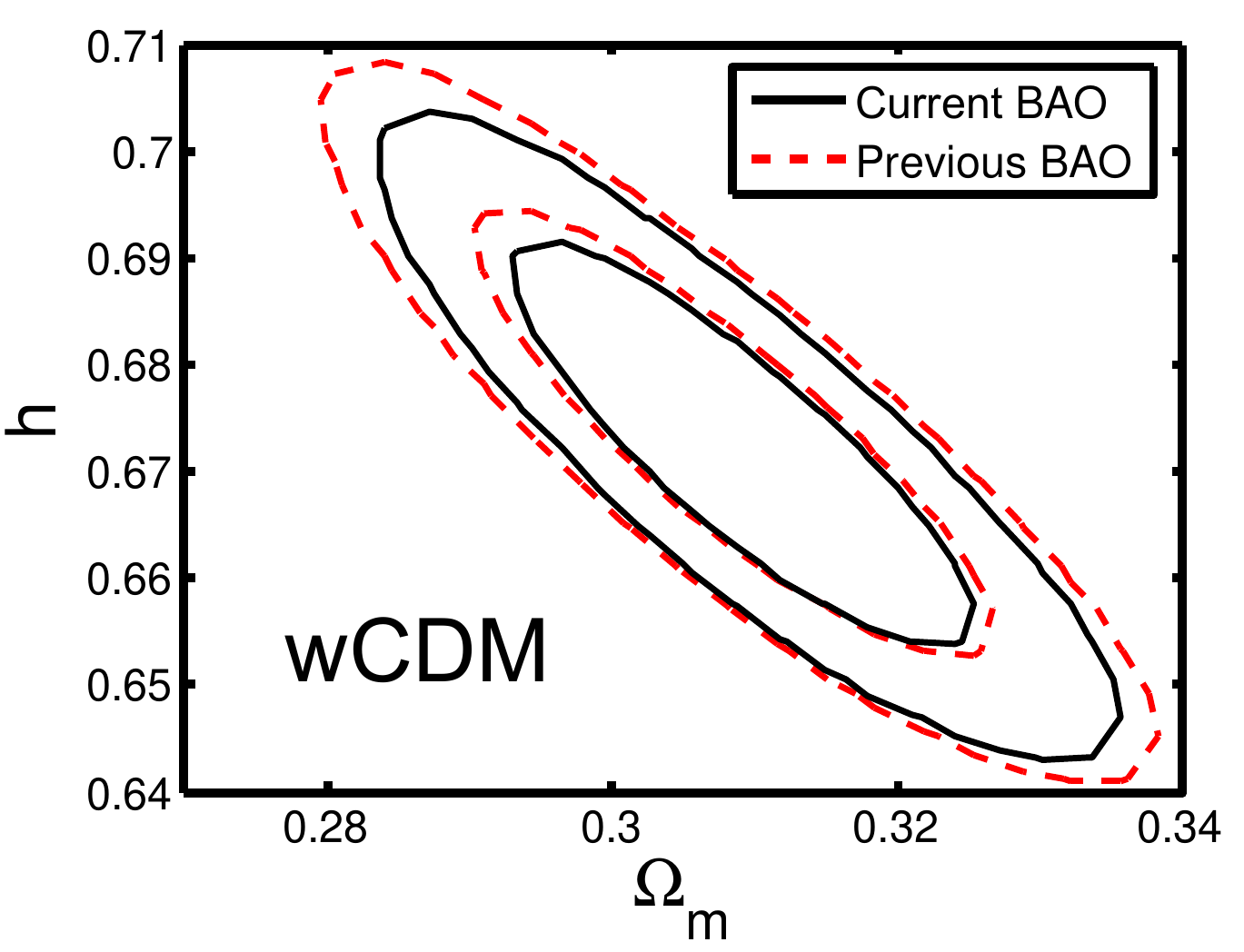}
\includegraphics[width=7cm]{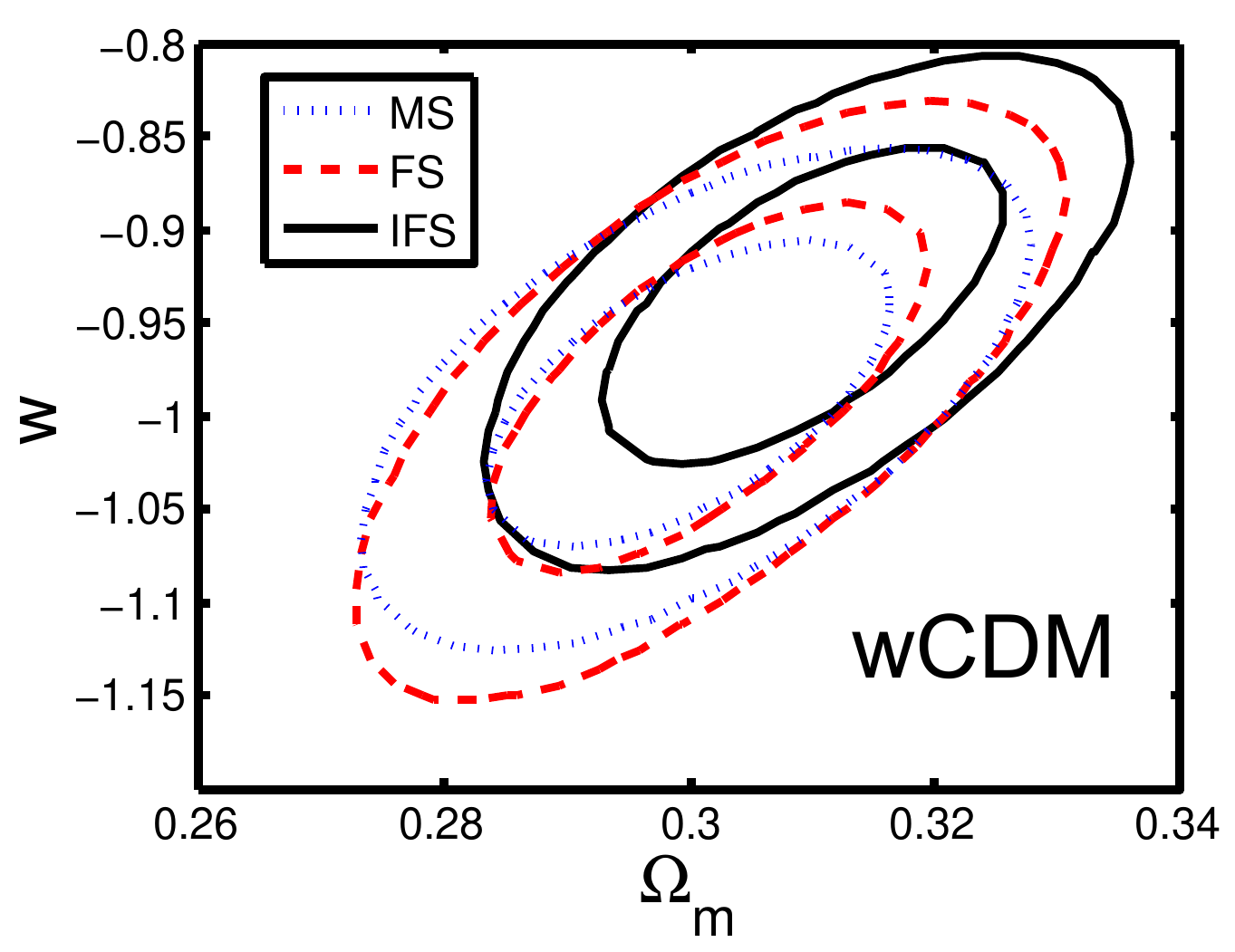}
\includegraphics[width=7cm]{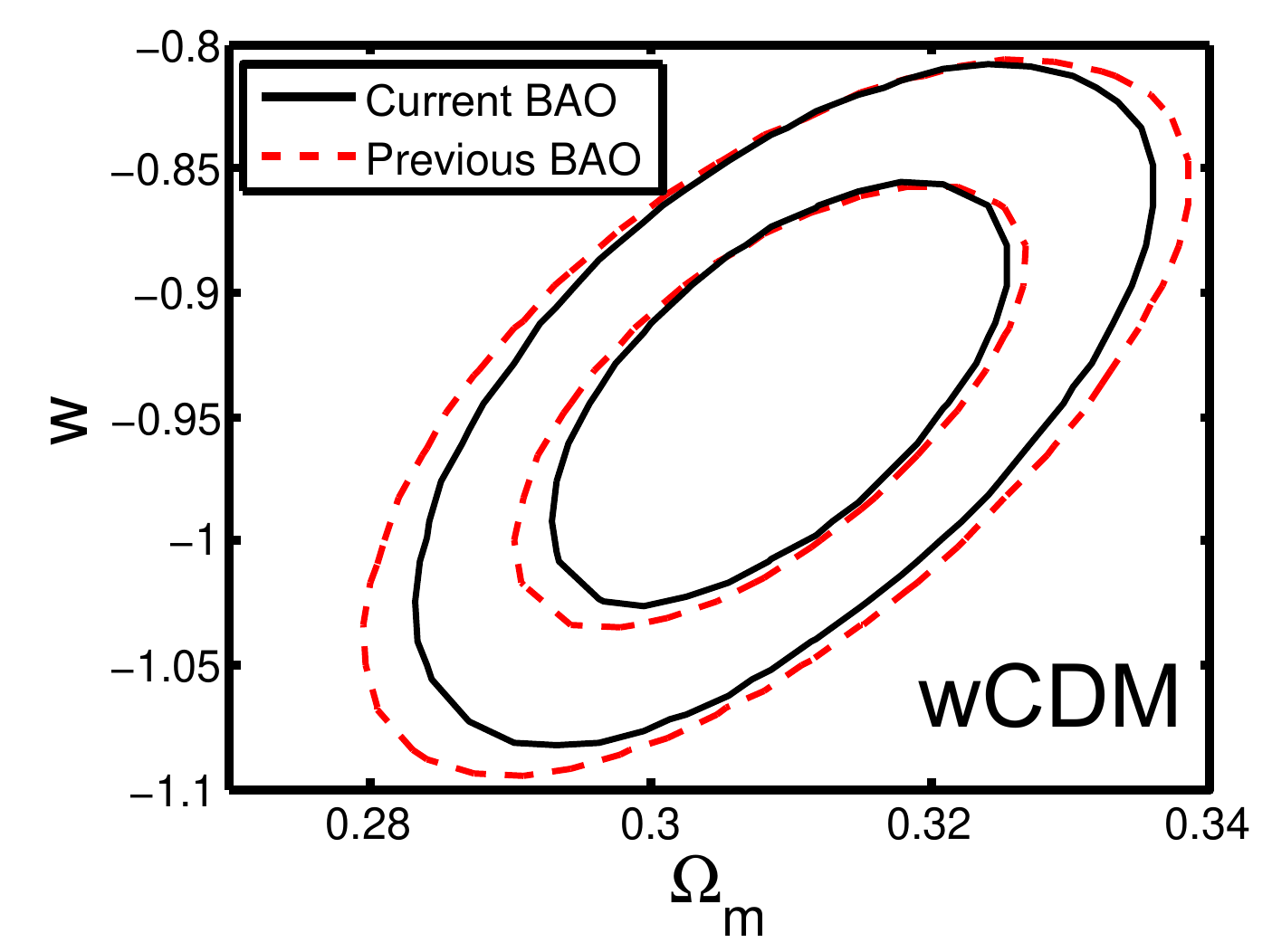}
\caption{\label{f21}$w$CDM parametrization: 1$\sigma$ and 2$\sigma$ confidence regions in the $\Omega_{\rm{m}}$--$h$ (upper panels) and $\Omega_{\rm{m}}$--$w$ (lower panels) planes.
The left panels show the effect of different statistic methods of SNIa, where current BAO data is used.
The blue dotted lines denote the results of MS, the red dashed lines represent the results of FS, and the black solid lines are the results of IFS.
The right panels show the effect of different BAO data, where the IFS is used.
The black solid lines denote the results of current BAO data, and the red dashed lines represent the results of previous BAO data.}
\end{figure*}
In Fig.~\ref{f21}, for $w$CDM parametrization, we plot 1$\sigma$ and 2$\sigma$ confidence regions in the $\Omega_{\rm{m}}$--$h$ and $\Omega_{\rm{m}}$--$w$ planes.
The left panels show the effect of different statistic methods of SNIa, where current BAO data is used in the analysis.
Similar to the case of $\Lambda $CDM, for the best-fit results, IFS yields a bigger $\Omega_{\rm{m}}$ and a smaller $h$.
Moreover, FS yields the biggest error bars for each parameter.
The right panels show the effect of different BAO data, where the IFS is used in the analysis.
We also find that, compared with the results of previous BAO data,
adopting current BAO data can give a tighter constraint for this model, but will not have significant effects on the best-fit values of parameters.
In addition, from the lower panels, we find that $w = -1$ lies in the 1$\sigma$ region of the $\Omega_{\rm{m}}$--$w$ plane. This implies that $\Lambda$CDM is favored.

\subsubsection {Chevallier-Polarski-Linder parametrization}

For CPL~\cite{CPL,Linder:2002et} parametrization, the EOS is parameterized as
\begin{equation}
\label{cpl1}
w(z)=w_0+w_a\frac{z}{1+z},
\end{equation}
where $w_0$ and $w_a$ are constants. The corresponding $E(z)$ can be
expressed as
\begin{equation}
\begin{aligned}
\quad E(z)^{2}&=\Omega_{\rm{m}}(1+z)^{3}+\Omega_{\rm{r}}(1+z)^{4}\\
\quad &+(1-\Omega_{\rm{m}}-\Omega_{\rm{r}})(1+z)^{3(1+w_{\rm{0}}+w_{\rm{a}})}\exp\left(-\frac{3w_{\rm{a}}z}{1+z}\right).
 \end{aligned}
\end{equation}

\begin{figure*}\centering
\includegraphics[width=7cm]{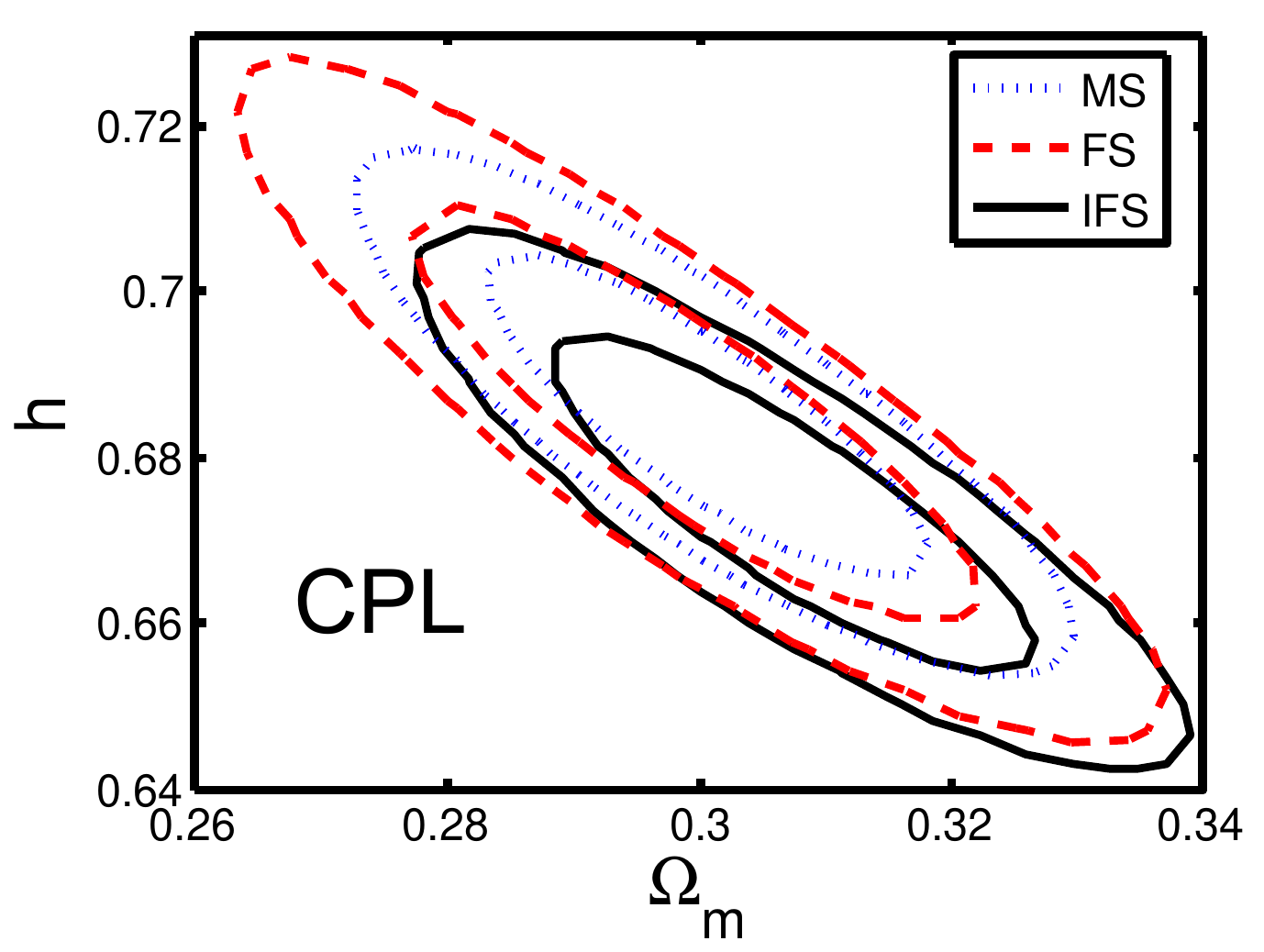}
\includegraphics[width=7cm]{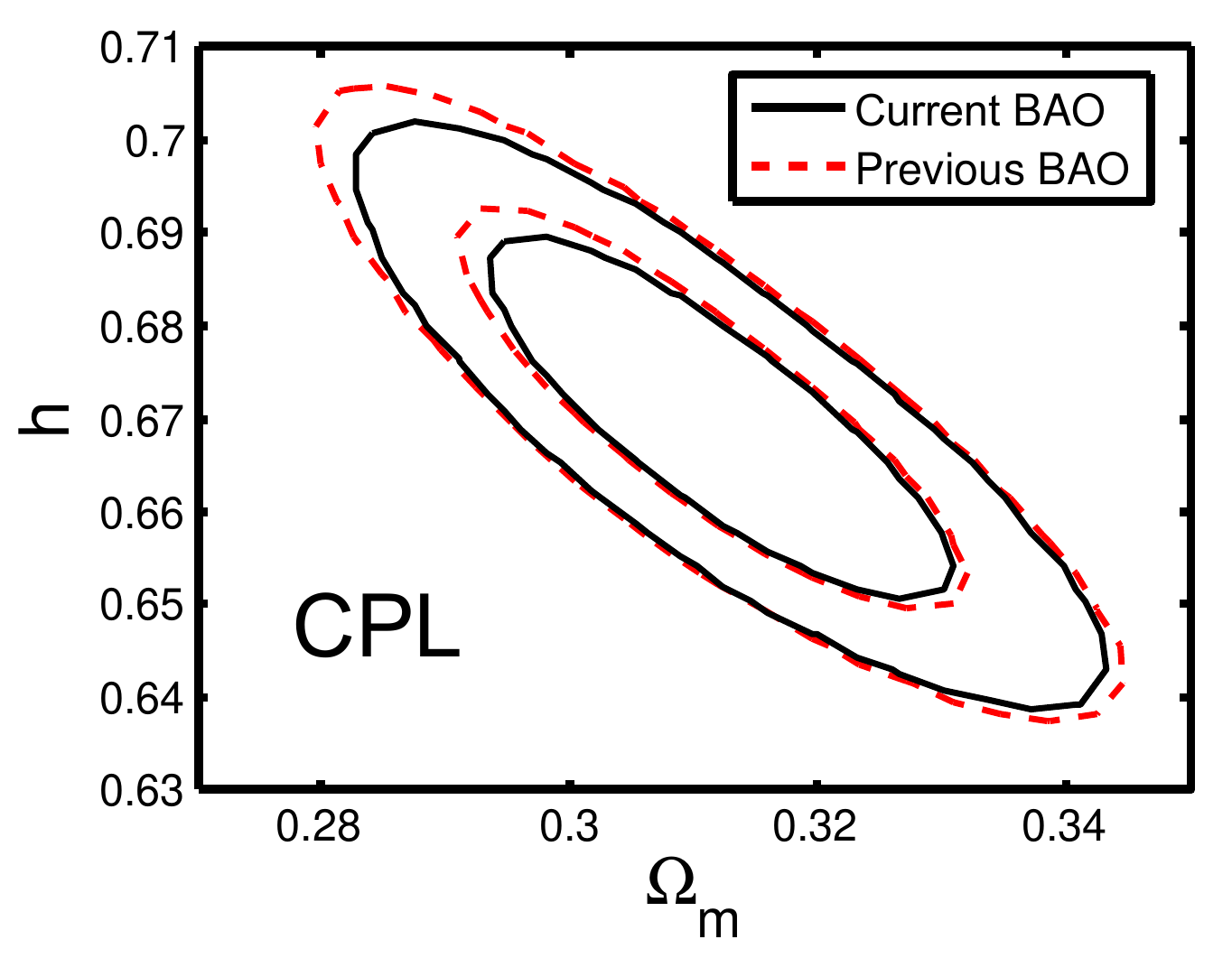}
\includegraphics[width=7cm]{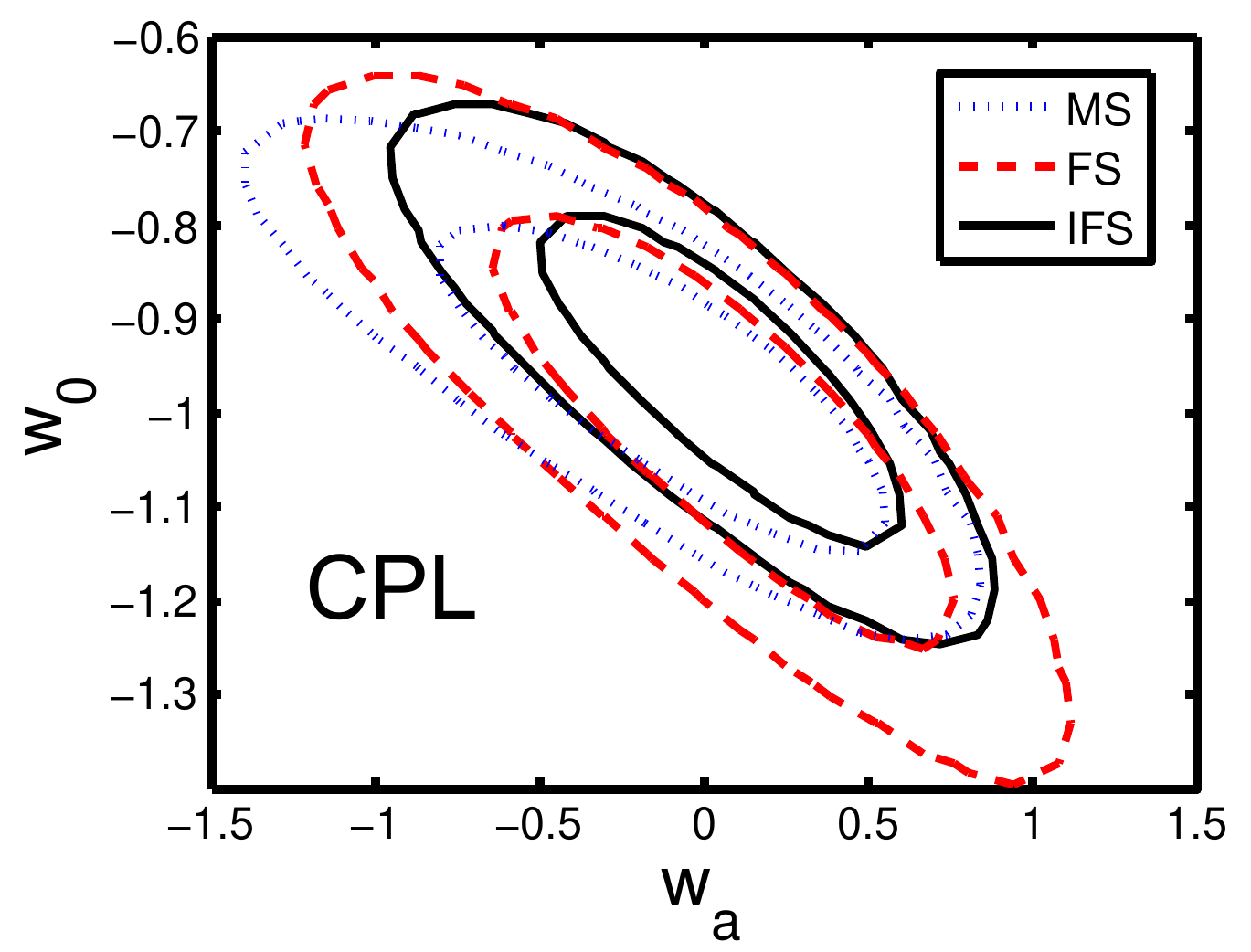}
\includegraphics[width=7cm]{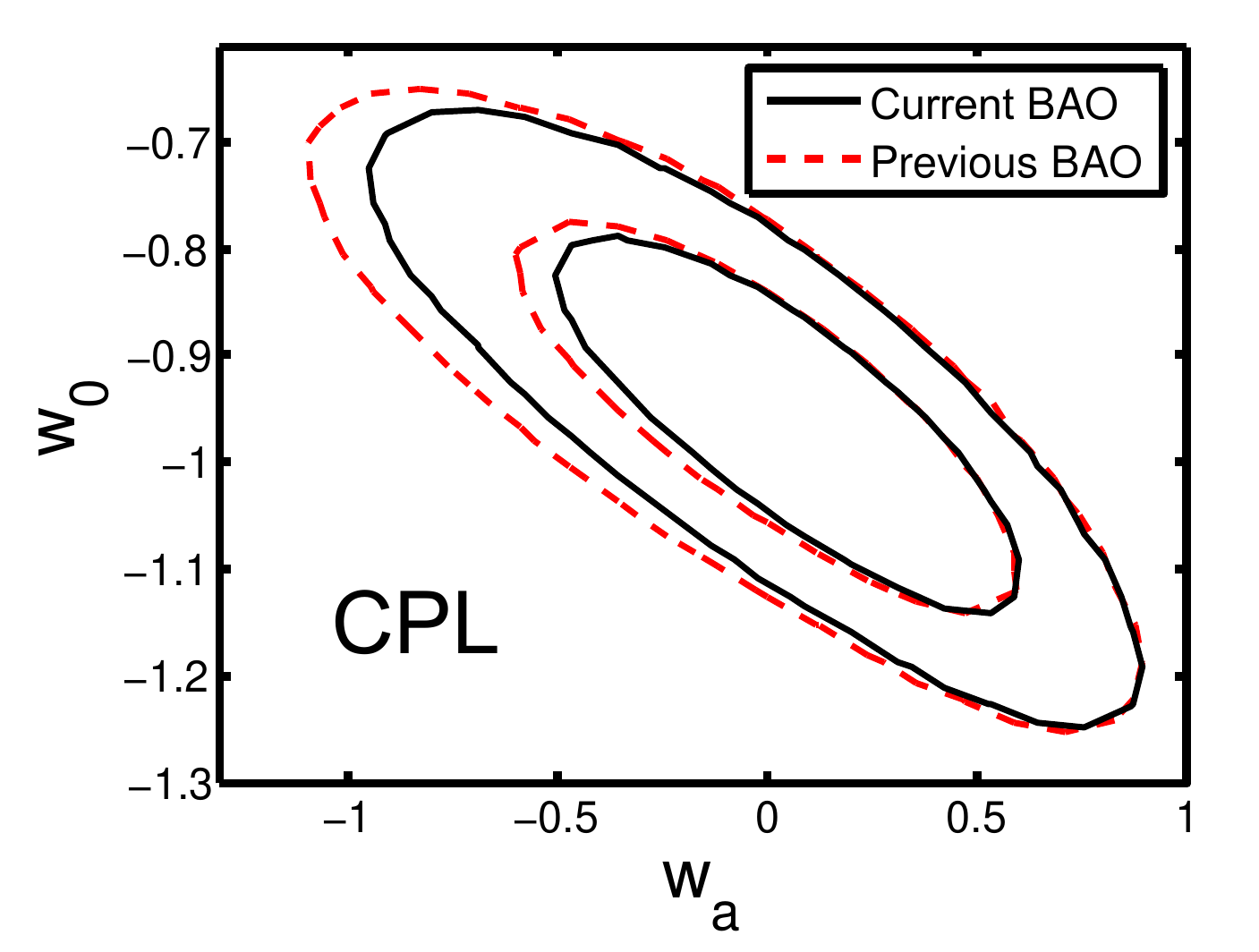}
\caption{\label{f31}CPL parametrization: 1$\sigma$ and 2$\sigma$ confidence regions in the $\Omega_{\rm{m}}$--$h$ (upper panels) and $w_a$--$w_0$ (lower panels) planes.
The left panels show the effect of different statistic methods of SNIa, where current BAO data is used.
The blue dotted lines denote the results of MS, the red dashed lines represent the results of FS, and the black solid lines are the results of IFS.
The right panels show the effect of different BAO data, where the IFS is used.
The black solid lines denote the results of current BAO data, and the red dashed lines represent the results of previous BAO data.}
\end{figure*}

In Fig.~\ref{f31}, for CPL, we plot 1$\sigma$ and 2$\sigma$ confidence regions in the $\Omega_{\rm{m}}$--$h$ and $w_a$--$w_0$ planes.
The left panels show the effect of different statistic methods of SNIa, where current BAO data is used in the analysis.
Similar to the case of $\Lambda $CDM, for the best-fit results, IFS yields a bigger $\Omega_{\rm{m}}$ and a smaller $h$.
Moreover, FS yields the biggest error bars for each parameter.
The right panels show the effect of different BAO data, where the IFS is used in the analysis.
We also find that, compared with the results of previous BAO data,
adopting current BAO data can give a tighter constraint for this model, but will not have significant effects on the best-fit values of parameters.
In addition, from the lower panels, we find that the point $(w_a=0, w_0=-1)$ lies in the 1$\sigma$ region of the $w_a$--$w_0$ plane.
This implies that $\Lambda$CDM is fairly consistent with current observational data.

\subsubsection {Wang parametrization}

For Wang parametrization~\citep{YunWang2008}, the EOS is
\begin{equation}
\label{ba}
w(z) = 3w_{a}-2w_{0} + \frac{3(w_{0}-w_{a})}{1+z},
\end{equation}
where $w_0$ and $w_a$ are constants. The Wang will reduce to wCDM if $w_0=w_a$, and reduce to $\Lambda$CDM if $w_0=w_a=-1$. The corresponding $E(z)$ can be
expressed as
\begin{equation}
\begin{aligned}
\quad E(z)^{2}&=\Omega_{\rm{m}}(1+z)^{3}+\Omega_{\rm{r}}(1+z)^{4}\\
\quad &+(1-\Omega_{\rm{m}}-\Omega_{\rm{r}})(1+z)^{3(1-2w_0+3w_a)}e^{\frac{9(w_0-w_a)z}{1+z}}.
 \end{aligned}
\end{equation}

\begin{figure*}\centering
\includegraphics[width=7cm]{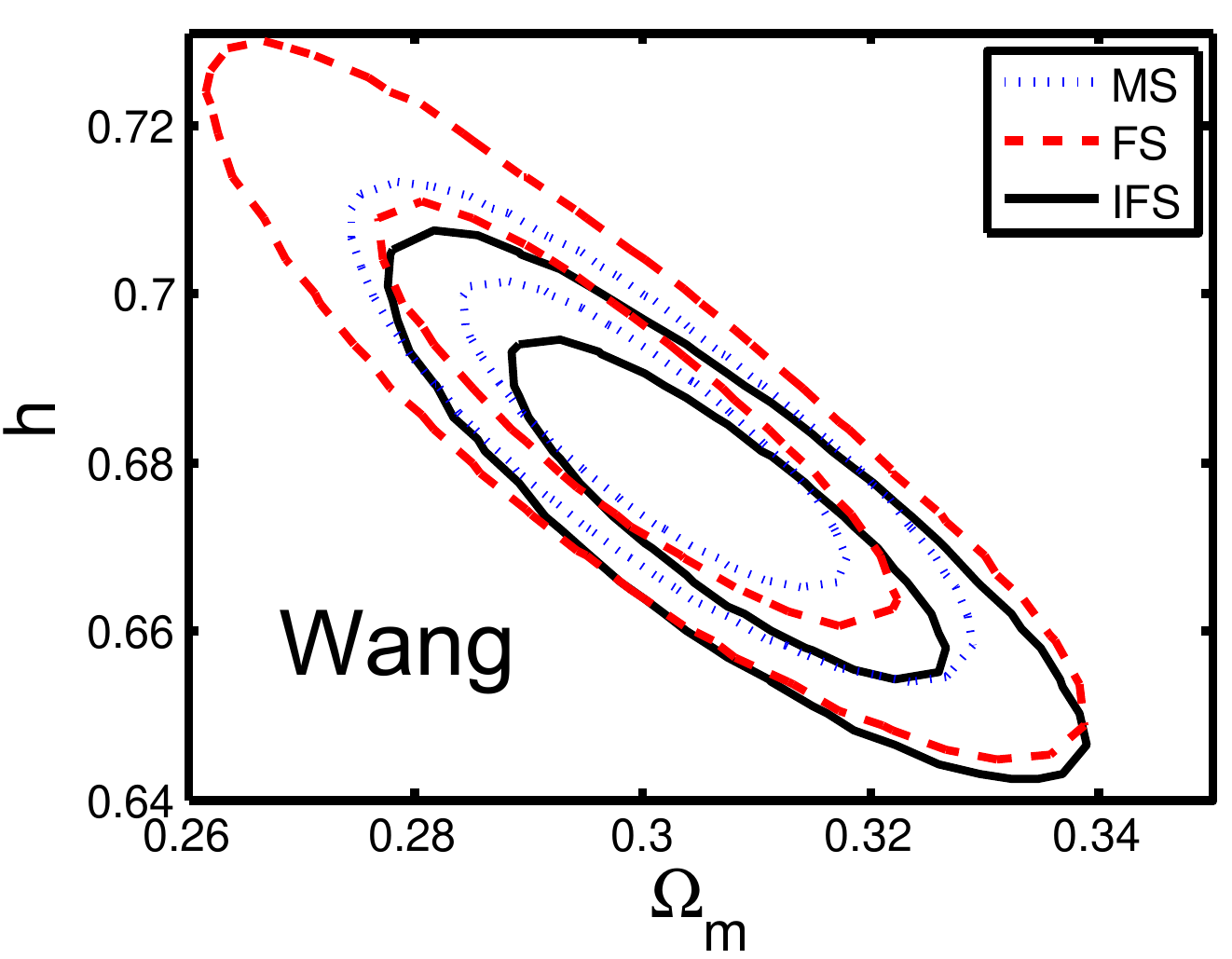}
\includegraphics[width=7cm]{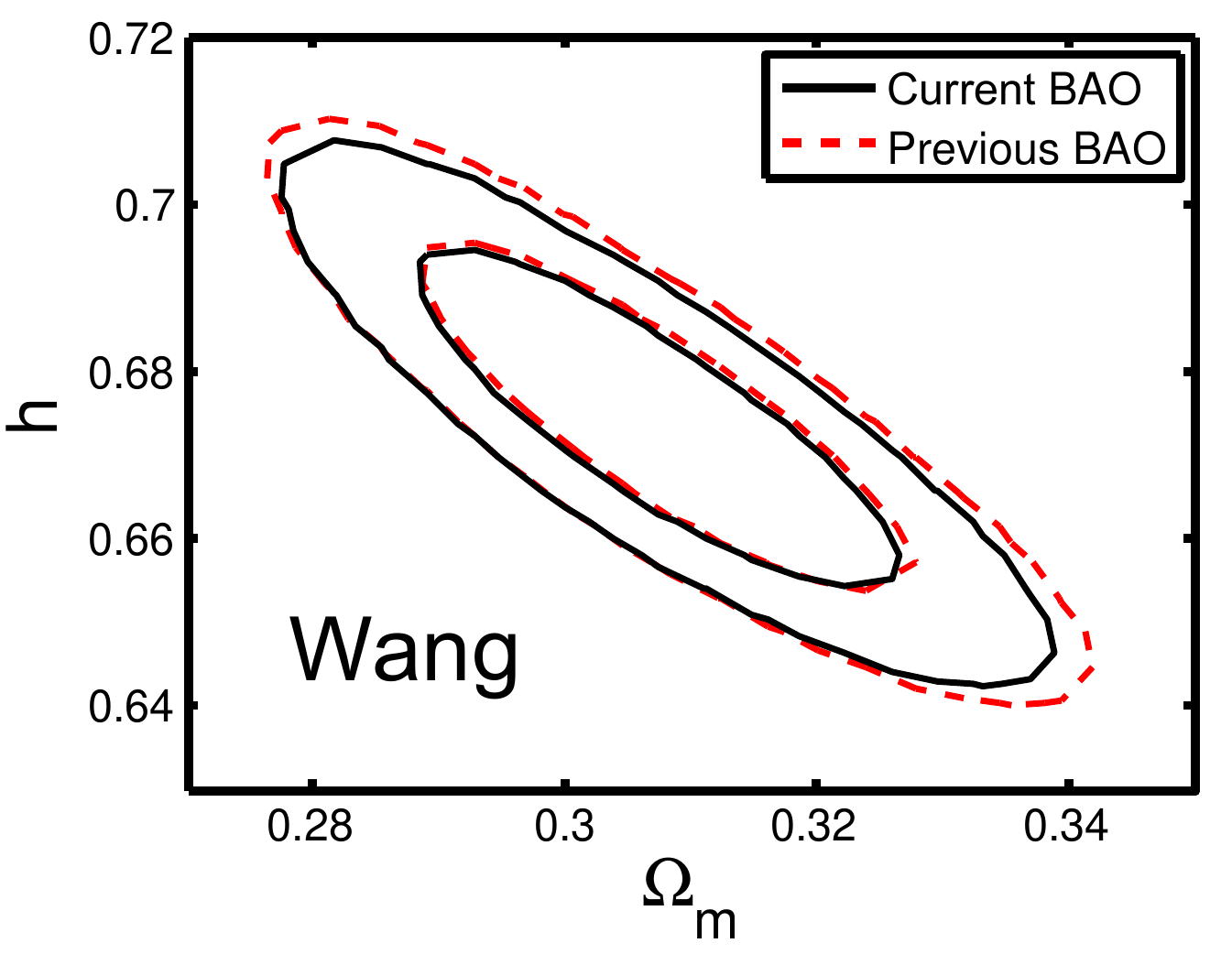}
\includegraphics[width=7cm]{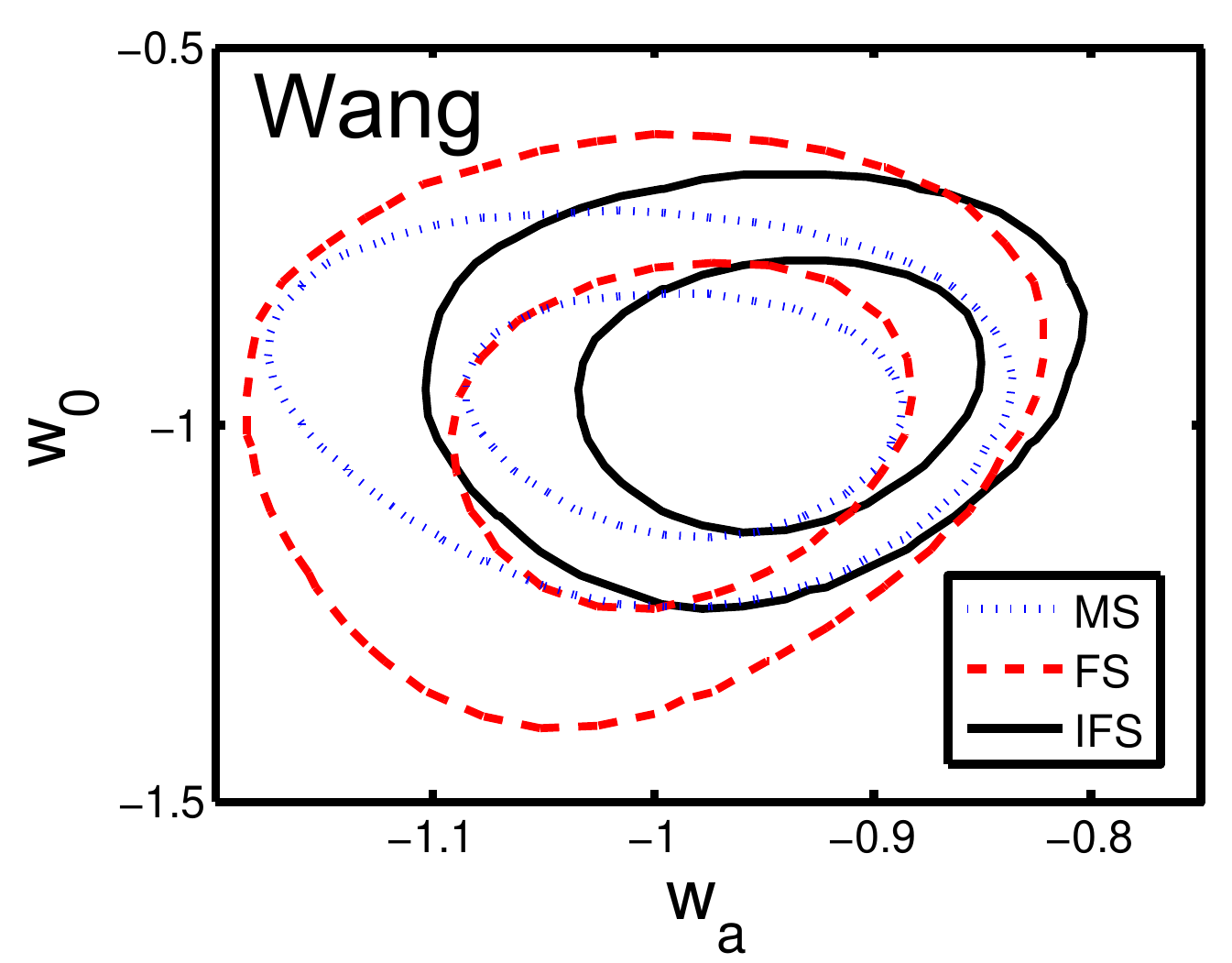}
\includegraphics[width=7cm]{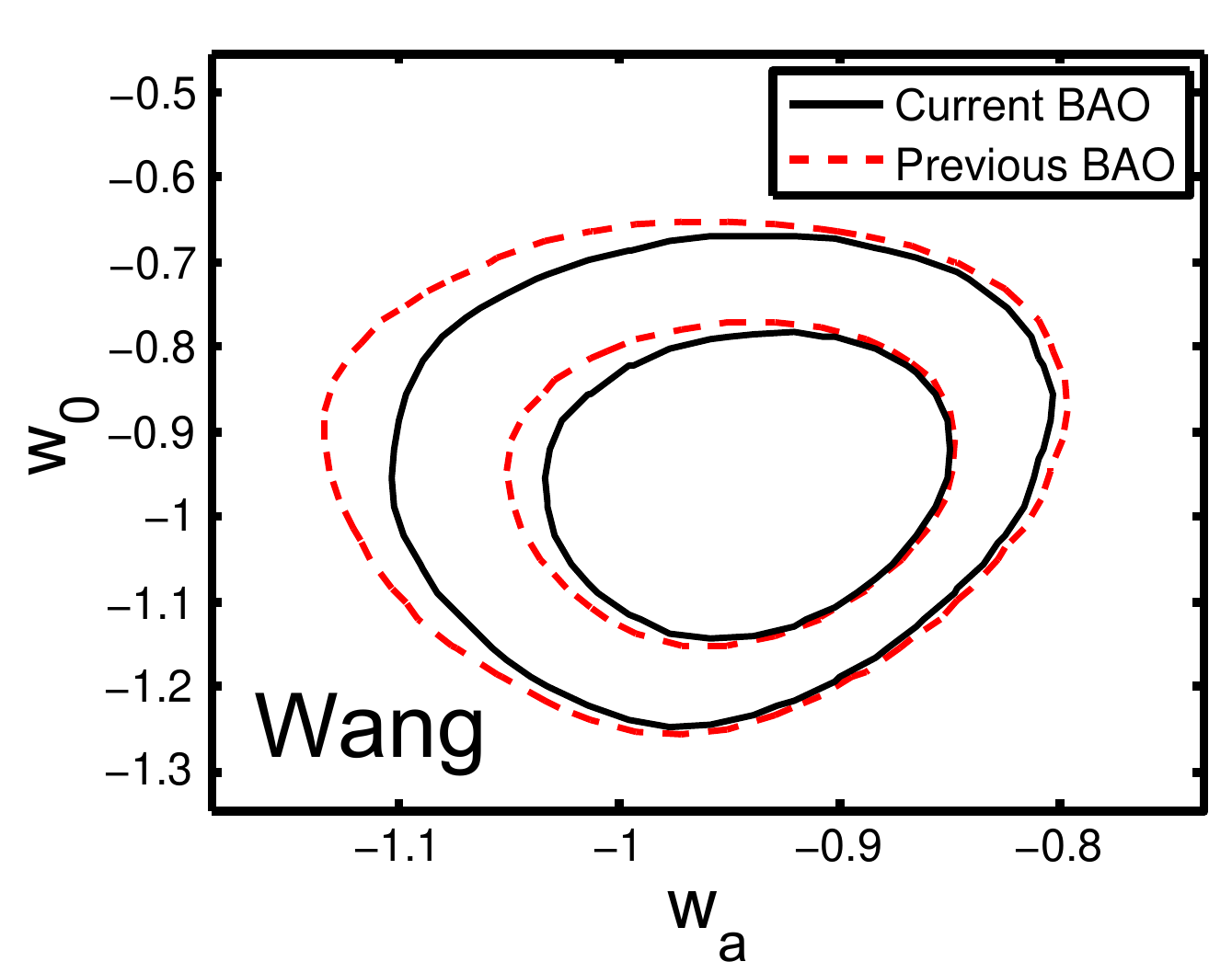}
\caption{\label{41}Wang parametrization: 1$\sigma$ and 2$\sigma$ confidence regions in the $\Omega_{\rm{m}}$--$h$ (upper panels) and $w_a$--$w_0$ (lower panels) planes.
The left panels show the effect of different statistic methods of SNIa, where current BAO data is used.
The blue dotted lines denote the results of MS, the red dashed lines represent the results of FS, and the black solid lines are the results of IFS.
The right panels show the effect of different BAO data, where the IFS is used.
The black solid lines denote the results of current BAO data, and the red dashed lines represent the results of previous BAO data.}
\end{figure*}

In Fig.~\ref{41}, for Wang parametrization, we plot 1$\sigma$ and 2$\sigma$ confidence regions in the $\Omega_{\rm{m}}$--$h$ and $w_a$--$w_0$ planes.
The left panel shows the effect of different statistic methods of SNIa, where current BAO data is used in the analysis.
For the best-fit results, IFS yields a bigger $\Omega_{\rm{m}}$ and a smaller $h$.
Moreover, for the fitting results of each parameter, FS yields the biggest error bars.
The right panels show the effect of different BAO data, where the IFS is used in the analysis.
We also find that, compared with the results of previous BAO data,
adopting current BAO data can give a tighter constraint for this model, but will not have significant effects on the best-fit values of parameters.
In addition, from the lower panels, we find that the point $(w_a=-1, w_0=-1)$ lies in the 1$\sigma$ region of the $w_a$--$w_0$ plane.
This implies that $\Lambda$CDM is fairly consistent with current observational data.

\subsection{Chaplygin gas models}
Chaplygin gas  model is a kind of fluid model. It is commonly viewed as arising from the d-brane theory.
The original Chaplygin gas \cite{cg} model has been excluded by the observational data \cite{Davis:2007na}. Here, we consider two kinds of these models: generalized Chaplygin gas (GCG) model \cite{gcg} and  new generalized Chaplygin gas (NGCG) model \cite{ngcg}.

\subsubsection {Generalized Chaplygin gas model}

The exotic EOS of GCG \cite{gcg} can be expressed as :
\be
\label{gcg}
p_{gcg}=-\frac{A}{\rho_{gcg}^{\xi}},
\ee
where $A$ is a positive
constant. Then, we can get the energy density of GCG:
\begin{equation}
\rho_{gcg}(a)=\rho_{gcg}(0)\left(A_s+{1-A_s\over
a^{3(1+\xi)}}\right)^{1\over 1+\xi},
\end{equation}
where $A_s\equiv A/\rho_{gcg}^{1+\xi}(0)$. Note that GCG
behaves like a dust matter if $A_s=0$, and GCG behaves like a
cosmological constant if $A_s=1$.  Considering a universe with GCG, baryon, and radiation, we have
\begin{equation}
\begin{aligned}
E(z)^{2}&=\Omega_{\rm{b}}(1+z)^{3}+\Omega_{\rm{r}}(1+z)^{4}\\
&+(1-\Omega_{\rm{b}}-\Omega_{\rm{r}})\left(A_{\rm{s}}+(1-A_{\rm{s}})(1+z)^{3(1+\xi)}\right)^{1\over 1+\xi}.
\end{aligned}
\end{equation}
Cosmological constant is
recovered for $\xi=0$ and
$\Omega_m=1-\Omega_{r}-A_s(1-\Omega_{r}-\Omega_{b})$.

\begin{figure*}\centering
\includegraphics[width=7cm]{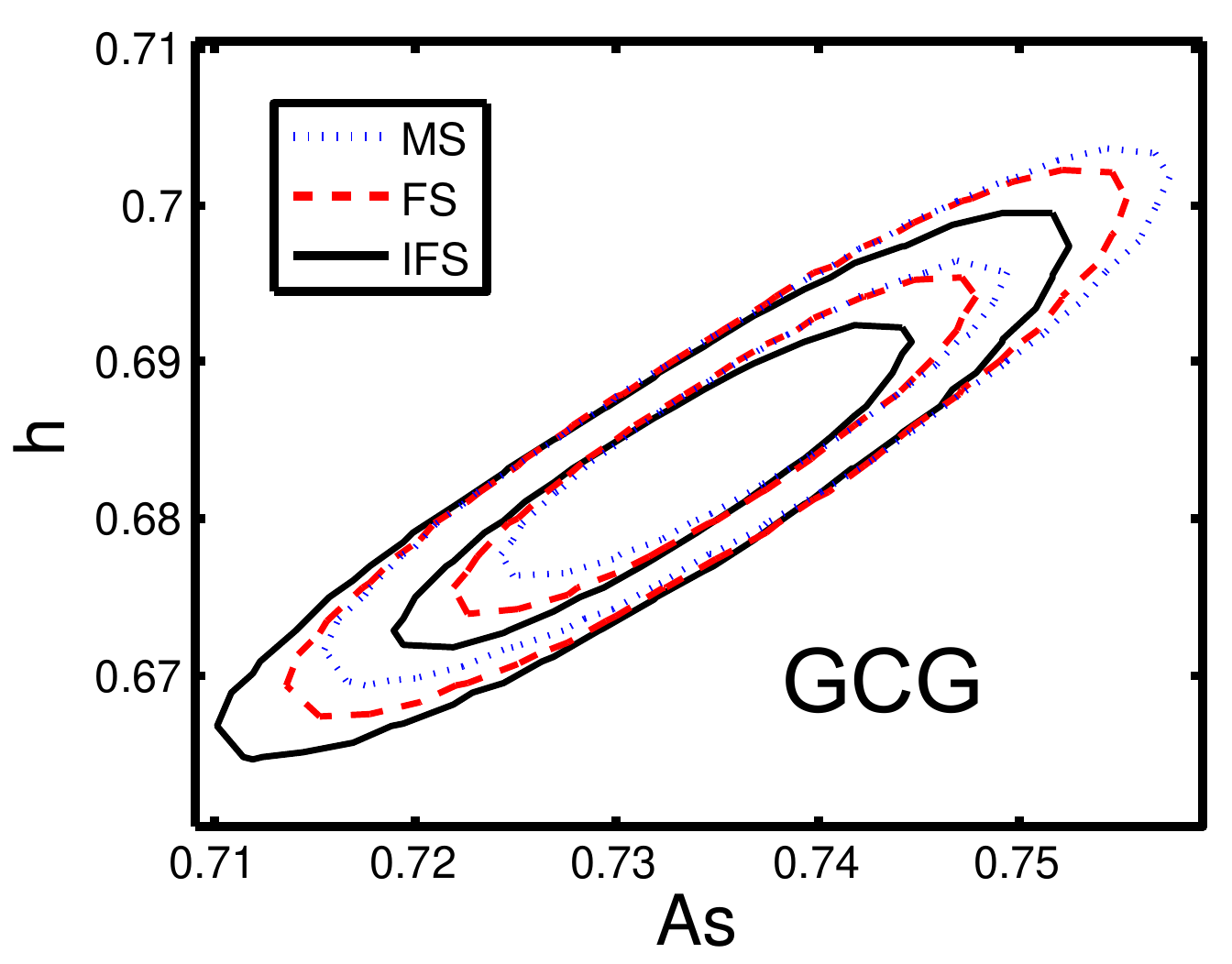}
\includegraphics[width=7cm]{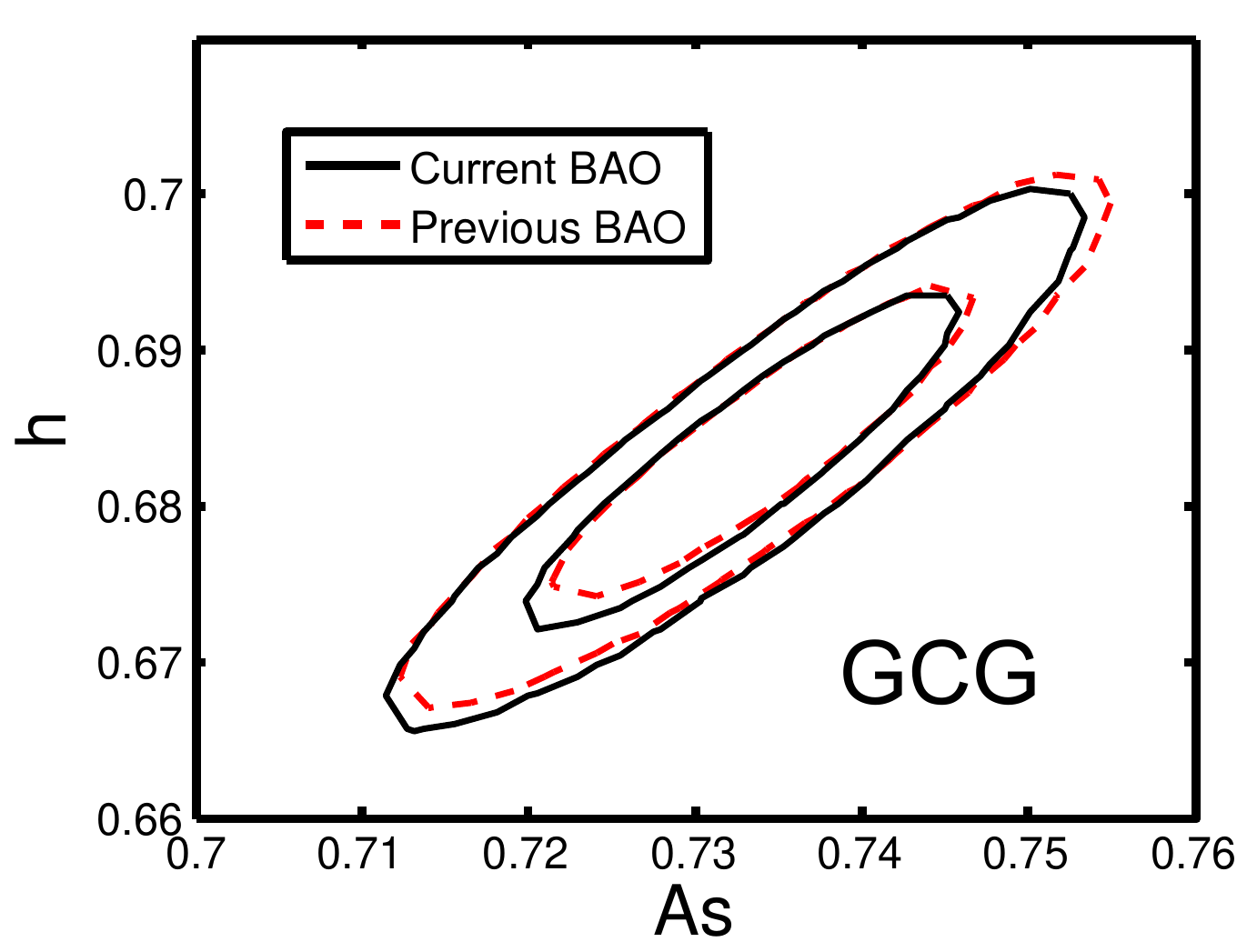}
\includegraphics[width=7cm]{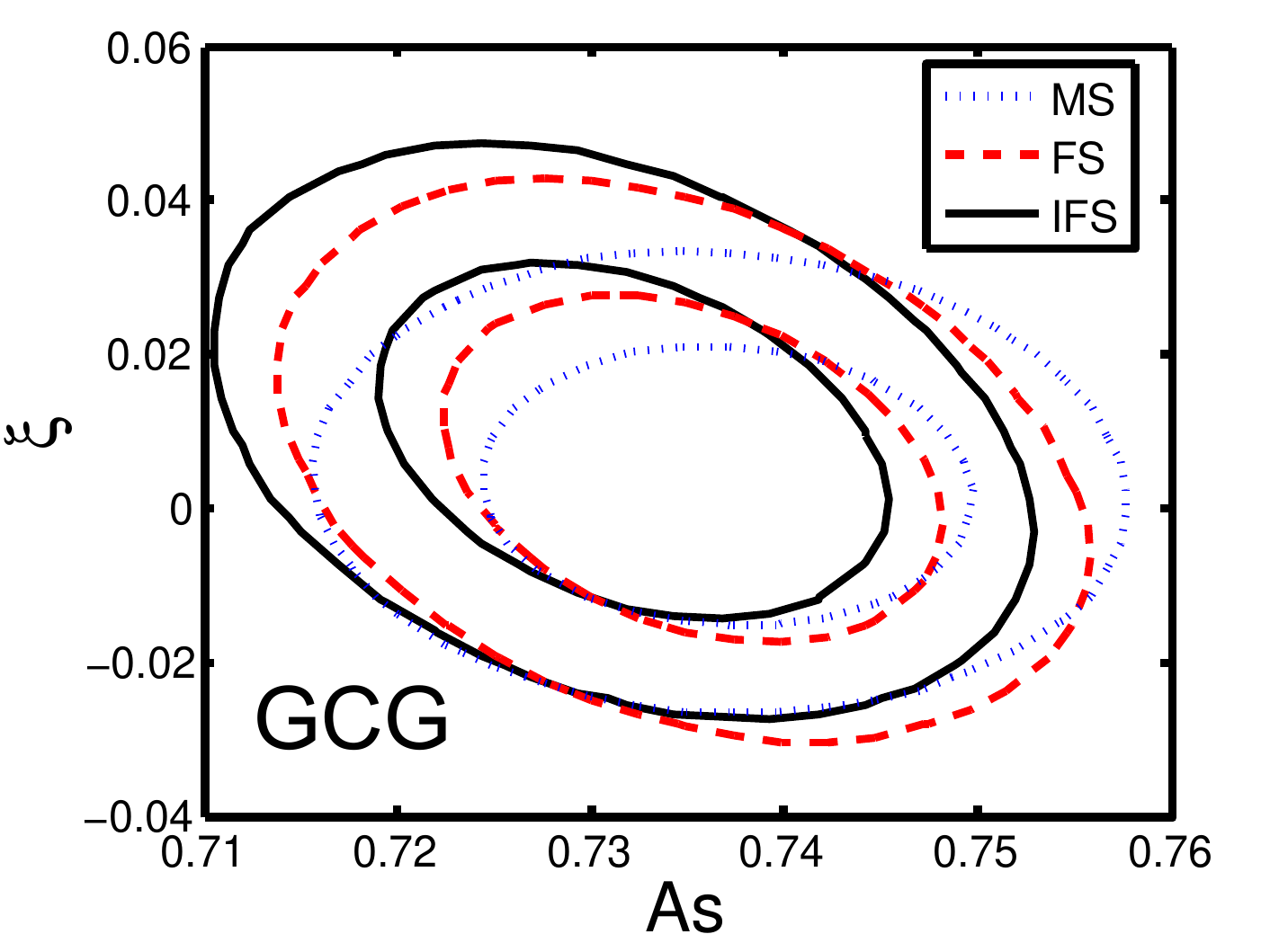}
\includegraphics[width=7cm]{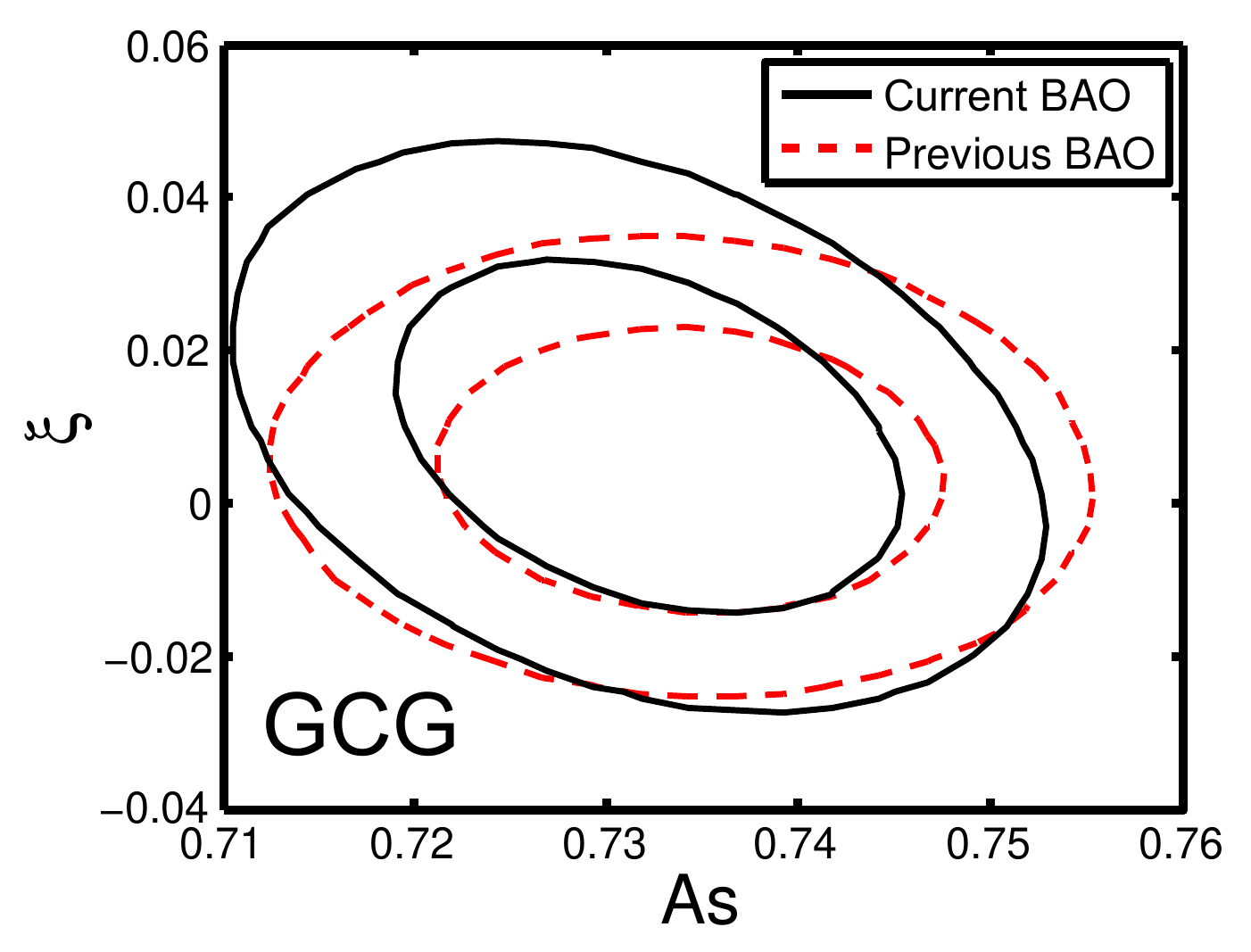}
\caption{\label{f51}GCG: 1$\sigma$ and 2$\sigma$ confidence regions in the $As$--$h$ (upper panels) and $As$--$\xi$ (lower panels) planes.
The left panels show the effect of different statistic methods of SNIa, where current BAO data is used.
The blue dotted lines denote the results of MS, the red dashed lines represent the results of FS, and the black solid lines are the results of IFS.
The right panels show the effect of different BAO data, where the IFS is used.
The black solid lines denote the results of current BAO data, and the red dashed lines represent the results of previous BAO data.}
\end{figure*}

We plot 1$\sigma$ and 2$\sigma$ confidence regions in the  $A_s$--$h$ (upper panels) and $A_s$--$\xi$ (lower panels) planes in Fig.~\ref{f51}.
The left panels show the effect of different statistic methods of SNIa, where current BAO data is used in the analysis.
For the best-fit results, IFS yields a smaller $A_s$ and a smaller $h$.
Note that a smaller $A_s$ results in a bigger $\Omega_m$.
The right panels show the effect of different BAO data, where the IFS is used in the analysis.
We also find that, different BAO data  will not have significant effects on the best-fit values of parameters.
In addition, from the lower panels, we find that $\xi = 0$ lies in the 1$\sigma$ region of the $A_s$--$\xi$ plane.
This implies that $\Lambda$CDM limit of this model is favored.

\subsubsection {New generalized chaplygin gas model}
The EOS of NGCG fluid \cite{ngcg} is given by
\begin{equation}
p_{\rm{ngcg}}=-\frac{\tilde{A}(a)}{\rho^{\zeta}_{\rm{ngcg}}},
\end{equation}
where $\tilde{A}(a)$ is a function of the scale factor $a$, and $\zeta$ is a free parameter. The energy density of NGCG can be expressed as
\begin{equation}
\rho_{\rm{ngcg}}=\left[Aa^{-3(1+\zeta)(1+\eta)}+Ba^{-3(1+\zeta)}\right]^{\frac{1}{1+\eta}},\label{ngcg1}
\end{equation}
where $A$ and $B$ are positive constants.
The form of the function $\tilde{A}(a)$ can be determined to be
\begin{equation}
\tilde{A}(a)=-\zeta Aa^{-3(1+\zeta)(1+\eta)}.
\end{equation}
NGCG reduces to GCG if $\zeta =-1$, reduces
to $w$CDM if $\eta = 0$, and reduces to $\Lambda$CDM if
($\zeta= -1$, $\eta = 0$). In a flat universe, we have
\begin{equation}
\begin{aligned}
E(z)^{2}&=\Omega_{\rm{b}}(1+z)^{3}+\Omega_{\rm{r}}(1+z)^{4}+(1-\Omega_{\rm{b}}-\Omega_{\rm{r}})(1+z)^{3}\\
&\left[1-\frac{\Omega_{\rm{de}}}{1-\Omega_{\rm{b}}-\Omega_{\rm{r}}}\left(1-(1+z)^{3\zeta(1+\eta)}\right)\right]^{1\over 1+\eta}.\label{ngcg}
\end{aligned}
\end{equation}

Here, we set $\Omega_{\rm{de}} = 1-\Omega_{\rm{m}}-\Omega_{\rm{r}}$.

\begin{figure*}\centering
\includegraphics[width=7cm]{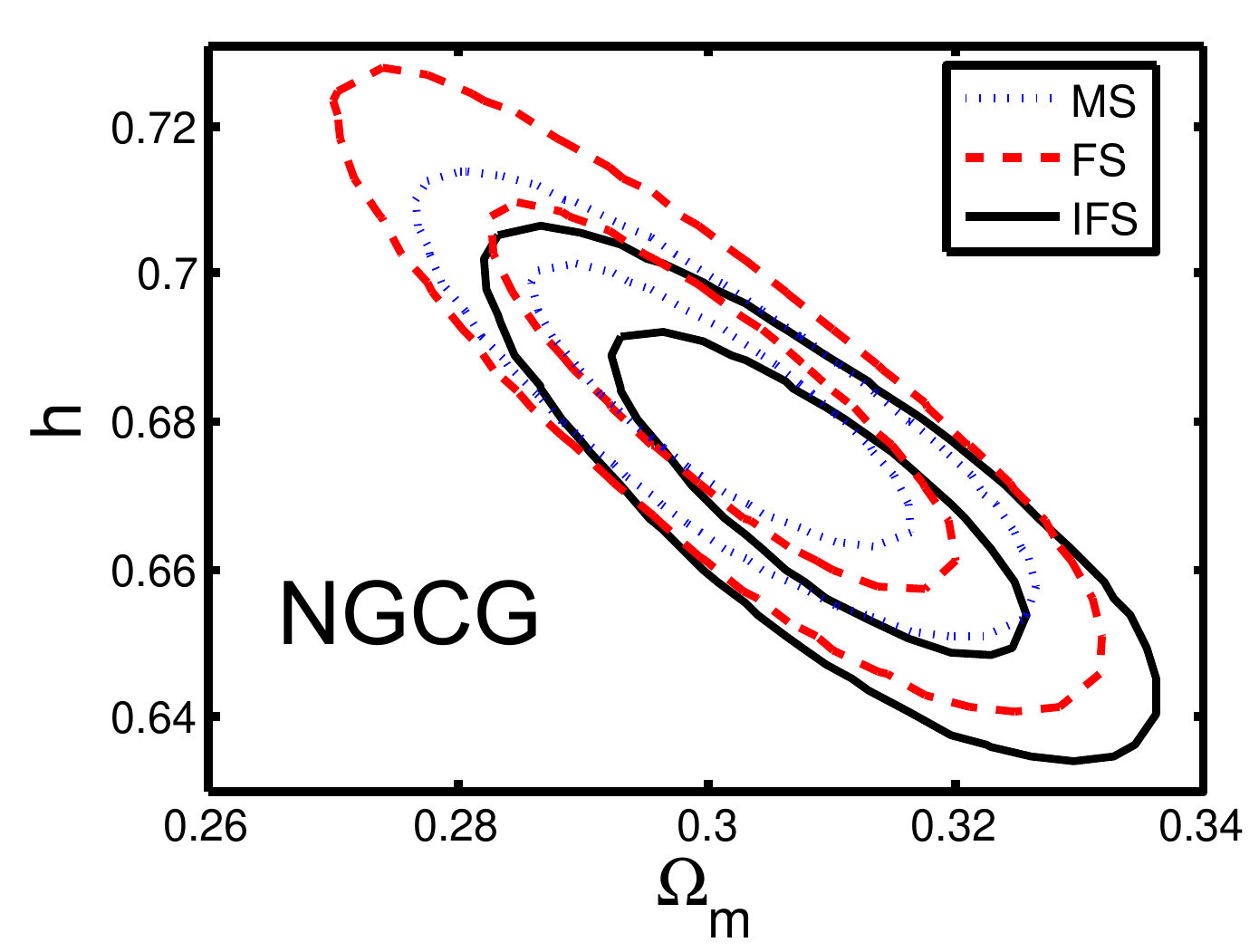}
\includegraphics[width=7cm]{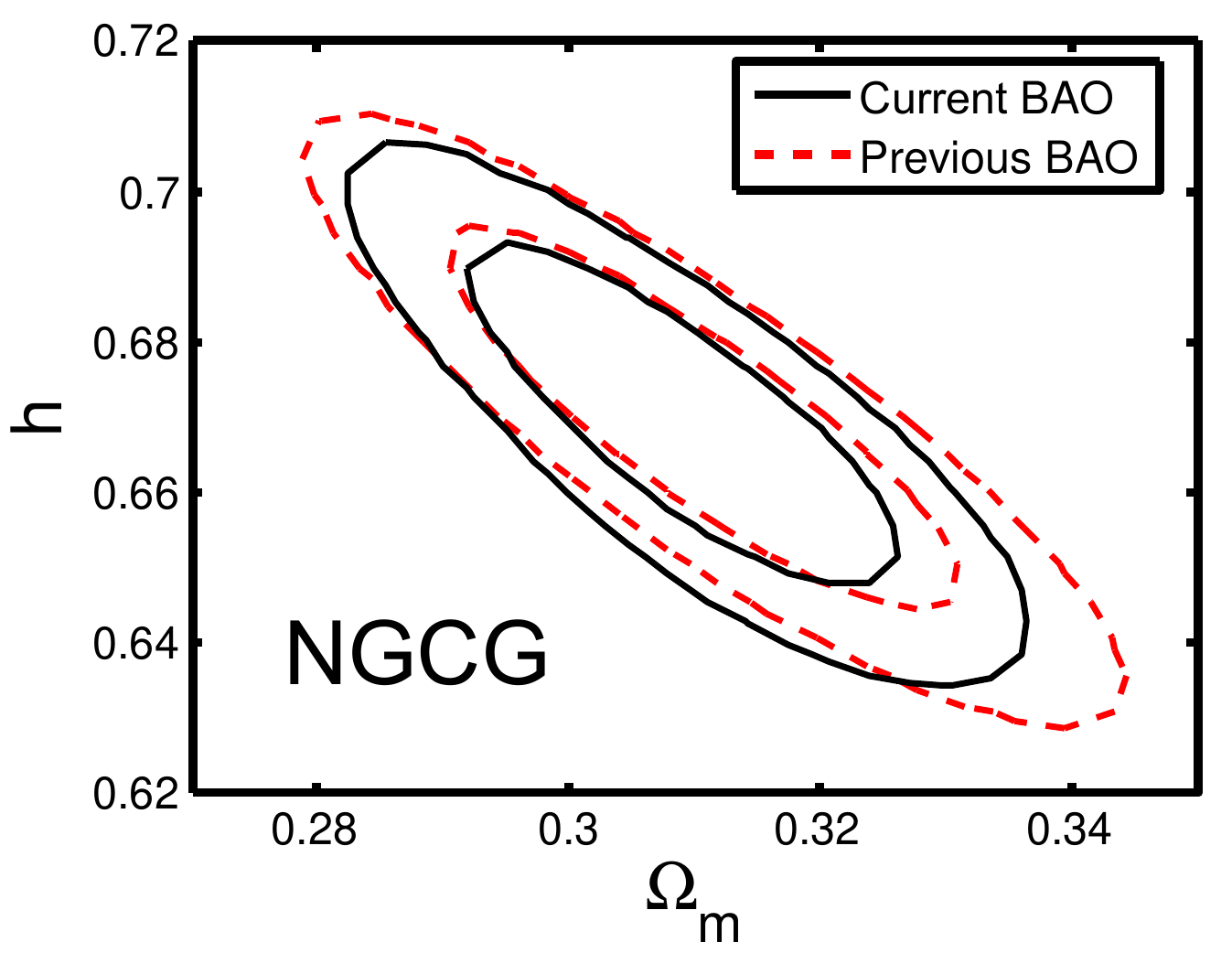}
\includegraphics[width=7cm]{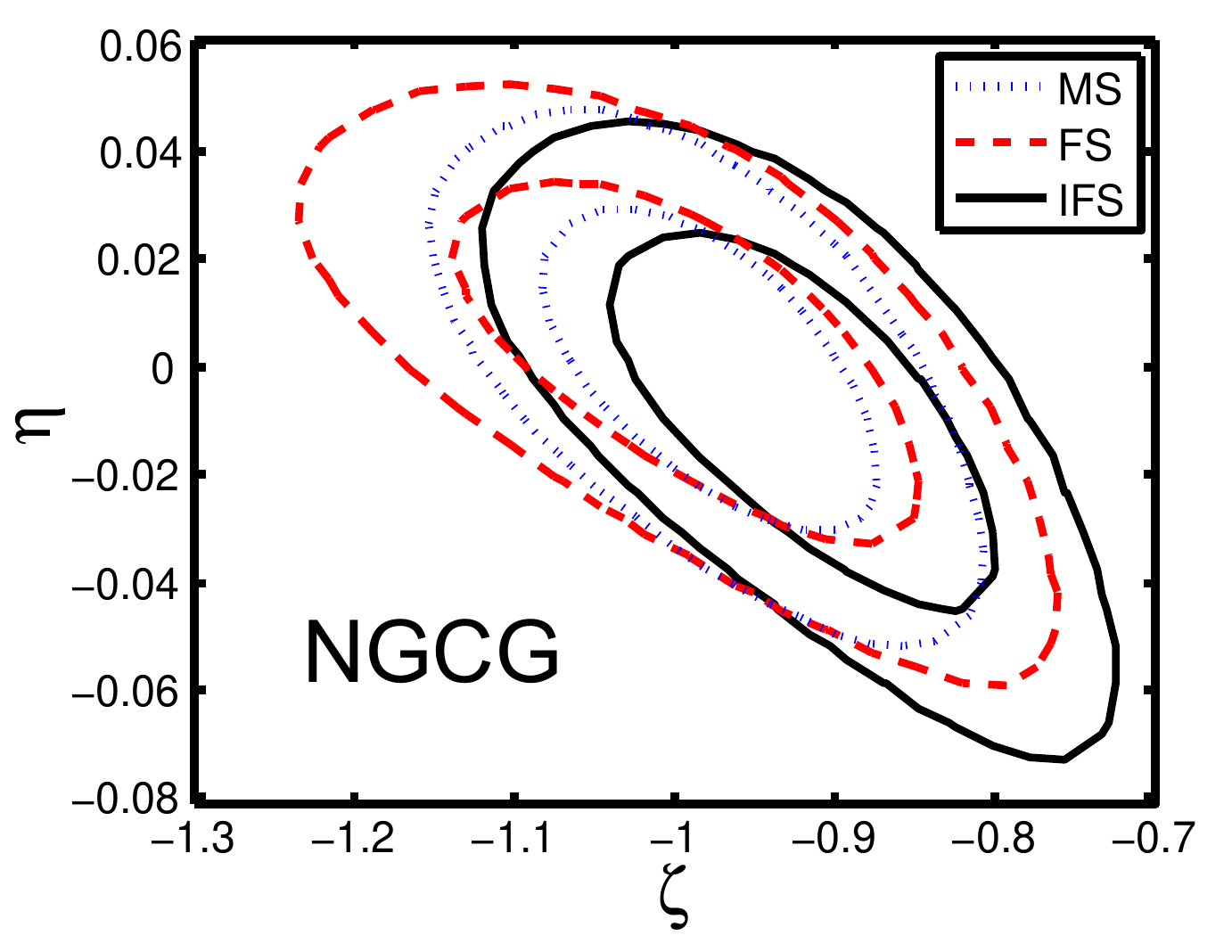}
\includegraphics[width=7cm]{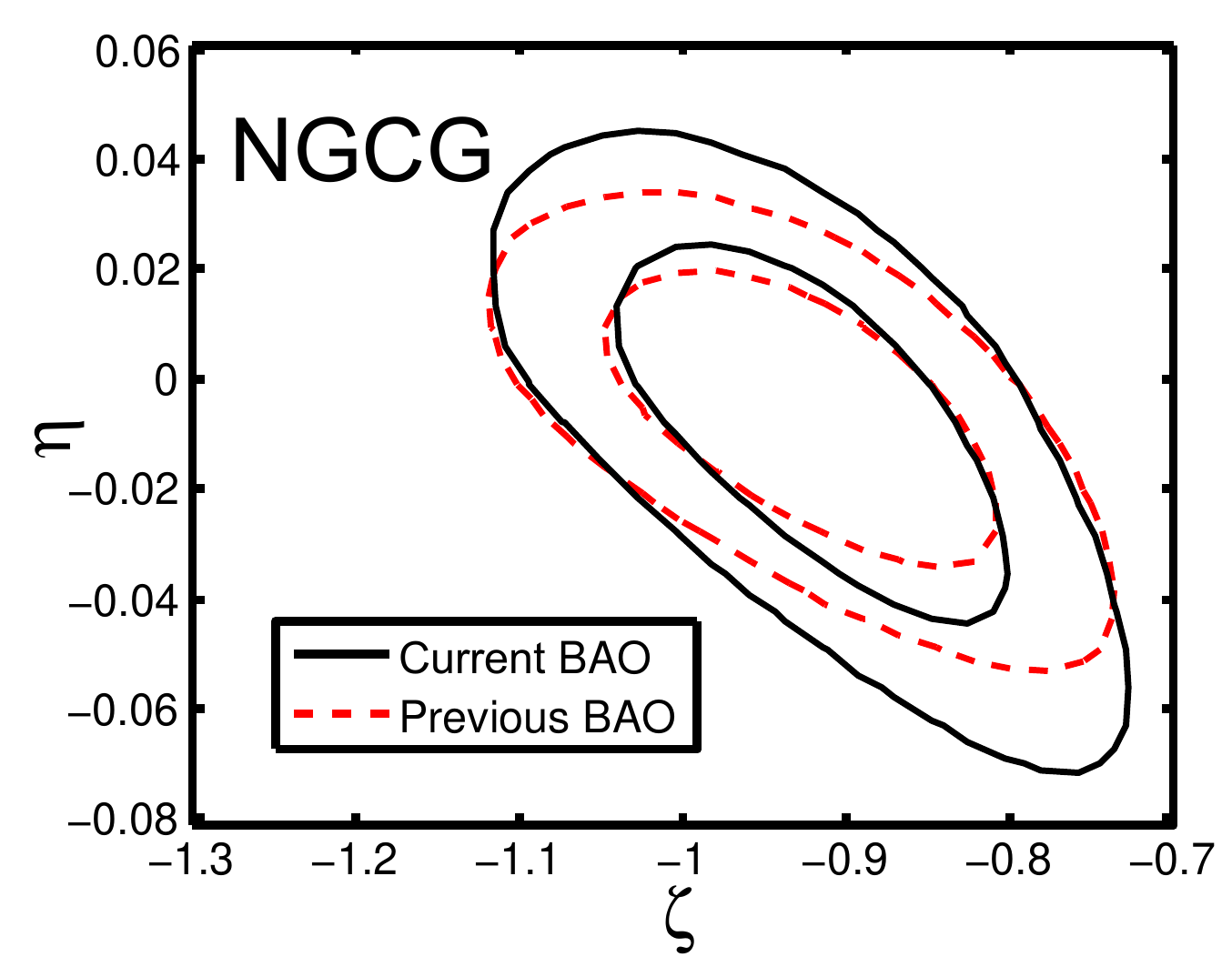}
\caption{\label{f61}NGCG: 1$\sigma$ and 2$\sigma$ confidence regions in the $\Omega_{\rm{m}}$--$h$ (upper panels) and $\zeta$--$\eta$ (lower panels) planes.
The left panels show the effect of different statistic methods of SNIa, where current BAO data is used.
The blue dotted lines denote the results of MS, the red dashed lines represent the results of FS, and the black solid lines are the results of IFS.
The right panels show the effect of different BAO data, where the IFS is used.
The black solid lines denote the results of current BAO data, and the red dashed lines represent the results of previous BAO data.}
\end{figure*}

In Fig.~\ref{f61}, for NGCG, we plot 1$\sigma$ and 2$\sigma$ confidence regions in the  $\Omega_{\rm{m}}$--$h$ (upper panels) and $\zeta$--$\eta$ (lower panels) planes.
The left panels show the effect of different statistic methods of SNIa, where current BAO data is used in the analysis.
For the best-fit results, IFS yields a bigger $\Omega_{\rm{m}}$ and a smaller $h$.
The right panels show the effect of different BAO data, where the IFS is used in the analysis.
We also find that, compared with the results of previous BAO data,
adopting current BAO data  will not have significant effects on the best-fit values of parameters.
In addition, from the lower panels, we find that the point $(\zeta = -1, \eta= 0)$ lies in the 1$\sigma$ region of the $\zeta$--$\eta$ contour,
indicating that $\Lambda$CDM limit of this model is favored.

\subsection {Holographic dark energy models}
Holographic dark energy models arise from the holographic principle. When the effect of gravity under the holographic principle is considered, the density of DE $\rho_{\rm{de}}$ is given by
\begin{equation}
\rho_{\rm{de}}=3c^2M_{pl}^2L^{-2},
\end{equation}
Different choices of the IR
cutoff L lead to different holographic DE models.
Here, we consider three holographic DE models: HDE \cite{hde}, ADE \cite{ade}, and RDE \cite{rde}.

\subsubsection{The original holographic dark energy}

HDE \cite{hde} chooses the future event horizon size:
\be
\label{Reh}
R_{eh} = a\int_{t}^{\infty}\frac{dt'}{a}=a\int_{a}^{\infty}\frac{da'}{Ha'^2},
\ee
as its IR cutoff
scale. The energy density of HDE reads
\begin{equation}
\rho_{de}=3c^2M^2_{Pl}R_{eh}^{-2},\label{hde}
\end{equation}
where $c$ is a constant which plays an important
role in determining properties of the holographic
DE. In this case, $E(z)$ is given by
\be
\label{Ez2}
E(z)=\left(\Omega_{m}(1+z)^3+\Omega_{r}(1+z)^4\over 1-\Omega_{de}(z)\right)^{1/2}.
\ee

The evolution of the dark energy density parameter
$\Omega_{de}(z)=\rho_{de}(z)/(3M_{Pl}H^2)$ is determined by a
differential equation:
\begin{equation}
\label{HDE}
\frac{d \Omega_{de}(z)}{dz}=-\frac{2\Omega_{de}(z)(1-\Omega_{de}(z))}{1+z} \left(
\frac{1+2\theta}{2(1+\theta)}+\frac{\sqrt{\Omega_{de}(z)}}{c} \right),
\end{equation}
where
\begin{equation}
\theta=\Omega_{r}(1+z)/\Omega_{m}.
\end{equation}

Solving Eq. (\ref{HDE}) numerically
and substituting the resultant $\Omega_{de}(z)$ into Eq.
(\ref{Ez2}), the corresponding $E(z)$ can be obtained.

\begin{figure*}\centering
\includegraphics[width=7cm]{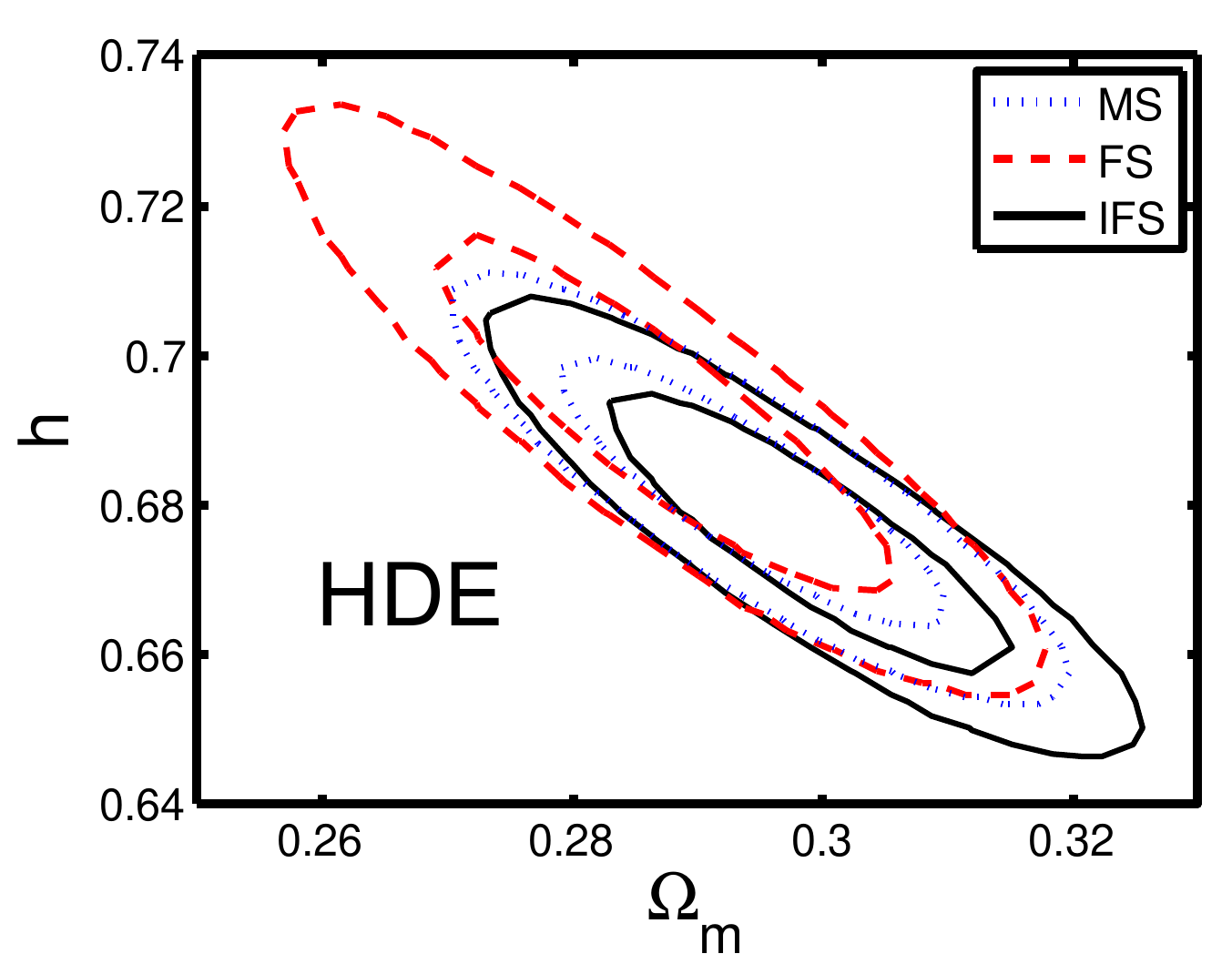}
\includegraphics[width=7cm]{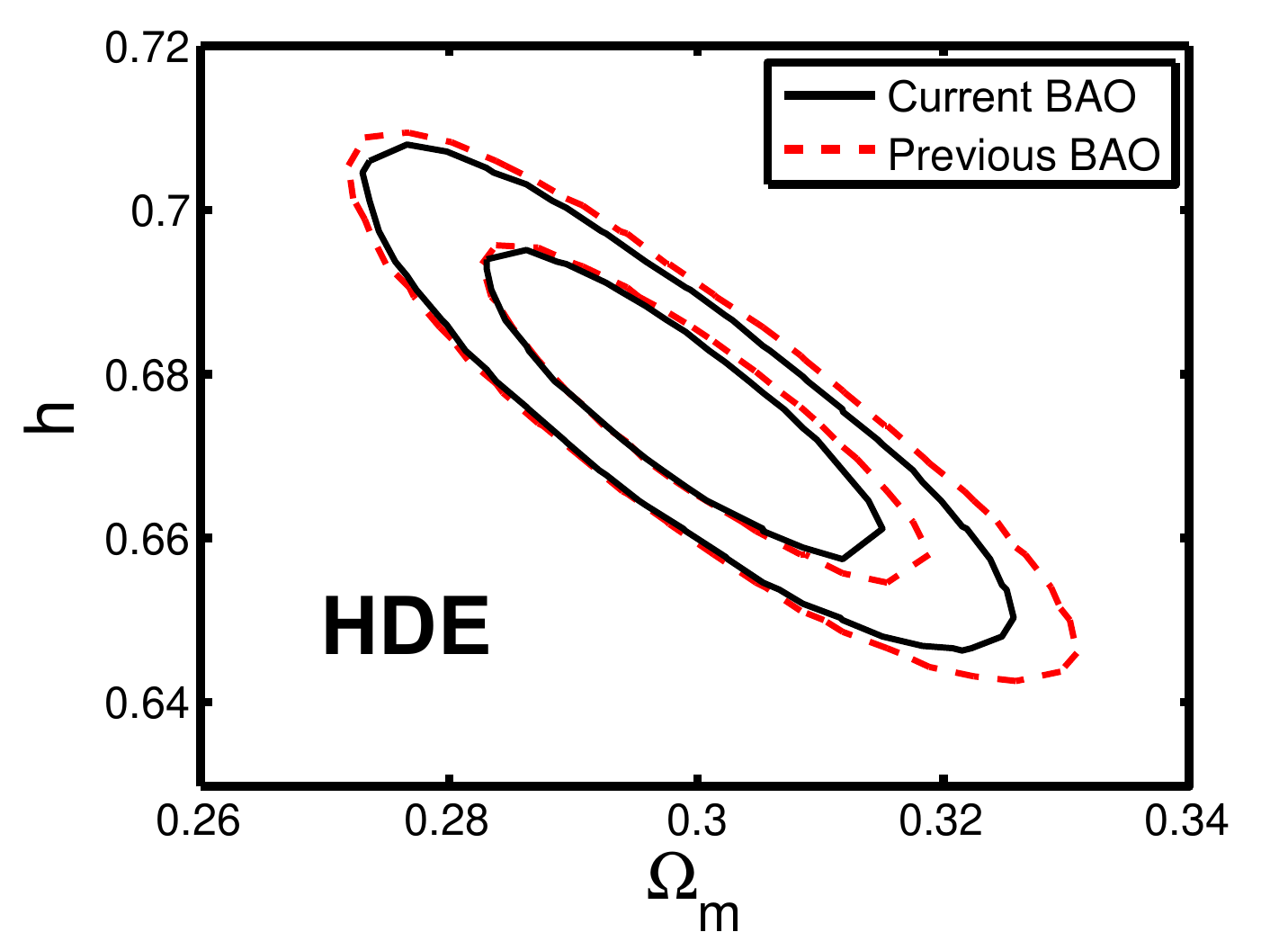}
\includegraphics[width=7cm]{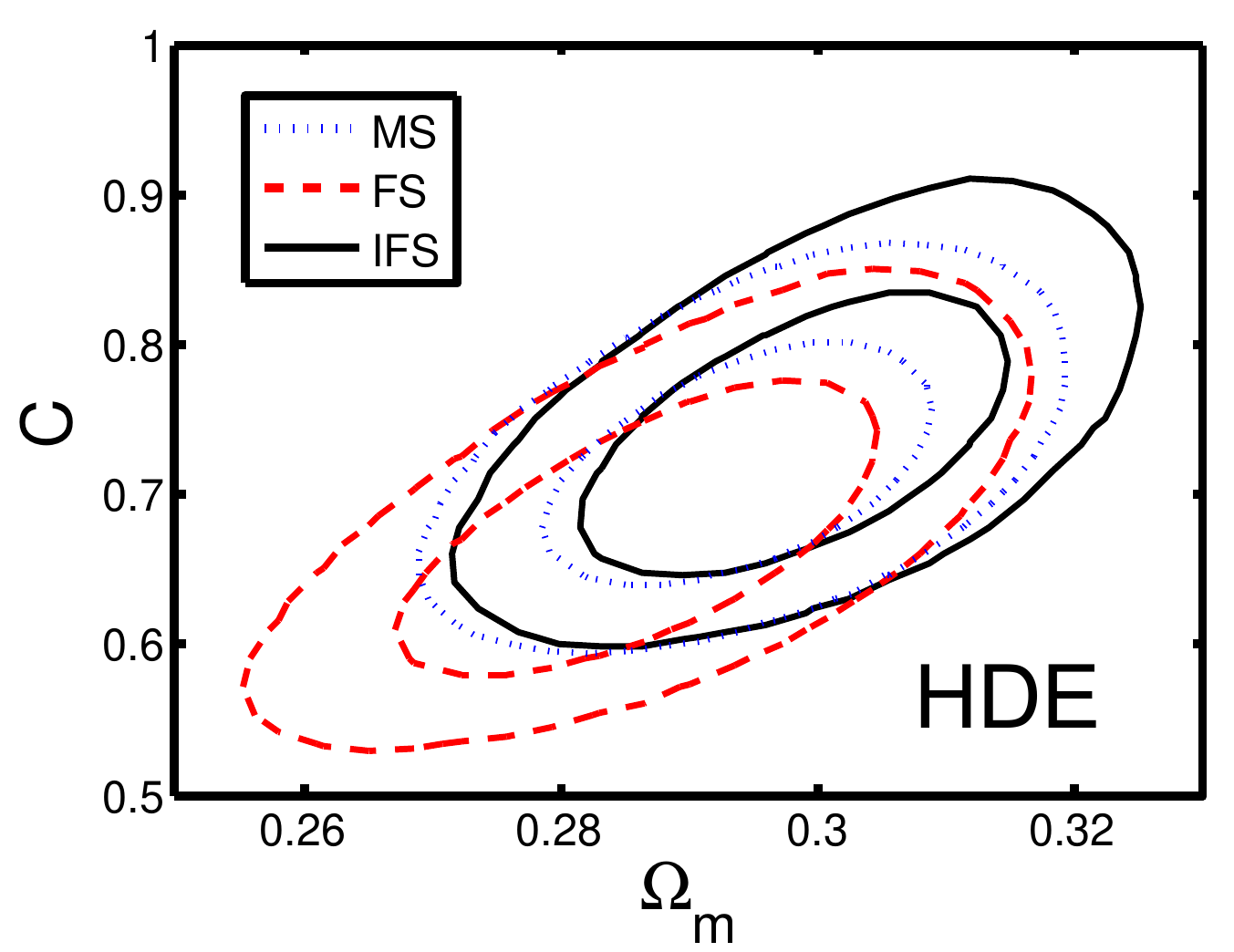}
\includegraphics[width=7cm]{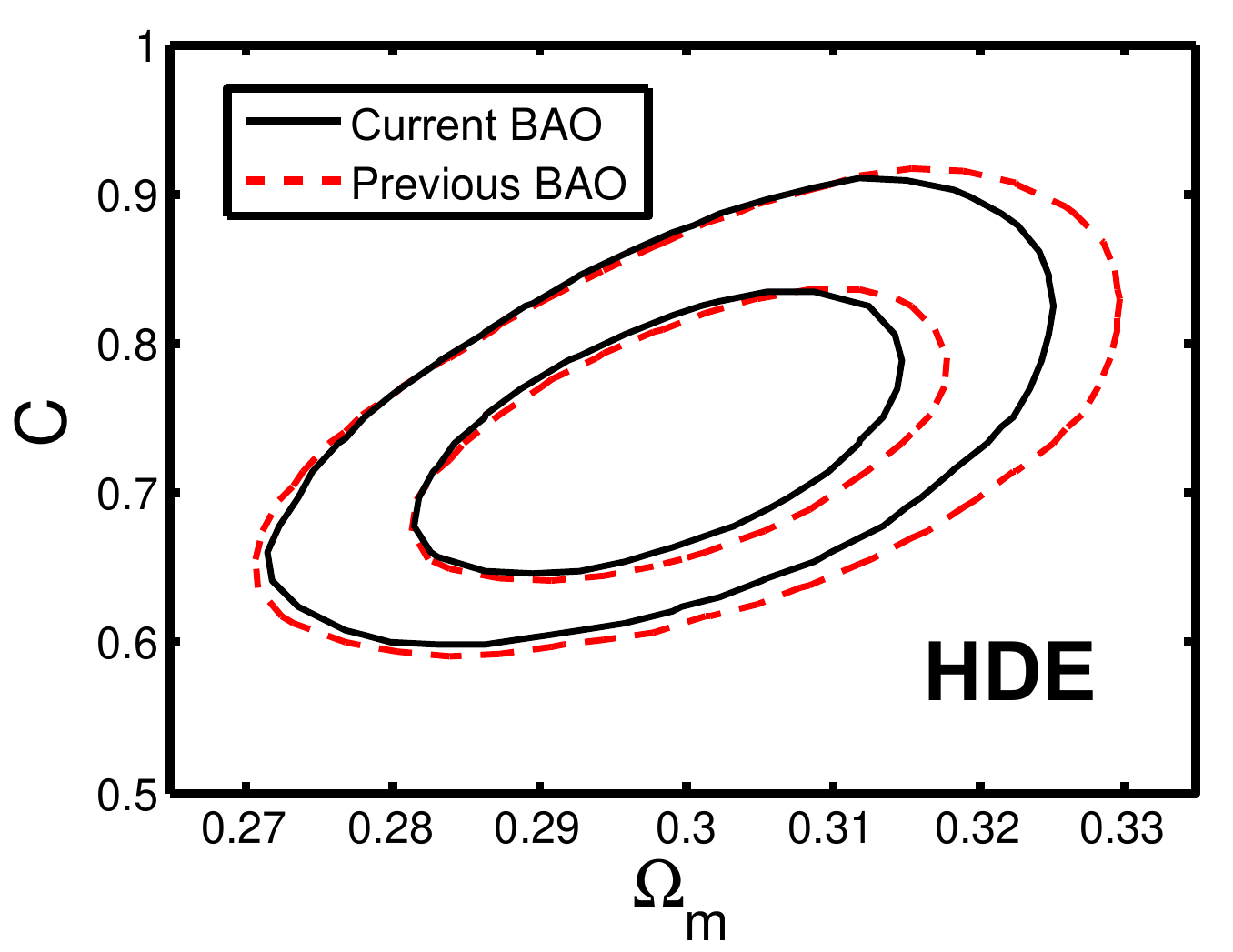}
\caption{\label{f71}HDE: 1$\sigma$ and 2$\sigma$ confidence regions in the $\Omega_{\rm{m}}$--$h$ (upper panels) and $\Omega_{\rm{m}}$--$c$ (lower panels) planes.
The left panels show the effect of different statistic methods of SNIa, where current BAO data is used.
The blue dotted lines denote the results of MS, the red dashed lines represent the results of FS, and the black solid lines are the results of IFS.
The right panels show the effect of different BAO data, where the IFS is used.
The black solid lines denote the results of current BAO data, and the red dashed lines represent the results of previous BAO data. }
\end{figure*}

In Fig.~\ref{f71}, we plot 1$\sigma$ and 2$\sigma$ confidence regions in the  $\Omega_{\rm{m}}$--$h$ (upper panels) and $\Omega_{\rm{m}}$--$c$ (lower panels) planes for this model.
The left panels show the effect of different statistic methods of SNIa, where current BAO data is used in the analysis.
For the best-fit results, compared with the results of MS, IFS yields a bigger $\Omega_{\rm{m}}$ and a smaller $h$, while FS yields a smaller $\Omega_{\rm{m}}$ and a bigger $h$.
The right panels show the effect of different BAO data, where the IFS is used in the analysis.
We also find that, compared with the results of previous BAO data,
adopting current BAO data can give a tighter constraint for this model, but will not have significant effects on the best-fit values of parameters.

\subsubsection {Agegraphic dark energy model }

ADE \cite{ade} chooses the conformal age of the universe
\begin{equation}
\eta=\int_0^t \frac{dt'}{a}=\int_0^a \frac{da'}{Ha'^{2}}
\end{equation}
as the IR cutoff, so the energy density of ADE is
\begin{equation}
\label{ade}
\rho_{de}=3n^{2}M_{Pl}^{2}\eta^{-2},
\end{equation}
where $n$ is a constant which plays the same role as $c$ in HDE.

As the same as HDE, $E(z)$ is also given by
Eq.~(\ref{Ez2}), where the function $\Omega_{de}(z)$ is governed by
the differential equation:
\begin{equation}
\label{ADE}
\frac{d \Omega_{de}(z)}{dz}=-\frac{2\Omega_{de}(z)}{1+z} \left(
\epsilon(z)-\frac{(1+z)\sqrt{\Omega_{de}(z)}}{n} \right),
\end{equation}
where
\begin{equation}
\label{epsilon}
\epsilon(z)={3\over 2}\left[{1+{4\over 3}\theta \over
1+\theta}(1-\Omega_{de}(z))+(1+w_{de}(z))\Omega_{de}(z)\right],
\end{equation}
with
\begin{equation}
\theta=\Omega_{r}(1+z)/\Omega_{m}, w_{de}(z)=-1+\frac{2(1+z)\sqrt{\Omega_{de}(z)}}{3n}.
\end{equation}
Following Ref.~\cite{Wei:2007xu}, we choose the initial condition,
$\Omega_{de}(z_{ini})=n^2(1+z_{ini})^{-2}/4$, at $z_{ini}=2000$, and
then Eq.~(\ref{ADE}) can be numerically solved.  Note that once $n$ is given, by
solving Eq.~(\ref{ADE}), $\Omega_{m}=1-\Omega_{de}(0)-\Omega_{r}$
can be derived accordingly.

\begin{figure*}\centering
\includegraphics[width=7cm]{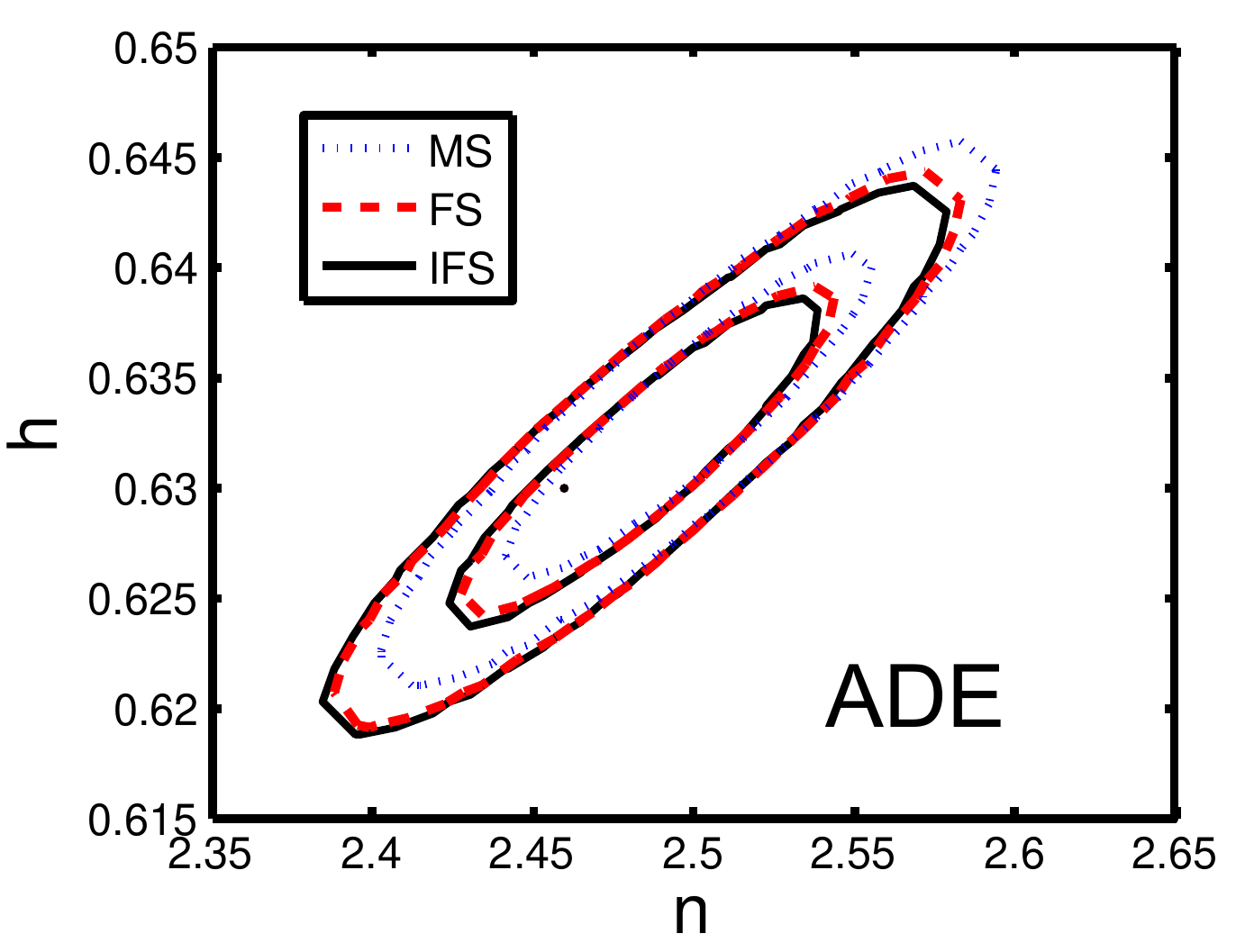}
\includegraphics[width=7cm]{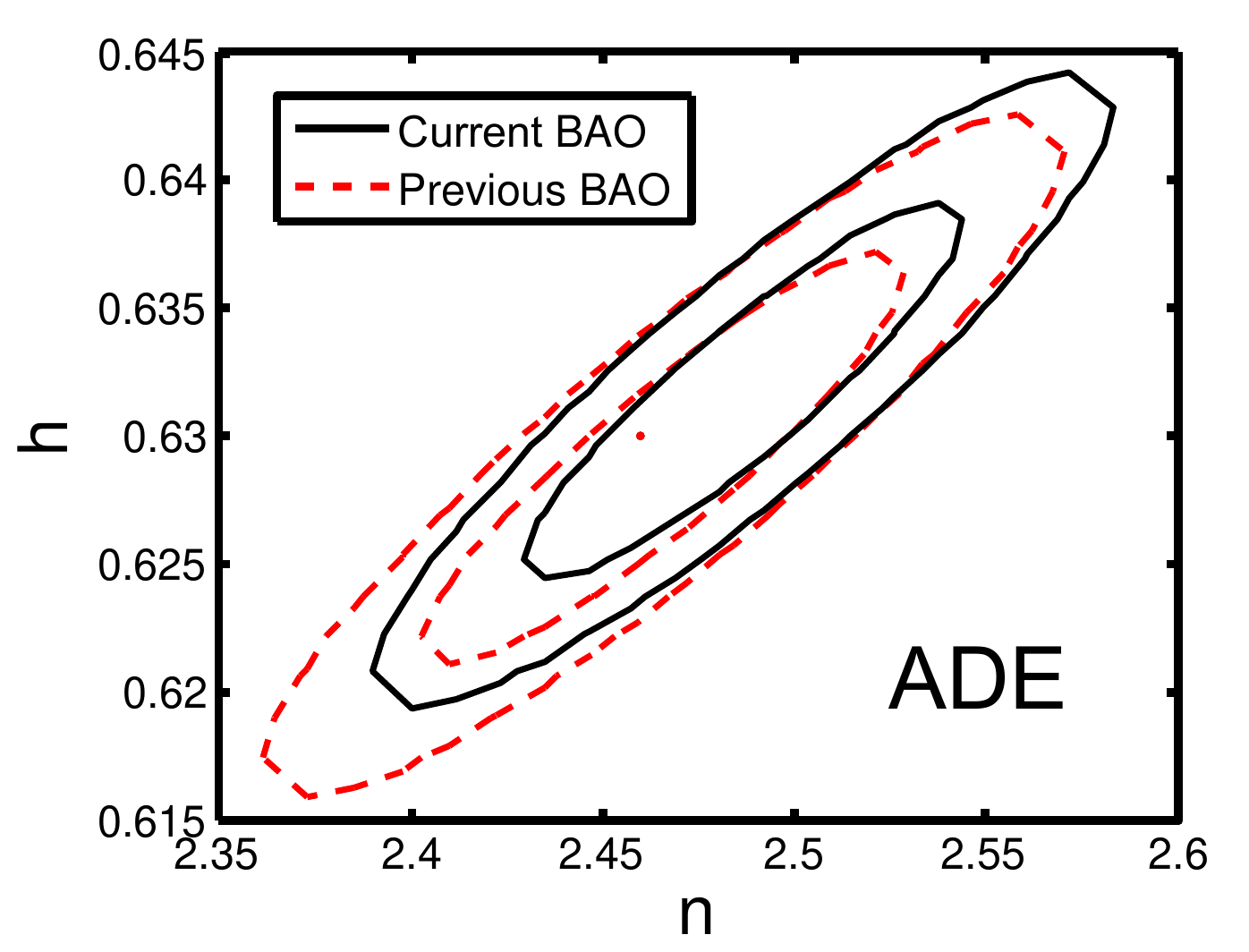}
\caption{\label{f81}ADE: 1$\sigma$ and 2$\sigma$ confidence regions in the $n$--$h$ plane.
The left panel shows the effect of different statistic methods of SNIa, where current BAO data is used.
The blue dotted lines denote the results of MS, the red dashed lines represent the results of FS, and the black solid lines are the results of IFS.
Right panel shows the effect of different BAO data, where the IFS is used.
The black solid lines denote the results of current BAO data, and the red dashed lines represent the results of previous BAO data.}
\end{figure*}

In Fig.~\ref{f81}, for ADE, we plot 1$\sigma$ and 2$\sigma$ confidence regions in the $n$--$h$ plane.
The left panel shows the effect of different statistic methods of SNIa, where current BAO data is used in the analysis.
For the best-fit results, IFS yields a smaller $n$ and a smaller $h$.
Note that a smaller $n$ is corresponding to a bigger $\Omega_m$.
The right panel shows the effect of different BAO data, where the IFS is used in the analysis.
We also find that, compared with the results of previous BAO data,
adopting current BAO data does not have significant effects on the best-fit values of parameters.

\subsubsection {Ricci dark energy model}

RDE~\cite{rde} chooses the average radius of the
Ricci scalar curvature as the IR cutoff length scale.
In a flat universe, the
Ricci scalar is ${\cal R}=-6(\dot{H}+2H^2)$, so the energy density of RDE can be expressed as,
\begin{equation}
\rho_{de}=3\gamma M_{Pl}^2(\dot{H}+2H^2),\label{rde}
\end{equation}
where $\gamma$ is a positive constant. The E(z) is determined by the following equation:
\begin{equation}
E^2=\Omega_{m}e^{-3x}+\Omega_{r}e^{-4x}+\gamma \left({1\over
2}{dE^2\over dx}+2E^2\right),
\end{equation}
where $x=\ln a$. Solving this equation, we get the
following form:
\begin{equation}
\begin{aligned}
E(z)^{2}&=\frac{2\Omega_{\rm{m}}}{2-\gamma}(1+z)^{3}+\Omega_{\rm{r}}(1+z)^{4}\\
&+\left(1-\Omega_{\rm{r}}-\frac{2\Omega_{\rm{m}}}{2-\gamma}\right)(1+z)^{(4-\frac{2}{\gamma})}.
\end{aligned}
\end{equation}

\begin{figure*}\centering
\includegraphics[width=7cm]{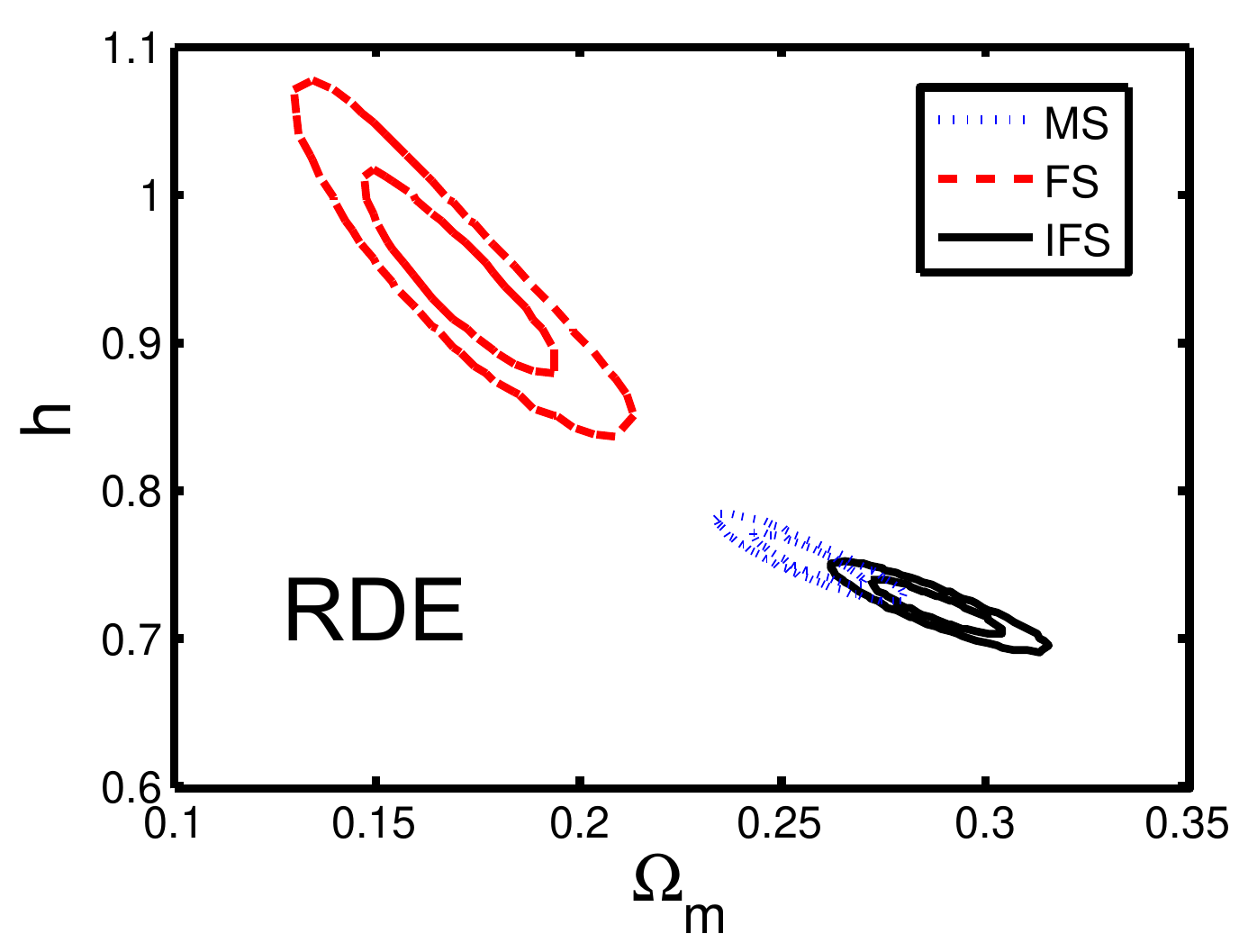}
\includegraphics[width=7cm]{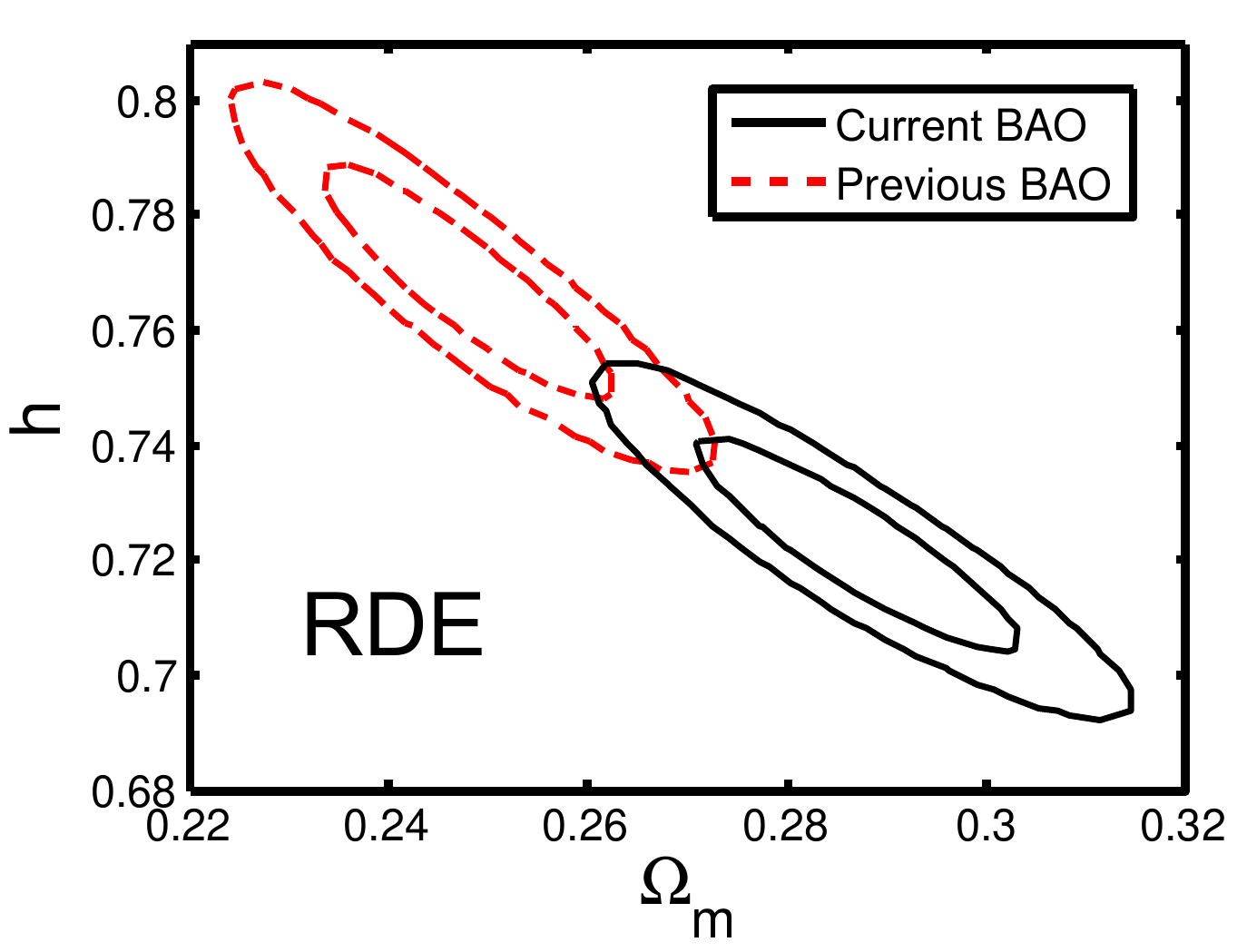}
\includegraphics[width=7cm]{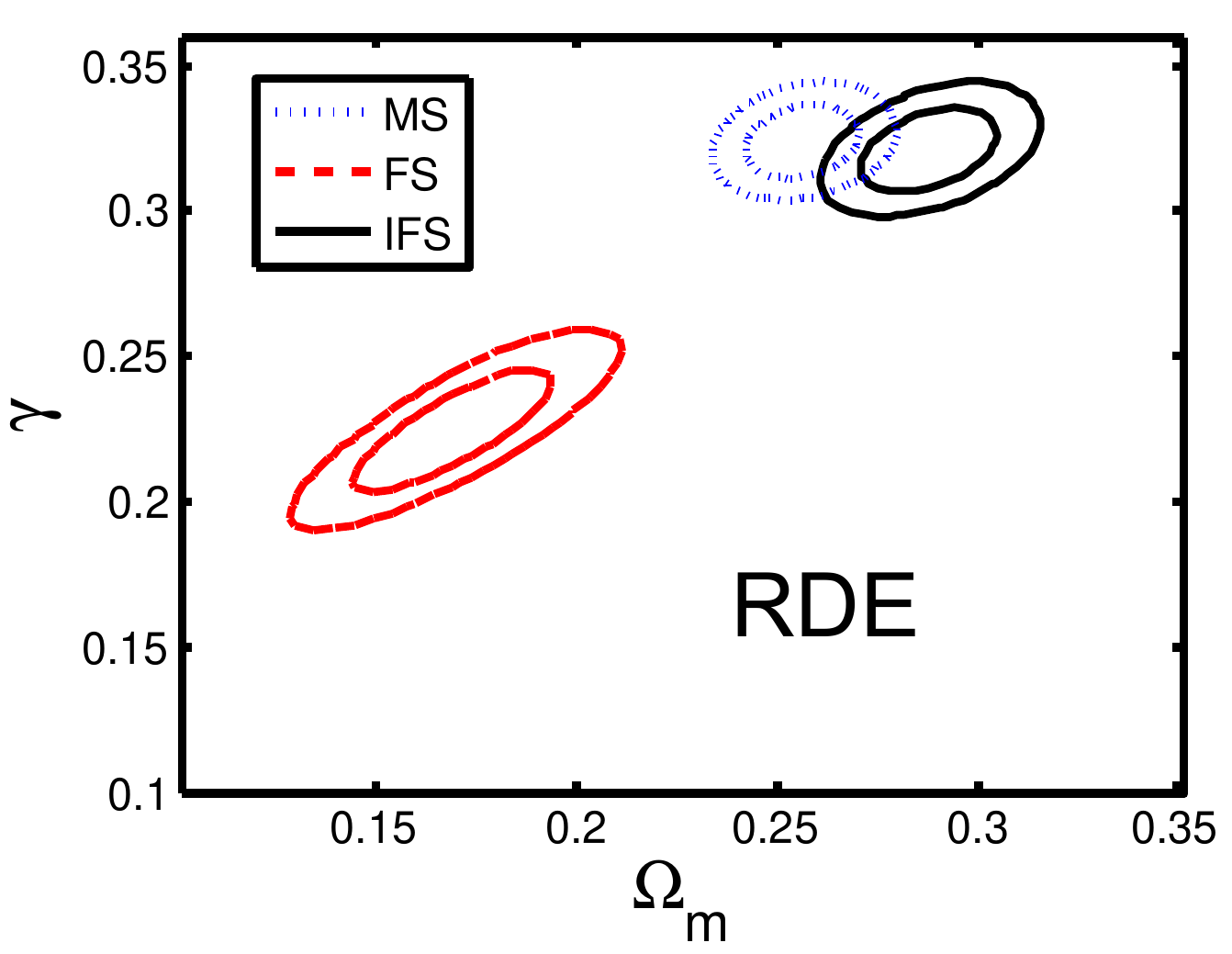}
\includegraphics[width=7cm]{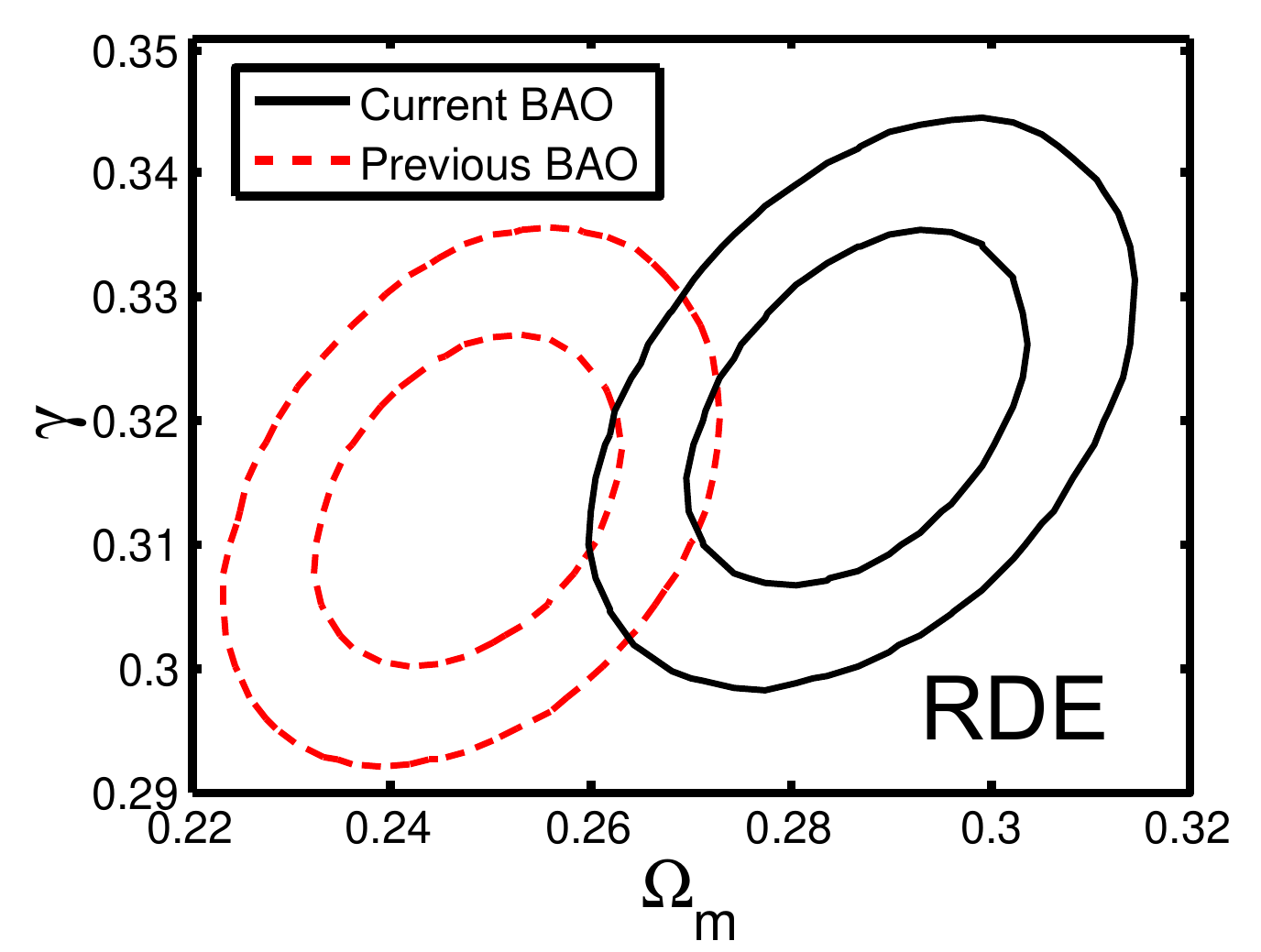}
\caption{\label{f91}RDE: 1$\sigma$ and 2$\sigma$ confidence regions in the $\Omega_{\rm{m}}$--$h$ (upper panels) and $\Omega_{\rm{m}}$--$\gamma$ (lower panels) planes.
The left panels show the effect of different statistic methods of SNIa, where current BAO data is used.
The blue dotted lines denote the results of MS, the red dashed lines represent the results of FS, and the black solid lines are the results of IFS.
The right panels show the effect of different BAO data, where the IFS is used.
The black solid lines denote the results of current BAO data, and the red dashed lines represent the results of previous BAO data.}
\end{figure*}

For RDE, 1$\sigma$ and 2$\sigma$ confidence regions in the $\Omega_{\rm{m}}$--$h$ (upper panels) and $\Omega_{\rm{m}}$--$\gamma$ (lower panels) planes are shown in Fig.~\ref{f91}.
We find that the fitting results given by FS deviate from the results of  MS  and IFS at least 4$\sigma$ C.L.\footnote{{\bf It is well known that, for a specific model, different observational data will give different parameter space. For many DE models, the difference among the parameter spaces given by different observational data are very small, so these models can fit current observations well. For the RED model, the difference among the parameter spaces given by different observational data are very large, so this model has been ruled out by current observations. This is the reason that there is a significant difference between the likelihoods in Fig.~\ref{f91}.}}
In addition, the fitting results given by current BAO data deviate from the results of  previous BAO data at least 1$\sigma$ C.L.
This implies that adopting different statistic methods of SNIa will cause serious tension for the parameter estimation of RDE.
Both different statistic method and different BAO data have great impact on  parameters estimation for RDE.

\subsection{Modified gravity models}
The MG theory can yield ``effective dark energy'' models
mimicking the real dark energy at the background cosmology level. It can lead to an accelerated universe without introducing the dark energy. In this subsection, we consider three models: DGP \cite{DGP},
$\alpha$DE\cite{alphaDE} and CMG \cite{Heisenberg2016a}.

\subsubsection {Dvali-Gabadadze-Porrati}

For DGP \cite{DGP}, the Friedmann equation is governed by
\begin{equation}
E(z)^2-{E(z)\over r_c}=\Omega_m(1+z)^3+\Omega_r(1+z)^4,
\end{equation}
where $r_c=(H_0(1-\Omega_m-\Omega_r))^{-1}$ is the crossover scale.
The reduced Hubble parameter $E(z)$ is given by \be \label{DGP}
E(z)=\sqrt{\Omega_{m}(1+z)^3+\Omega_{r}(1+z)^4+\Omega_{r_c}}+\sqrt{\Omega_{r_c}},
\ee where $\Omega_{rc}=1/(4r_c^2H_0^2)$ is a constant.

\begin{figure*}\centering
\includegraphics[width=7cm]{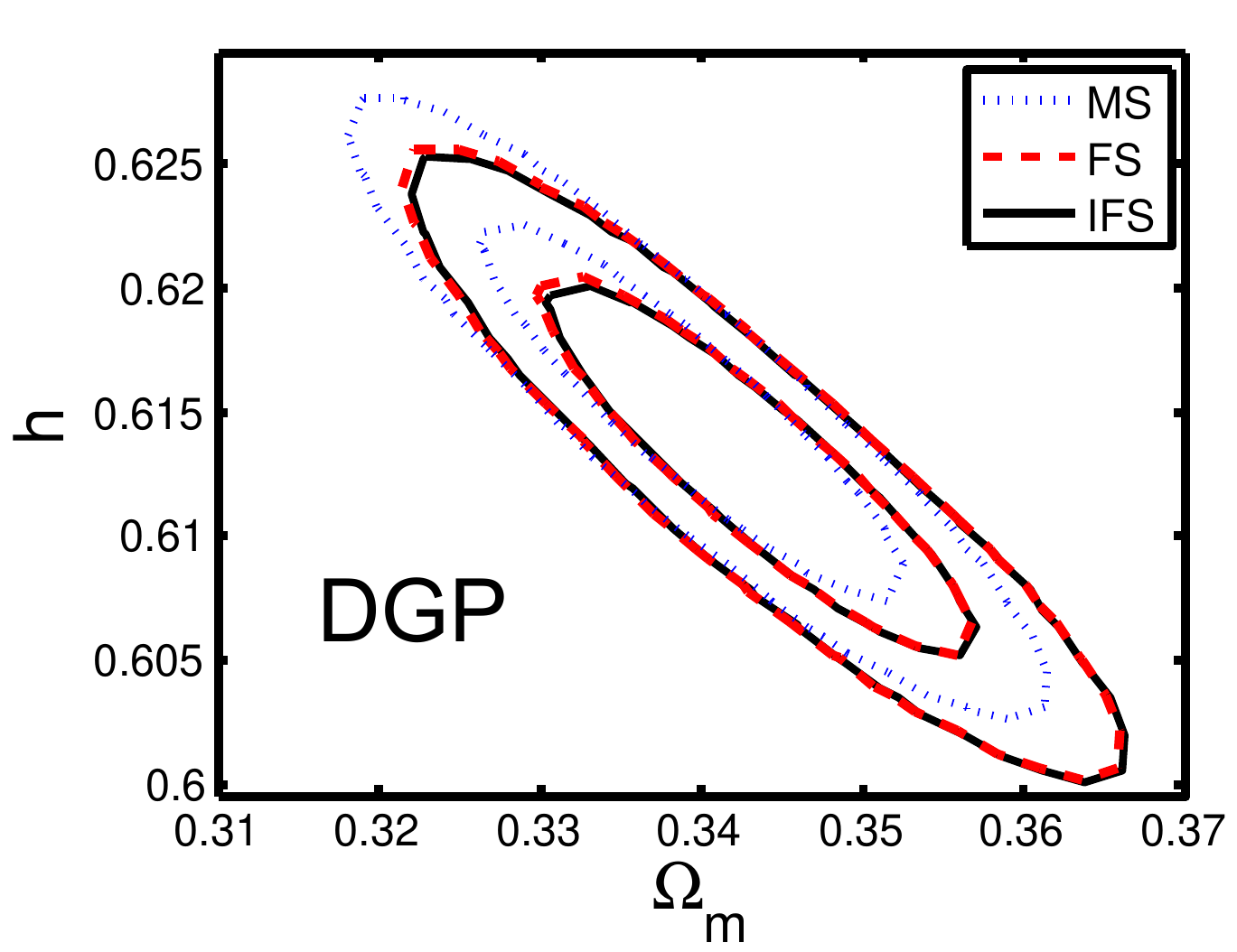}
\includegraphics[width=7cm]{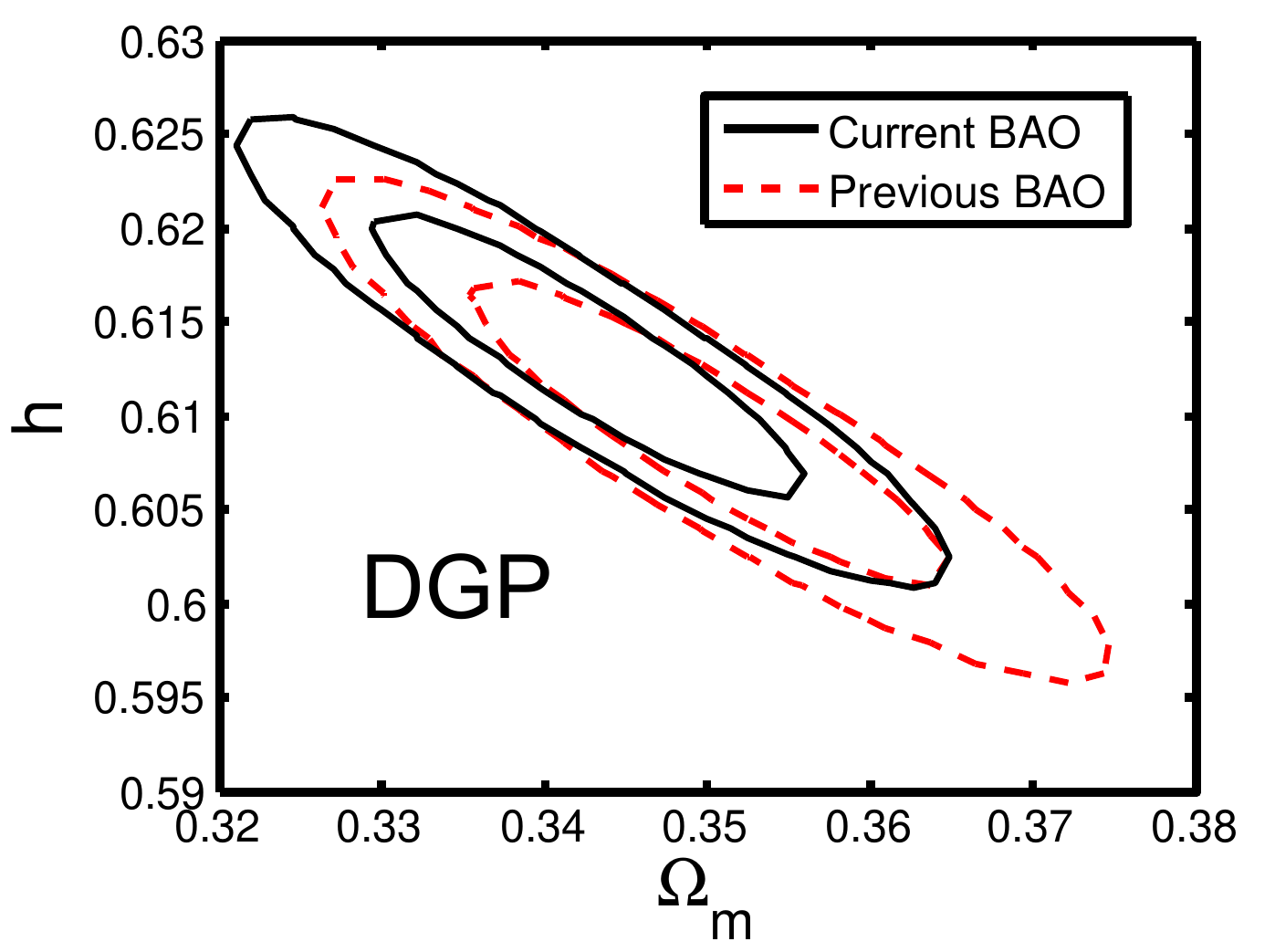}
\caption{\label{f101}DGP: 1$\sigma$ and 2$\sigma$ confidence regions in the $\Omega_{\rm{m}}$--$h$ plane.
The left panel shows the effect of different statistic methods of SNIa, where current BAO data is used.
The blue dotted lines denote the results of MS, the red dashed lines represent the results of FS, and the black solid lines are the results of IFS.
Right panel shows the effect of different BAO data, where the IFS is used.
The black solid lines denote the results of current BAO data, and the red dashed lines represent the results of  previous BAO data.}
\end{figure*}

In Fig.~\ref{f101}, for DGP, we plot 1$\sigma$ and 2$\sigma$ confidence regions in the $\Omega_{\rm{m}}$--$h$ plane.
The left panel shows the effect of different statistic methods of SNIa, where current BAO data is used in the analysis.
For the best-fit results, IFS yields a bigger $\Omega_{\rm{m}}$ and a smaller $h$.
The right panel shows the effect of different BAO data, where the IFS is used in the analysis.
We find that different BAO data have obviously impact on parameter estimation for DGP.

\subsubsection {DGP's phenomenological extension }

The $\alpha$DE is an extension of DGP's phenomenological. It is proposed by Dvali and Turner \cite{alphaDE}.
By adding a parameter $\alpha$, it can interpolate between the DGP and the $\Lambda$CDM.
In this model, the Friedmann equation is modified as
\begin{equation}
E(z)^2-{E(z)^\alpha\over
r_c^{2-\alpha}}=\Omega_m(1+z)^3+\Omega_r(1+z)^4,
\end{equation}
where $\alpha$ is a phenomenological parameter, and
$r_c=(1-\Omega_m-\Omega_r)^{1/(\alpha-2)}H_0^{-1}$. According to
this Friedmann equation, $E(z)$ is determined by the following
equation:
\begin{equation}
E(z)^2=\Omega_m(1+z)^3+\Omega_{r}(1+z)^4+E(z)^\alpha(1-\Omega_m-\Omega_{r}).
\end{equation}
Note that it collapses to the DGP, when $\alpha=1$, and
collapses to $\Lambda$CDM, when $\alpha=0$.

\begin{figure*}\centering
\includegraphics[width=7cm]{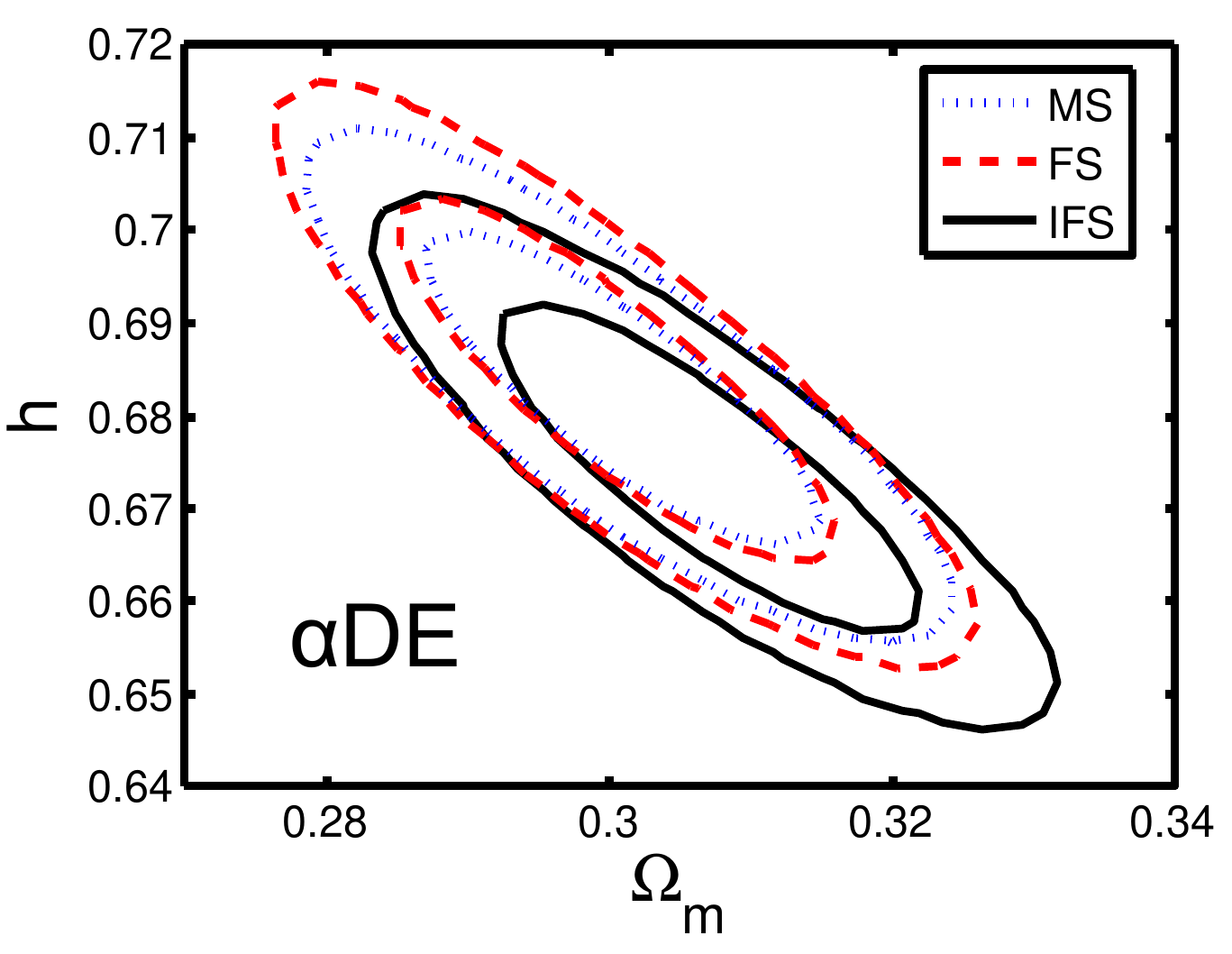}
\includegraphics[width=7cm]{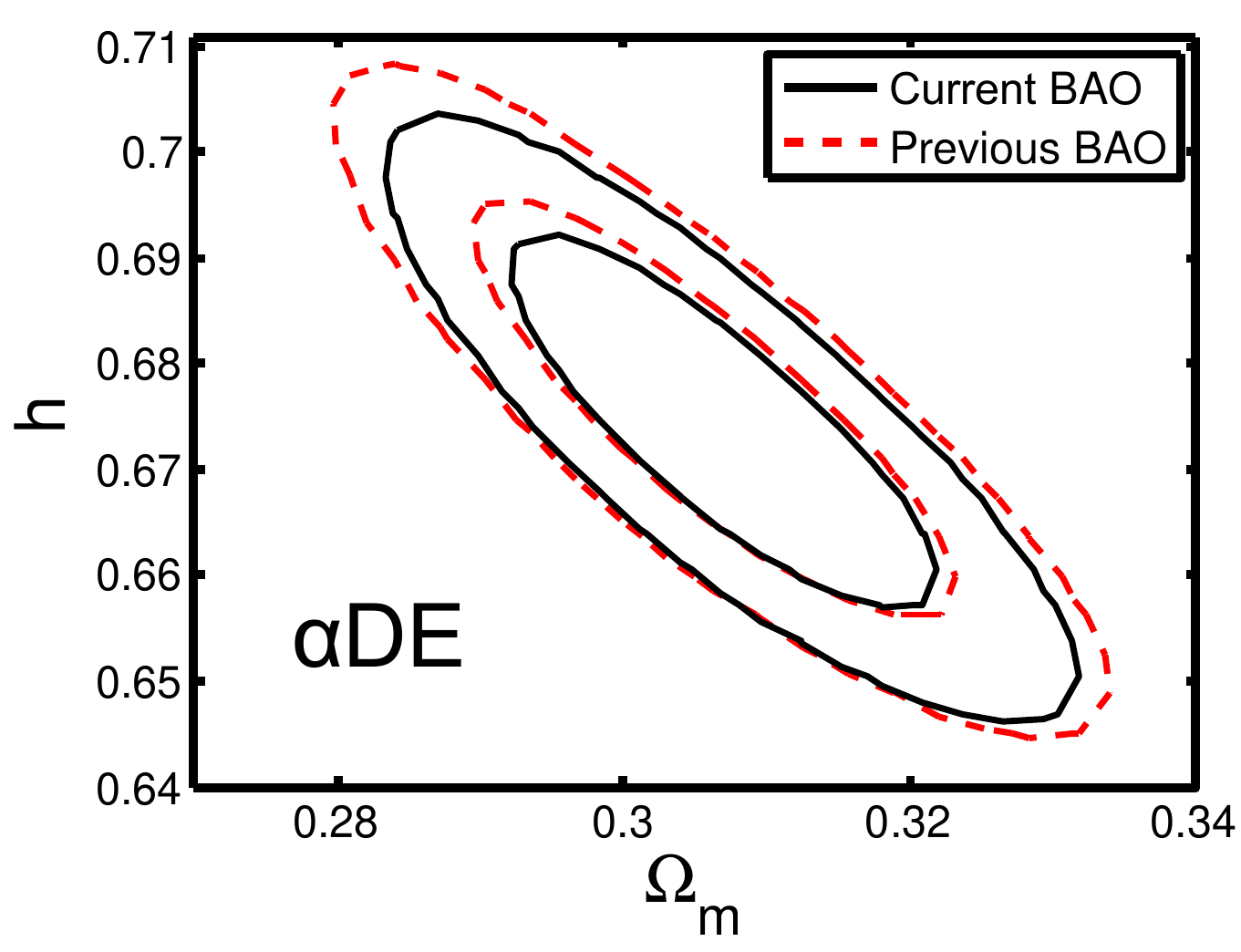}
\includegraphics[width=7cm]{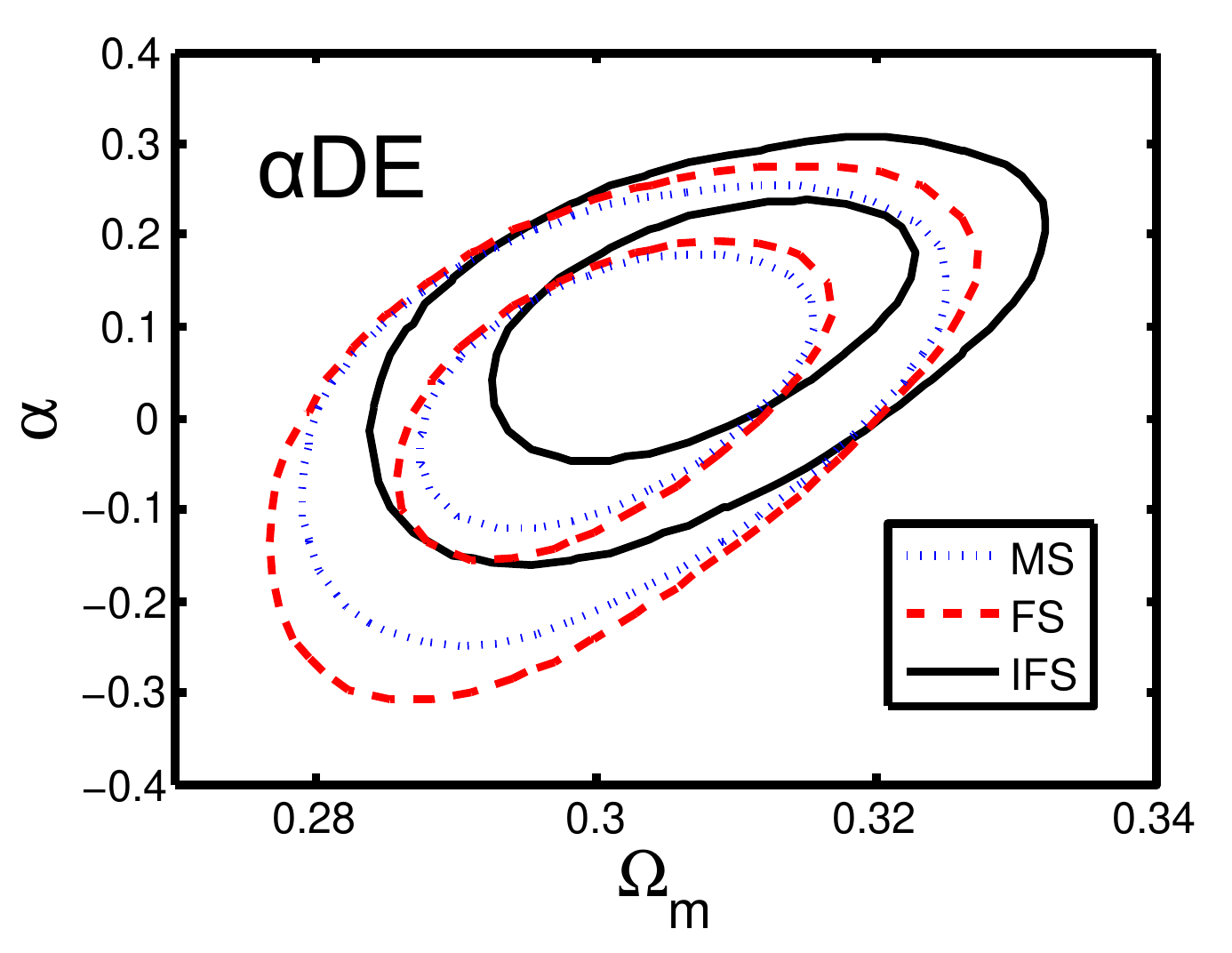}
\includegraphics[width=7cm]{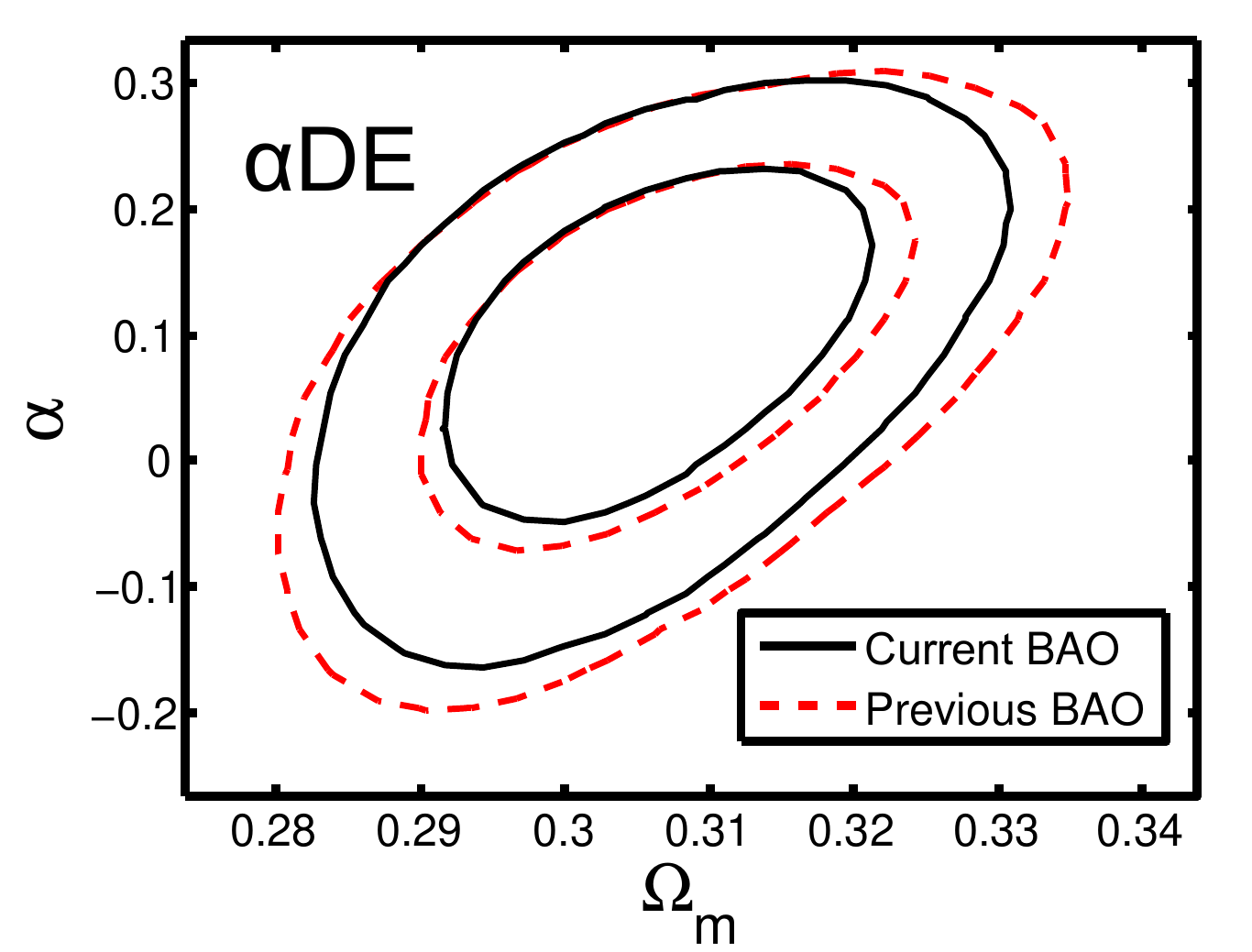}
\caption{\label{f111}The $\alpha$DE: 1$\sigma$ and 2$\sigma$ confidence regions in the $\Omega_{\rm{m}}$--$h$ (upper panels) and $\Omega_{\rm{m}}$--$\alpha$ (lower panels) planes.
The left panels show the effect of different statistic methods of SNIa, where current BAO data is used.
The blue dotted lines denote the results of MS, the red dashed lines represent the results of FS, and the black solid lines are the results of IFS.
The right panels show the effect of different BAO data, where the IFS is used.
The black solid lines denote the results of current BAO data, and the red dashed lines represent the results of previous BAO data.}
\end{figure*}

For this model, we plot the 1$\sigma$ and 2$\sigma$ confidence regions in the $\Omega_{\rm{m}}$--$h$ and $\Omega_{\rm{m}}$--$\alpha$ planes in Fig.~\ref{f111}.
From the left panels, we find that, for the best-fit value, adopting IFS will yield a bigger $\Omega_{\rm{m}}$ and a smaller $h$.
From the right panels, we find that, compared with the results of previous BAO data,
adopting current BAO data can give a tighter constraint for this model, but will not have significant effects on the best-fit values of parameters.
In addition, from the lower panels, we find that $\alpha = 0$ lies in the 1$\sigma$ region of the $\Omega_{\rm{m}}$--$\alpha$ plane.
This indicated that the $\Lambda$CDM limit of this model is favored. Thus, current observational data prefer to $\Lambda$CDM and exclude the DGP.

\subsubsection {Doubly Coupled Massive Gravity}
For this model, a matter field of the dark sector is coupled to an effective composite metric while a standard matter field couples to the dynamical metric \cite{Heisenberg2016a}.
The Hubble function is governed by the following equation:
\begin{eqnarray}
&&12M_{Pl}^2\beta(1+z)H\frac{dH}{dz}=3m^2M_{Pl}^2(\beta(\kappa_1(2-4z) \nonumber\\
&&-z(2+z)\kappa_2+2(\kappa_2+\kappa_3))) +2\rho_{r}(1+z)^4\beta\nonumber\\
&&+2(\kappa_1+(1+z)(\kappa_2+\kappa_3+z\kappa_3)+9M_{Pl}^2H^2)
\end{eqnarray}
where $\kappa_1$, $\kappa_2$
and $\kappa_3$ are combination parameters, and $\beta$ is the coupling parameter. By integrating the above
equation, we obtain the Hubble function, which results in
\begin{eqnarray}\label{HubbleFunction}
H^2&=&\frac{1}{6M_{Pl}^2\beta}(2\rho_{r}\beta(1+z)^4+M_{Pl}^2(-m^2(-3z\beta(2\kappa_1 \nonumber\\
&+&(2+z)\kappa_2)+2\beta\kappa_3+(2\kappa_1+3(1+z)(\kappa_2 \nonumber\\
&+&2(1+z)\kappa_3)))+6\beta(1+z)^3c_1)) \,,
\end{eqnarray}
with $c_1$ being an integration constant. Comparing this equation with the Friedmann equation given at \cite{Heisenberg2016b}, the $c_1$ can be fixed. Rewriting the equation, we get the deduced Hubble parameter
\begin{equation}
E(z)=\sqrt{\Omega_m(1+z)^3+\Omega_{r}(1+z)^4+c_2z^2+c_3z+c_4}.
\end{equation}
where
\begin{eqnarray}\label{HubbleFunction}
c_2&=&-\frac{m^2}{6\beta H_0^2}(6k_3-3\beta k_2), \nonumber\\
c_3&=&-\frac{m^2}{6\beta H_0^2}(12k_3+3k_2-6\beta(k_1+k_2)), \nonumber\\
c_4&=&-\frac{m^2}{6\beta H_0^2}((2+2\beta)k_3+2k_1+3k_2). \
\end{eqnarray}

Here, we take the $c_2$ and $c_3$ as new model parameters.
Note that $c_4$ need to satisfy $c_4 = 1-\Omega_{r}-\Omega_m$.

\begin{figure*}\centering
\includegraphics[width=7cm]{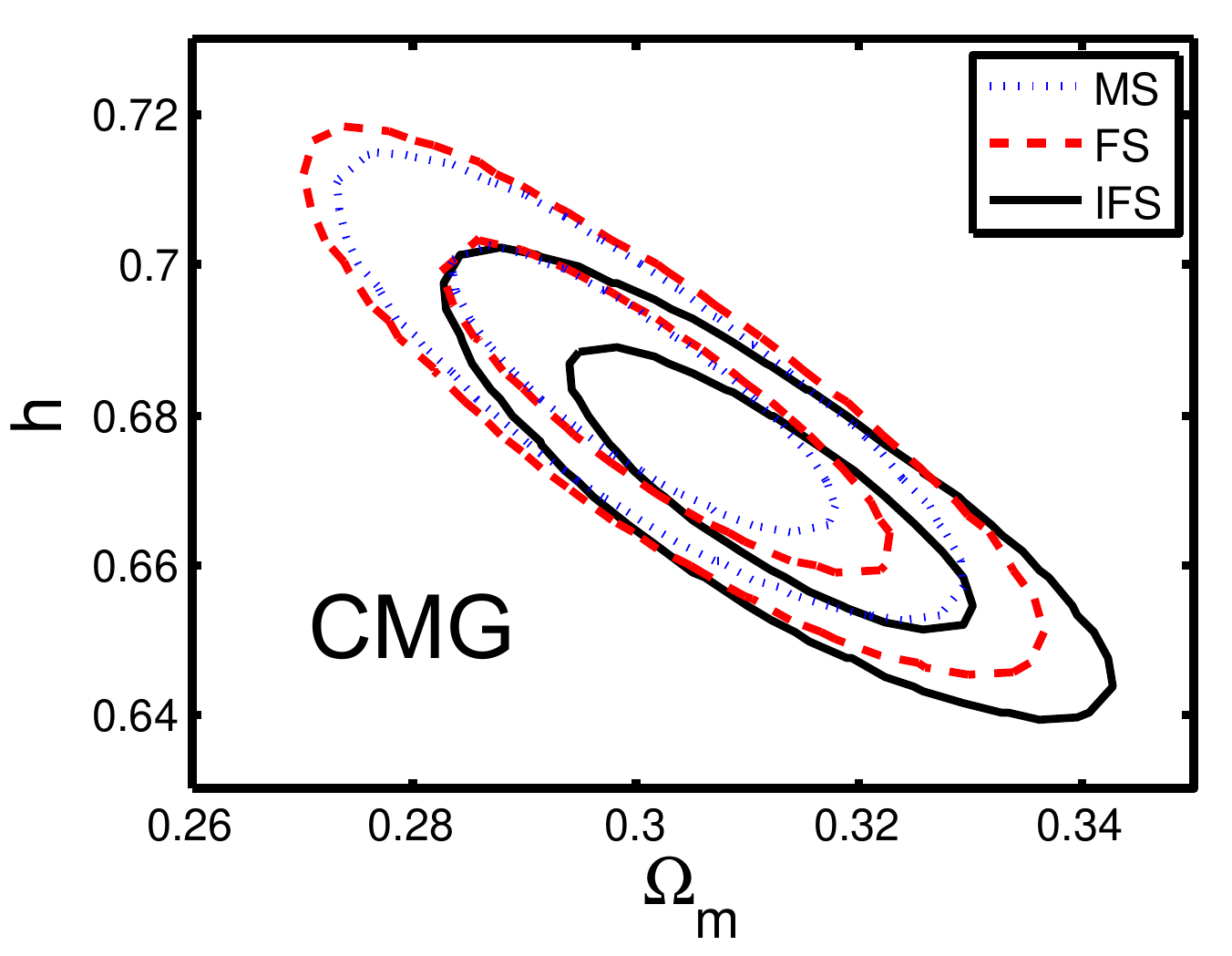}
\includegraphics[width=7cm]{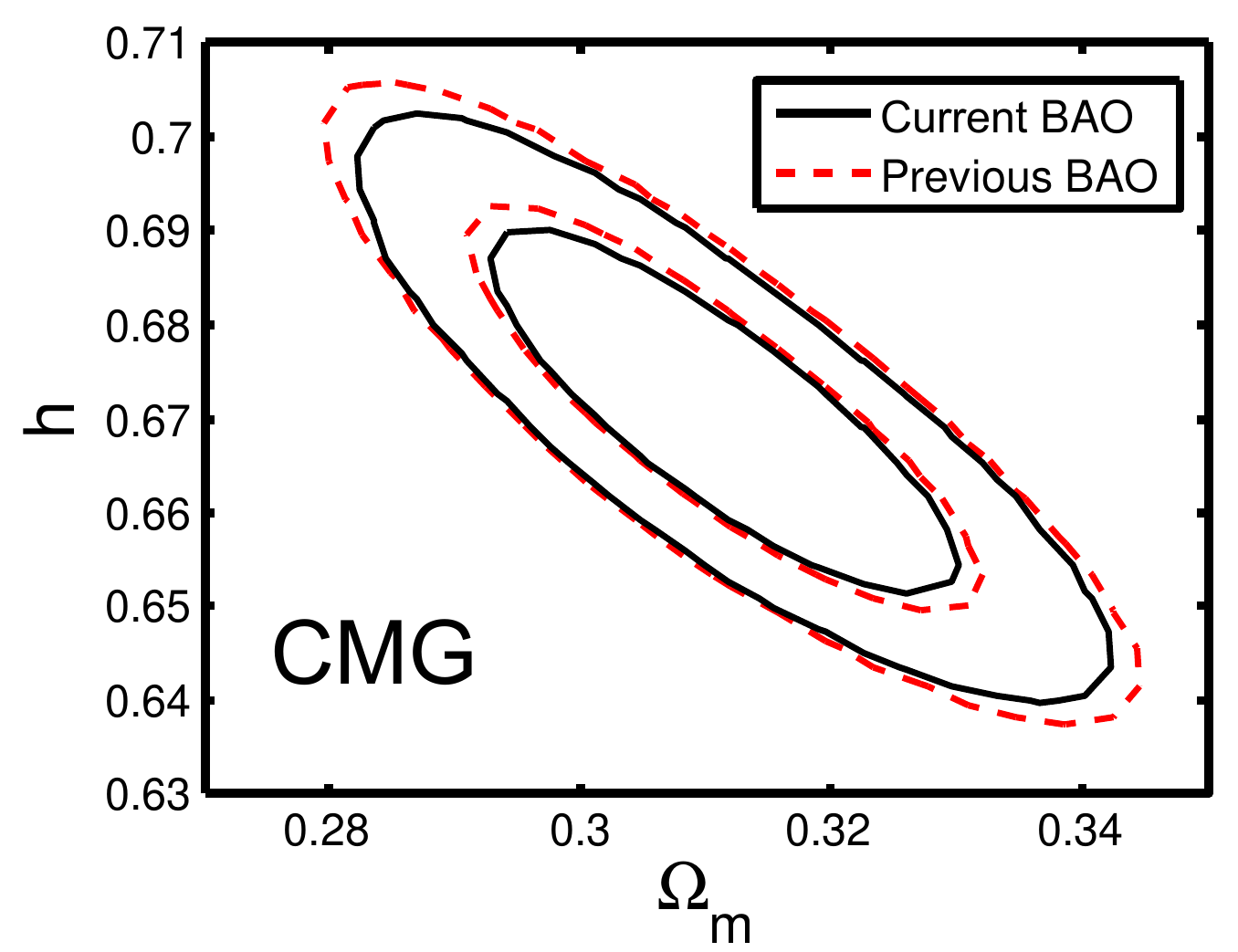}
\includegraphics[width=7cm]{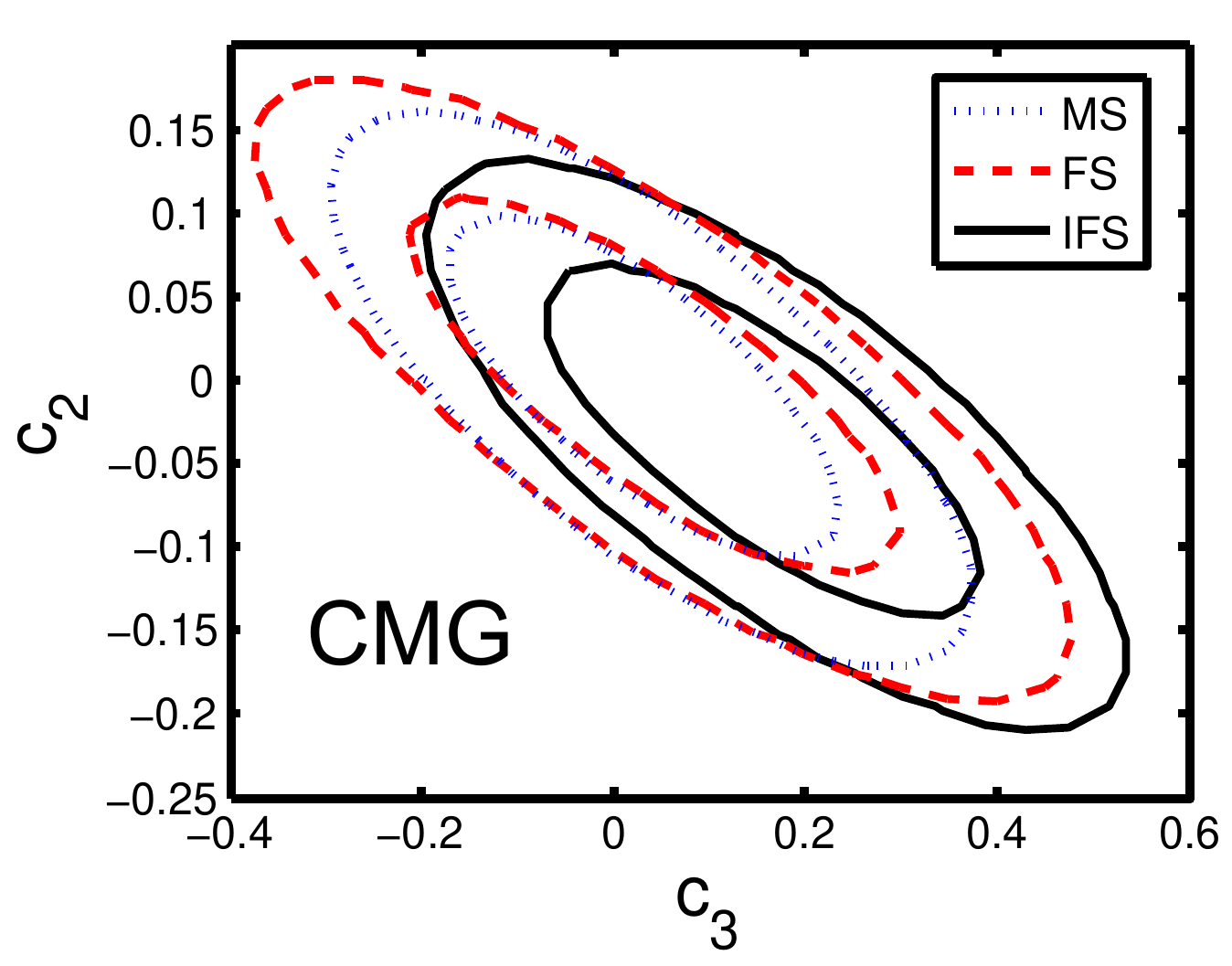}
\includegraphics[width=7cm]{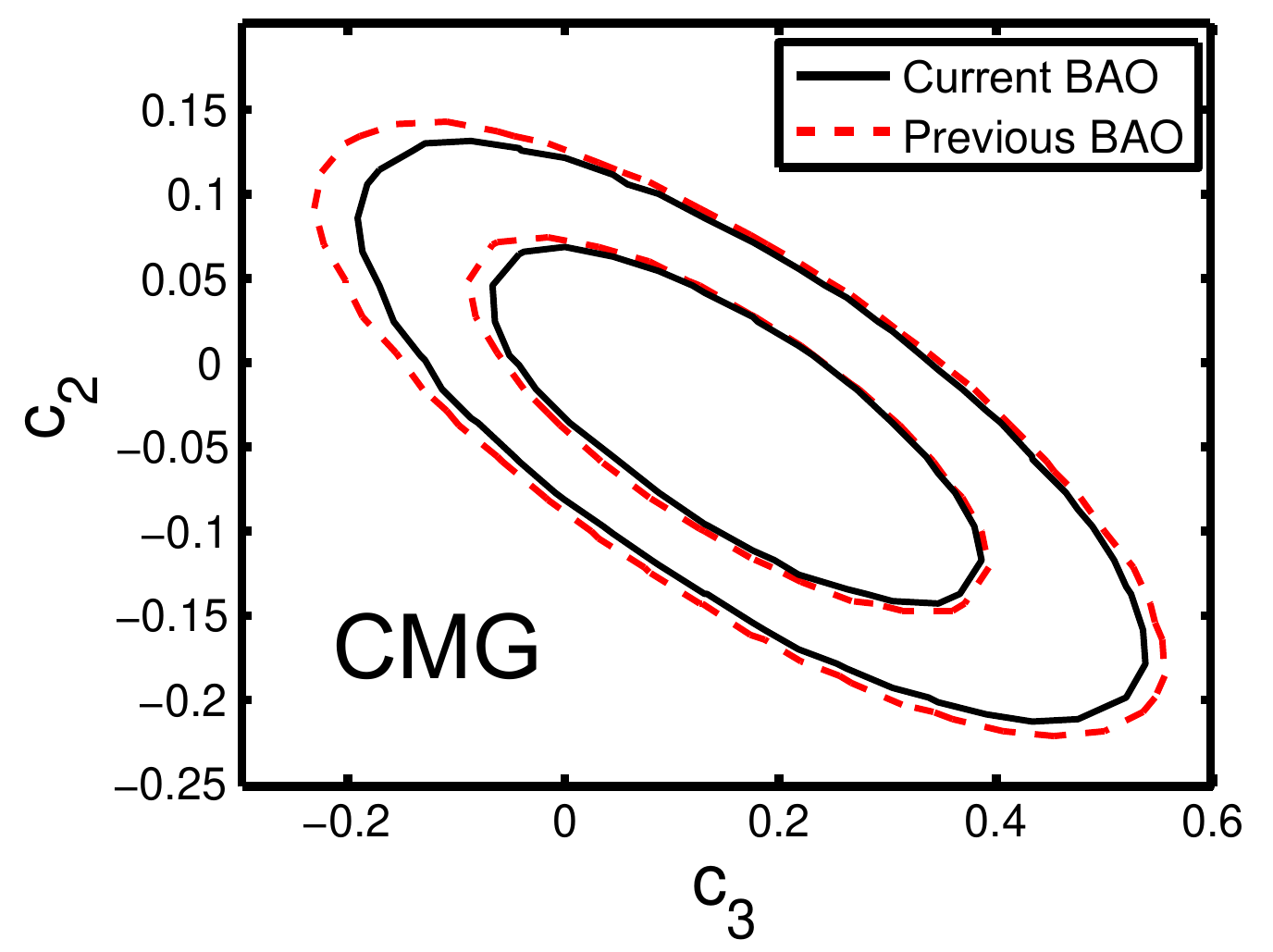}
\caption{\label{f121}CMG: 1$\sigma$ and 2$\sigma$ confidence regions in the $\Omega_{\rm{m}}$--$h$ (upper panels) and $c_2$--$c_3$ (lower panels) planes.
The left panels show the effect of different statistic methods of SNIa, where current BAO data is used.
The blue dotted lines denote the results of MS, the red dashed lines represent the results of FS, and the black solid lines are the results of IFS.
The right panels show the effect of different BAO data, where the IFS is used.
The black solid lines denote the results of current BAO data, and the red dashed lines represent the results of previous BAO data.}
\end{figure*}
We  plot 1$\sigma$ and 2$\sigma$ confidence regions in the $\Omega_{\rm{m}}$--$h$ (upper panels) and $c_2$--$c_3$ (lower panels) planes in Fig.~\ref{f121}.
The impact of different statistic methods and different BAO data for the this model on parameters estimation are similar to the case of the $\Lambda$CDM model.
IFS, which yields bigger $\Omega_{\rm{m}}$, can help to deduce the tension between the SNIa and the other measurements.
From the lower panels,
we find that ($c_2=0$, $c_3=0$) is in the 1$\sigma$ range of the contours, indicating that
the $\Lambda$CDM limit of this model is favored.

\subsubsection {Vacuum metamorphosis model}

\begin{figure*}\centering
\includegraphics[width=7cm]{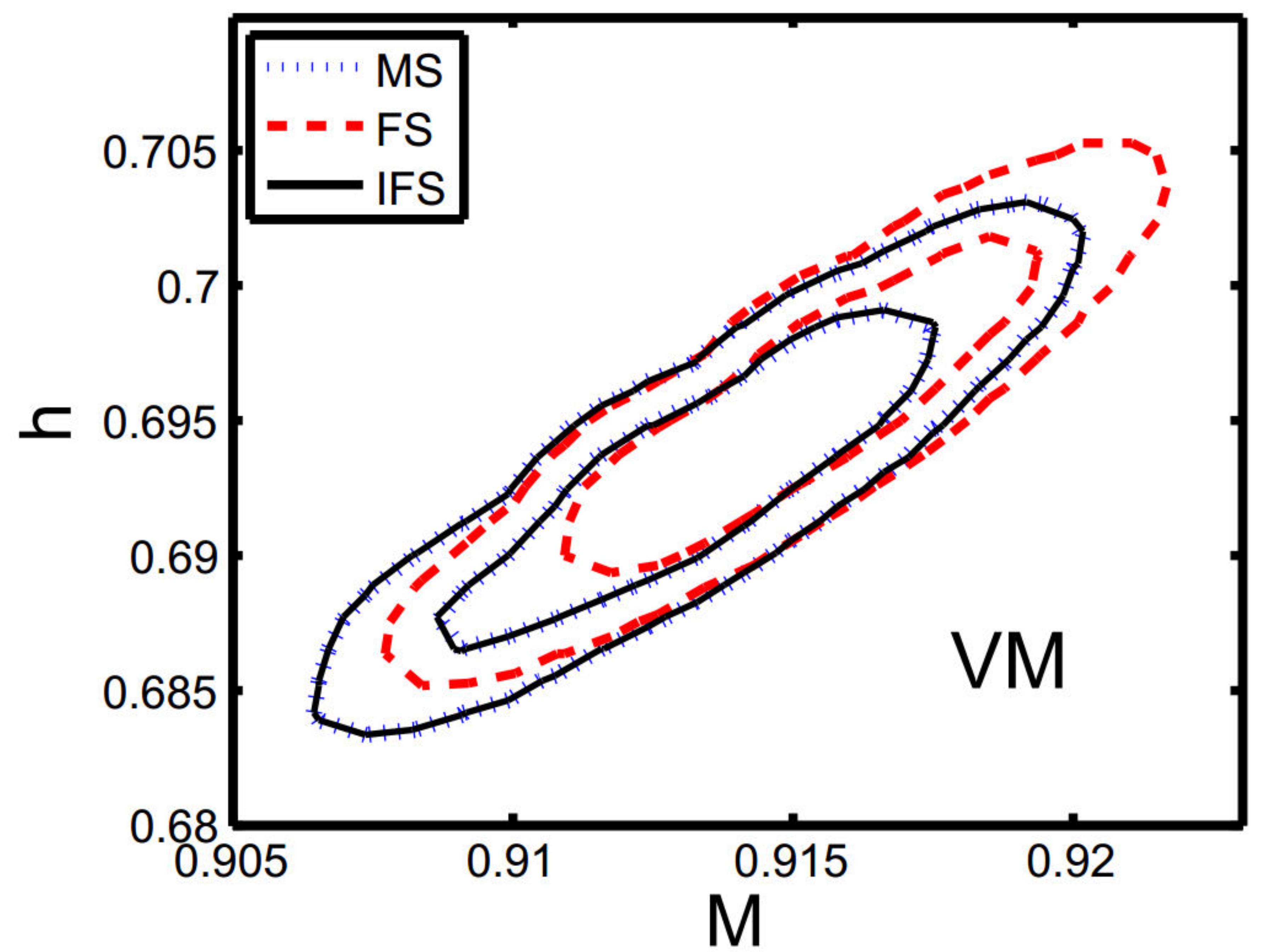}
\includegraphics[width=7cm]{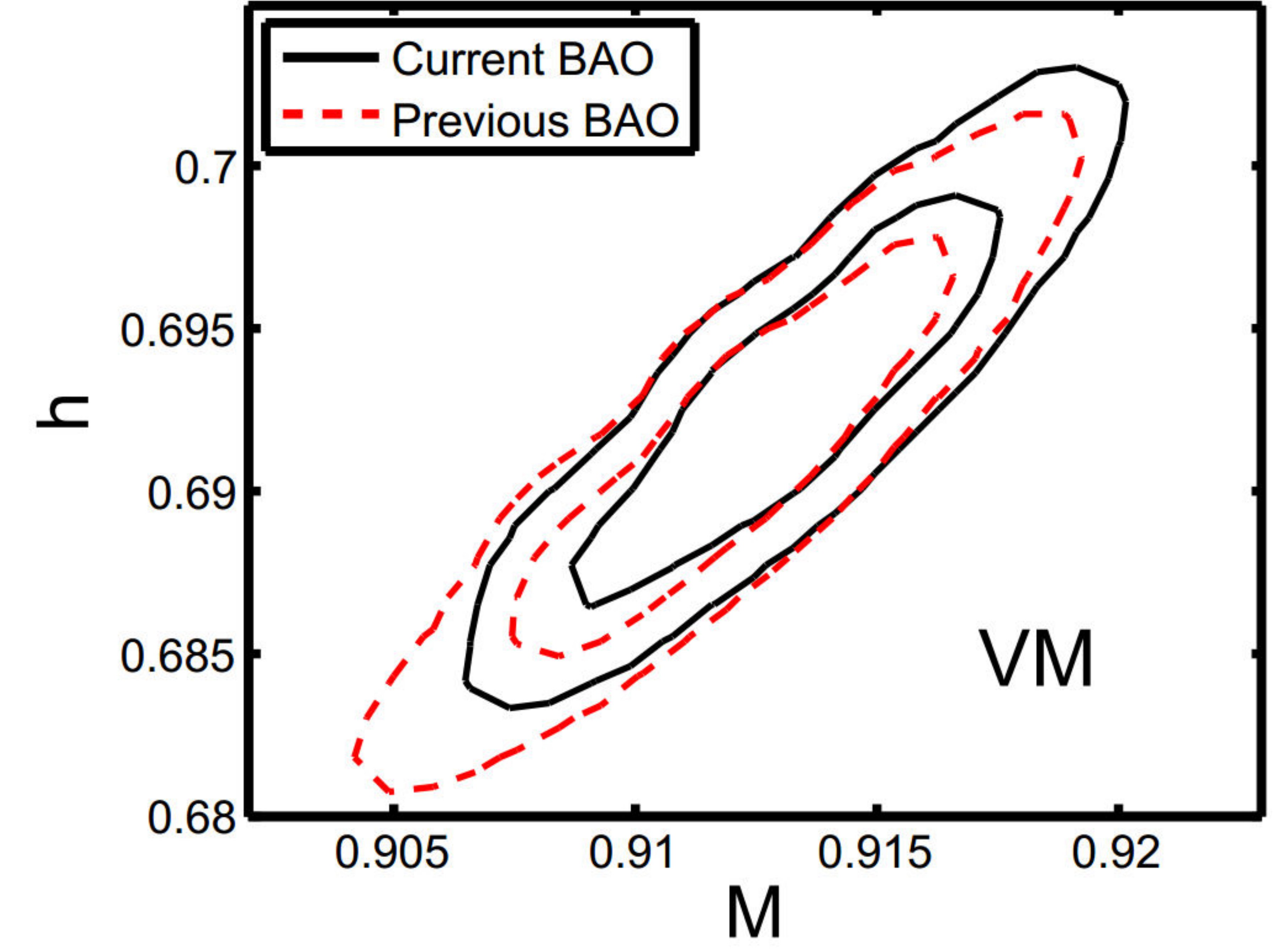}
\caption{\label{f14} VM: 1$\sigma$ and 2$\sigma$ confidence regions in the $M$--$h$ plane.
The left panels show the effect of different statistic methods of SNIa, where current BAO data is used.
The blue dotted lines denote the results of MS, the red dashed lines represent the results of FS, and the black solid lines are the results of IFS.
The right panels show the effect of different BAO data, where the IFS is used.
The black solid lines denote the results of current BAO data, and the red dashed lines represent the results of previous BAO data.}
\end{figure*}

{\bf The vacuum metamorphosis model take into account quantum loop corrections to gravity in the presence of a massive scalar field\cite{Valentino2018}.The phase transition is induced once the Ricci scalar curvature R has evolved to become of order the mass squared of the field, and thereafter R is frozen to be of order $m^2$.As the result,the phase transition criticality condition is
\begin{equation}
R=6(\dot{H}+H^2)=m^2,
\end{equation}
with the difining $M = m^2/(12H_0^2)$,the Friedmann equation is modified as
\begin{equation}
   E(z)^2= \left\lbrace
   \begin{array}{ll}
    \Omega_m(1+z)^3+\Omega_r(1+z)^4+M\{1-[3(\frac{4}{3\Omega_m})^4M(1-M)^3]^{-1}\} &\quad \text{if}\quad  z>z_t\\
    (1-M)(1+z)^4+M &\quad \text{otherwise.}
    \end{array}
    \right.
\end{equation}
The phase transition occurs at
\begin{equation}
z_t=-1+\frac{3\Omega_m}{4(1-M)},
\end{equation}
in this model,we can calculate
\begin{equation}
\Omega_m=\frac{4}{3}[3M(1-M)^3]^{1/4}.
\end{equation}
So, there is only one free parameter in the original model. Here we use more common parameter M and plot 1$\sigma$ and 2$\sigma$ confidence regions in the $M$--$h$ plane for different statistic methods(left panek) and different BAO data(right panel) in Fig.\ref{f14}.}

\section{ Model comparison and more discussions about different statistic methods of SNIa }\label{sec:concl}

In this section,
we assess the {\bf thirteen} DE models by applying the information criteria, the AIC and the BIC.
Moreover, we further study the effect of different statistic methods of SNIa,
and check which one can give tighter cosmological constraints, by making use of AIC, BIC, and FoM.

\subsection{ Model comparison }

\begin{figure*}\centering
\includegraphics[width=7cm]{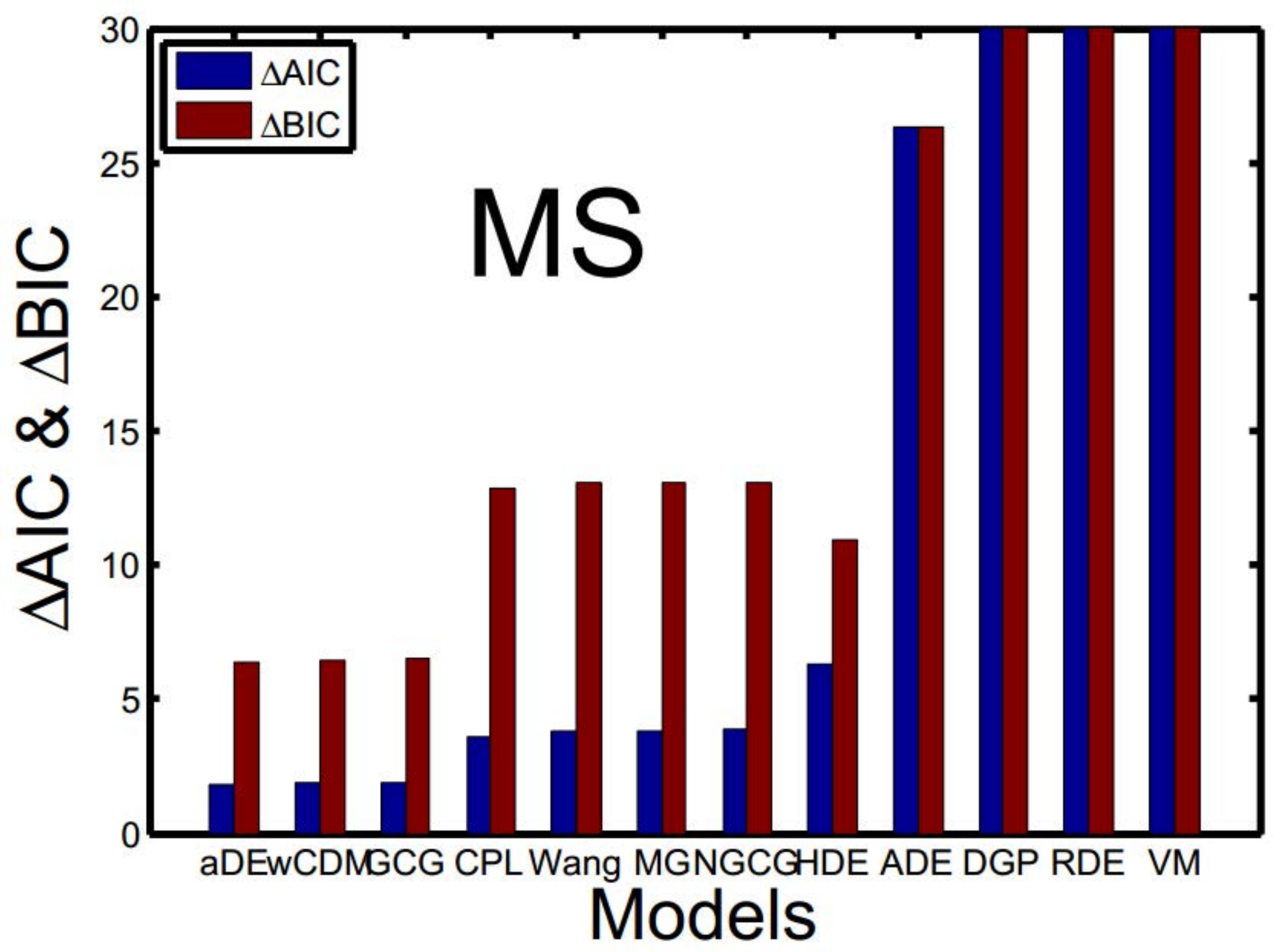}
\includegraphics[width=7cm]{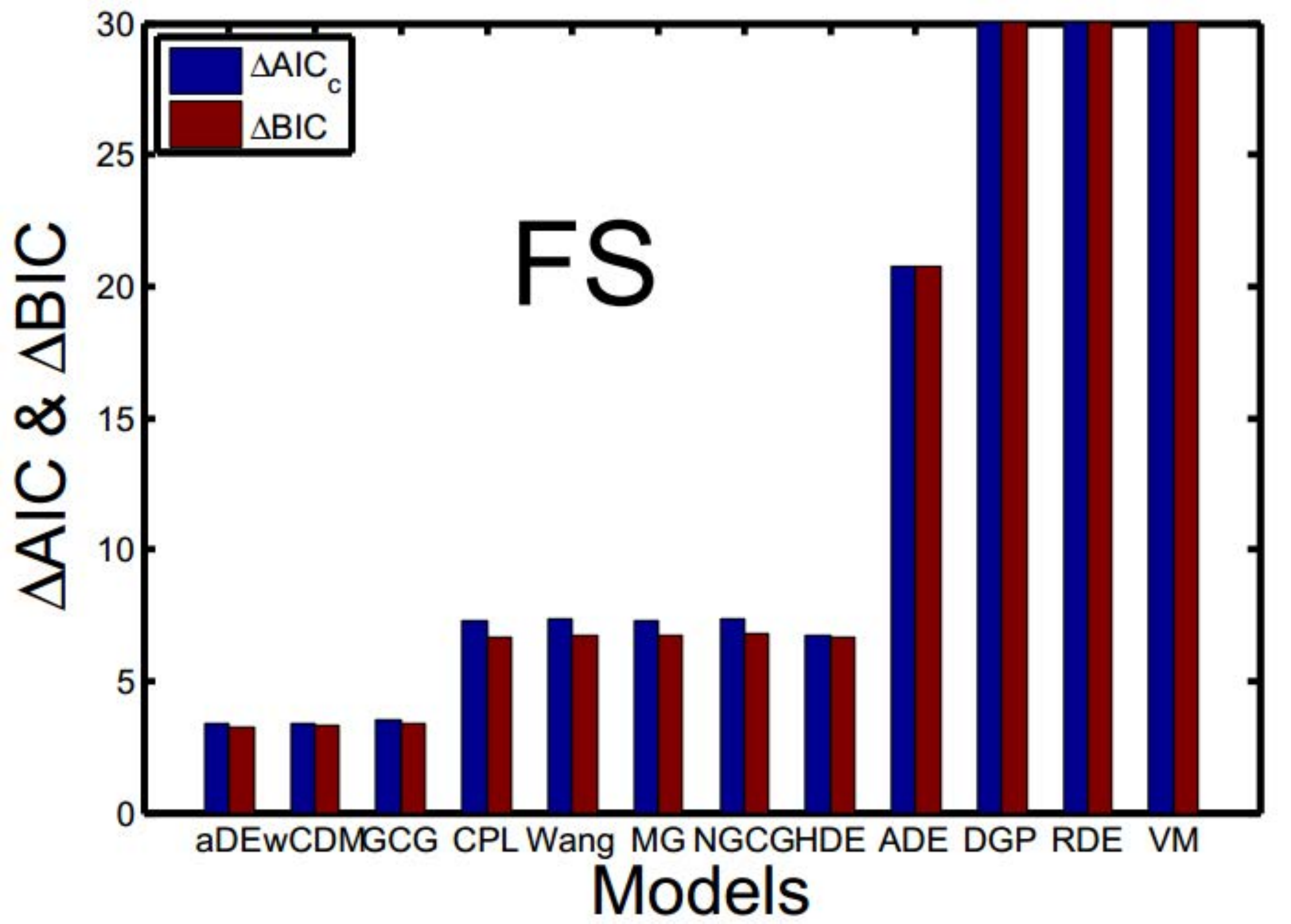}
\hspace{0.1\columnwidth}
\includegraphics[width=7cm]{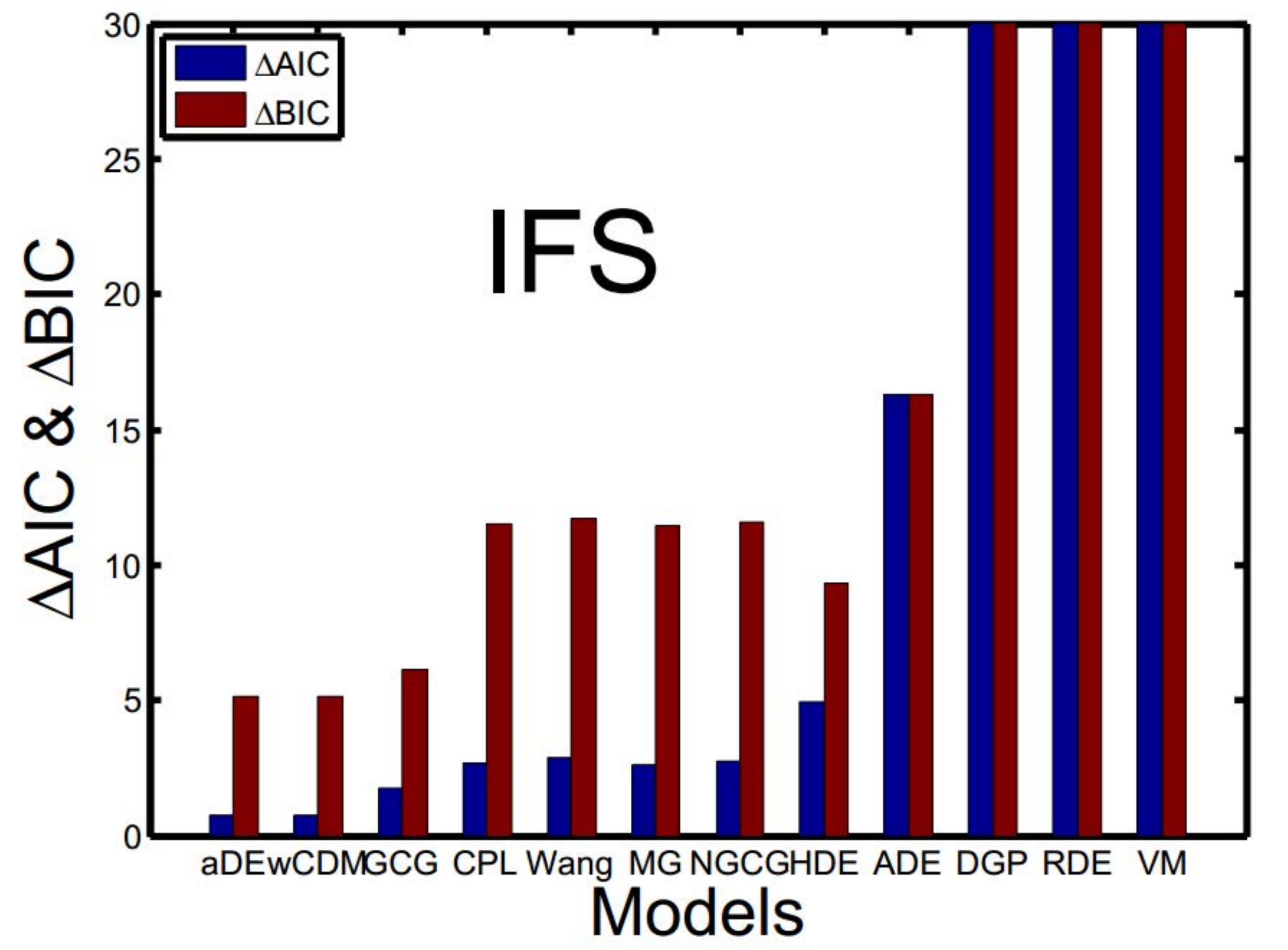}
\caption{\label{AICBIC}Graphical representation of model comparison results. Note that current BAO data is used.
The values of $\Delta$AIC  and $\Delta$BIC for each model correspond to the results given in table~\ref{tab:result}.
The upper left panel represents the results of MS, the upper right correspond to the results of FS, and the lower panel denotes the results of IFS.}
\end{figure*}

Now, let us make a comparison for the {\bf thirteen} DE models.
In Fig.\ref{AICBIC}, we plot the histogram of  $\Delta$AIC and $\Delta$BIC results for each model.
The upper left panel represents the results of MS, the upper right corresponds to the results of FS, and the lower panel denotes the results of IFS.

From Fig.\ref{AICBIC}, we find the {\bf thirteen} DE models can be divided into four grades.
$\Lambda$CDM is in grade one, because it yields the lowest value of AIC and BIC among the {\bf thirteen} DE models.
Therefore, in this work, the $\Delta$AIC and $\Delta$BIC values for all other models are measured with respect to $\Lambda$CDM.
Three models, including $w$CDM,$\alpha$DE and GCG, belong to grade two.
Due to one extra parameter, they are punished by the information criteria.
Five models, including CPL, Wang, CMG, NGCG and HDE, are in grade three.
Note that CPL, Wang, CMG and NGCG can all be reduced to $\Lambda$CDM;
due to two extra parameters, they are further punished by the information criteria.
HDE can not be reduced to $\Lambda$CDM, but it fits the data as well as CPL, Wang, NGCG and CMG, based on BIC results.
{\bf Finally, ADE, DGP, VM and RDE are in grade four.
These four models are excluded by current observational data.}
This conclusion is also in accordance with Ref. \cite{zhangxin2016}.
It is clear that the grades of these model are insensitive to using different statistic methods of SNIa or using different BAO data.

\subsection{ Effect of different statistic methods of SNIa.}
Now, let us turn to the further discussions about the effect of different statistic methods of SNIa.

In Table~\ref{tab:result}, we list the results of $\chi^2_{min}$, $\Delta$AIC and $\Delta$BIC for all the models, where three kinds of statistic methods of SNIa are all taken into account.
Note that current BAO data is always used in the analysis.
Similar to Fig. \ref{AICBIC}, here $\Delta$AIC and $\Delta$BIC
values for all other models are measured with respect to $\Lambda$CDM.
Moreover, these models are arranged in an order of increasing $\Delta$AIC for the case of MS.
From this table, we find that adopting the IFS technique yields the smallest value of $\Delta$AIC for all the models.
In addition, adopting the FS technique yields the smallest value of $\Delta$BIC (except for ADE and DGP),
because of the dramatic decrease of the numbers of SNIa for the FS case.

Moreover, we also discuss the effect of different statistic methods of SNIa by considering the FoM.
FoM are mainly used to assess the observational constraints on the EoS parameters of DE parametrization.
Therefore, in table \ref{tab:fom}, we list the value of  FoM for CPL and Wang parametrization.
Note that current BAO measurement is used in this analysis.
From the table we can find that, for both CPL and Wang parametrization, adopting the IFS yields the greatest value of FoM.
This means that the IFS can give tighter DE constraints, compared with the case of MS and FS.
It should be mentioned that, using previous BAO measurement, this conclusion still come into existence.

\begin{table*}
\caption{Summary of the information criteria results. Note that current BAO data is used in this analysis.  $\Lambda$CDM is preferred
by both AIC and BIC. Thus, $\Delta$AIC and $\Delta$BIC
values for all other models in the table are measured with respect
to this model. These models are arranged in an order of increasing
$\Delta$AIC for the case of MS. It should be stressed that, for the case of FS, the values of $\Delta$$AIC_c$ are calculated, instead of $\Delta$AIC.}
\label{tab:result}
\begin{tabular*}{\textwidth}{@{}l*{15}{@{\extracolsep{0pt plus12pt}}l}}
\hline\hline &\multicolumn{3}{c}{MS}&\multicolumn{3}{c}{FS}&\multicolumn{3}{c}{IFS} \\
 \cline{2-4}\cline{5-7}\cline{8-10}
Model  & $\chi^2_{min}$ &  $\Delta$AIC   &  $\Delta$BIC  &  $\chi^2_{min}$ &  $\Delta$$AIC_{c}$   &  $\Delta$BIC &  $\chi^2_{min}$ &  $\Delta$AIC   &  $\Delta$BIC  \\
\hline
$\Lambda$CDM      & 684.880   & 0 &   0 & 15.066  & 0 &   0  & 517.151   & 0 &   0  \\
$\alpha$DE   & 684.666   & 1.786	&  6.405 & 14.940	&	3.389 &3.275 & 515.890   & 0.739  &  5.160\\

$w$CDM       & 684.736	   & 1.856	 &  6.475 & 14.968 &	3.417 &	3.304 & 515.901    & 	0.750  &  5.172 \\
GCG        & 684.792		& 1.913		&  6.532	 & 15.066&	3.515 &	3.401 & 516.893    & 1.741  &  6.163 \\

CPL      & 684.490	 & 3.610	   & 12.848	  & 14.947	 &	7.263  &6.684 & 515.812    & 2.661  &  11.504 \\

Wang      & 684.695	 & 3.815	   & 13.053	  & 15.014	 &	7.330 &6.751 & 516.040    & 2.889  &  11.732 \\

CMG       & 684.724	& 3.845		 &  13.082	 & 14.969	&	7.285 &6.706 	& 515.780   & 2.629  &  11.472 \\

NGCG       & 684.745	   & 3.866		  &  13.103		& 15.078	&	7.349 &6.814 & 515.920    & 2.769 &  11.612\\

HDE        & 689.221   & 6.341	 &  10.960	 &18.302 &	6.751 &	6.637  & 520.082   & 4.930  &  9.352 \\
ADE        & 711.226	 & 26.347  &  26.347  & 35.847 &	20.781	&20.781 & 533.432    & 16.280&  16.280\\
DGP         & 742.109   & 57.230	 & 57.230  & 63.648&	48.582&	48.582  & 560.285    & 43.134  &  43.134 \\
VM       &783.335      & 98.455     &    98.455        &97.415   &82.571&82.349   &619.254    &102.103     &102.103\\
RDE      & 806.530 	   & 123.651	  &  128.269	& 86.736 &75.186 &	75.072 & 630.948   & 115.796  &  120.218 \\

\hline
\end{tabular*}
\end{table*}

\begin{table*}
\tiny
  \caption{The value of FoM for CPL and Wang parametrization. Current BAO data is used in this analysis. }

\label{tab:fom}
\scriptsize
\setlength\tabcolsep{10.8pt}
\renewcommand{\arraystretch}{1.5}
\centering
\begin{tabular*}{\textwidth}{@{}l*{15}{@{\extracolsep{0pt plus12pt}}l}}
\\
\hline\hline &\multicolumn{3}{c}{CPL}&\multicolumn{3}{c}{Wang} \\
 \cline{2-4}\cline{5-7}
  & MS & FS & IFS & MS & FS & IFS \\ \hline

 $FoM$             & $40.27$
                   & $33.59$
                   & $49.11$
                   & $146.73$
                   & $79.51$
                   & $148.68$\\ 

\hline
\end{tabular*}
\end{table*}

From Table \ref{tab:result} and Table \ref{tab:fom}, one can see that, adopting IFS not only yields the lowest value of $\Delta$AIC for all the models, but also gives tighter DE constraints for CPL and Wang parametrization.
This implies that adopting the IFS technique yields the strongest constraints on DE.
In other words, the IFS technique has the strongest constraint ability.
Therefore, it is very important to take into account the systematic uncertainties of SNIa during the cosmology-fits seriously.

\section{Conclusions and Discussions}\label{sec:conc2}

In this work, we have tested {\bf thirteen} DE models against current observational data.
The observational data include the JLA samples of SNIa observation \citep{Betoule2014}, the BAO observation from the  SDSS DR12 \citep{Alam2016}, and the CMB distance priors from the Planck 2015 \citep{Planck201514}.
To make a more systematic and comprehensive comparison, the following two factors are considered:
for the SNIa data, we have taken into account three kinds of statistics methods of SNIa, including the MS, FS and IFS technique;
for the BAO data, we have made a comparison by using two previous BAO measurement extracted from the SDSS at $z = 0.35$ (SDSS DR7 \citep{CW12}) and $z = 0.57$ (BOSS DR11 \citep{Anderson14}).

We have introduced the basic information of {\bf thirteen} DE models, which can be divided into the following five classes: 1) Cosmological constant model; 2) DE models with parameterized EOS;
3) Chaplygin gas models; 4) Holographic dark energy models; 5) Modified gravity models.
For each DE model, we have plotted the 2$\sigma$ confidence regions of  various model parameters to study the impacts of different statistics methods of SNIa and different BAO data on parameter estimation.
Moreover, in order to assess the worth of each model, we have plotted the histogram of $\Delta$AIC and $\Delta$BIC results for all the models.
Finally, to study the effect of different statistics methods of SNIa, we have listed the results of $\chi^2_{min}$, $\Delta$AIC and $\Delta$BIC for each model, as well as the FoM results for CPL and Wang parametrization.
Based on our analysis, we find that:
\begin{itemize}

\item
Cosmological observations can divide the {\bf thirteen} models into four grades. (see Fig.\ref{AICBIC}).
$\Lambda$CDM,  which is most favored by current observations, belongs to grade one;
$\alpha$DE, $w$CDM and GCG belong to grade two;
CPL, Wang, CMG, NGCG and HDE belong to grade three;
{\bf ADE, DGP,VM and RDE, which are excluded by current observation, belong to grade four}\footnote{{\bf There is a debate that less parameters may not mean necessarily a less complicate model. For example a cosmological constant could be explained by invoking a multiverse, while a varying equation of state could be explained by a scalar field. Therefore, if one compare these models by using their $\chi^2$ values only, CPL will become the best model.}}.

\item
For parameter estimation, adopting the IFS technique yields
the biggest $\Omega_m$ and the smallest $h$ for all the models.
In contrast, Using different BAO data does not cause significant effect (see Fig. \ref{f11}-\ref{f121}).

\item
The IFS technique has the strongest constraint ability on various DE models.
For examples, adopting the IFS technique yields the smallest value of $\Delta$AIC for all the models (see Table \ref{tab:result});
in addition, making use of this technique yields the biggest FoM values for CPL and Wang parametrizations (see Table \ref{tab:fom}).
\end{itemize}

In summary, $\Lambda$CDM still has the best capability to explain current observations;
this conclusion is independent of adopting different statistics methods of SNIa or using different BAO data.
In addition, the IFS technique can give the strongest constraints on various DE models and can cause an obvious impact on parameter estimation,
this implies that it is very important to take into account the systematic uncertainties of SNIa during the cosmology-fits seriously.

In addition to analysing specific DE models, another popular way of studying DE is considering the model-independent DE reconstructions.
The most commonly used reconstructions include the binned parametrization \citep{Huterer2003,Huang2009,wll2011,LiXD2011}
and the polynomial fitting \citep{Alam2003,YunWang2009}.
It is interesting to constrain these DE reconstructions by taking into account the systematic uncertainties of SNIa.

In a recent work, Ma, Corasaniti and Bassett proposed a new  method to explore the JLA data,
by applying Bayesian graphs \citep{Bassett2016}.
They found that the error bars of various model parameters can be significantly reduced by using this analysis technique.
It will be very interesting to revisit the statistical method of Ref. \citep{Bassett2016}
by taking into account the systematic uncertainties of SNIa simultaneously.
This will be done in a future work.

\begin{acknowledgments}
{\bf We are grateful to the referee for the helpful suggestions.}
SW is supported by the National Natural Science Foundation of China under Grant No. 11405024
and the Fundamental Research Funds for the Central Universities under Grant No. 16lgpy50.

\end{acknowledgments}


\label{lastpage}

\end{document}